\documentclass[aps,pra,superscriptaddress,floatfix,10pt,floatfix,twocolumn]{revtex4-2}
\usepackage{graphicx,graphics}
\usepackage{amsmath,amssymb,amsfonts}
\usepackage{latexsym,verbatim}
\usepackage{bm}
\usepackage{color}
\usepackage{cancel}
\usepackage{xcolor}
\usepackage{ulem}
\usepackage{multirow}
\usepackage{comment}
\usepackage{overpic}
\usepackage{array}
\usepackage{nicefrac}
\usepackage{setspace}
\usepackage{physics}
\usepackage{bbold}
\usepackage{soul}
\usepackage{enumerate}
\usepackage{float}
\usepackage{comment}
\usepackage{mathtools}
\usepackage{dcolumn}
\usepackage{titlesec}
\usepackage{placeins}
\usepackage{dblfloatfix}

\titlespacing*{\section}{0pt}{1.5em}{0.5em}
\titlespacing*{\subsection}{0pt}{1.5em}{0.5em}
\titlespacing*{\subsubsection}{0pt}{1.5em}{0.5em}

\usepackage[breaklinks=true,colorlinks,citecolor=blue,linkcolor=blue,urlcolor=blue]{hyperref}

\newcommand{\icol}[1]{
	\left(\begin{smallmatrix}#1\end{smallmatrix}\right)%
}

\begin{document}
	
	\title{
		Formal Integration of Electron Scattering Processes via Separation of Dynamical and Geometric Contributions
	}

	\author{Lorenzo Bagnasacco}
	\email{lorenzo.bagnasacco@sns.it}
	\affiliation{NEST, Scuola Normale Superiore and Istituto Nanoscienze-CNR, Piazza dei Cavalieri 7, I-56126 Pisa, Italy}
	
	\author{Fabio Taddei}
	\affiliation{NEST, Scuola Normale Superiore and Istituto Nanoscienze-CNR, Piazza dei Cavalieri 7, I-56126 Pisa, Italy}
	
	\author{Vittorio Giovannetti}
	\affiliation{NEST, Scuola Normale Superiore and Istituto Nanoscienze-CNR, Piazza dei Cavalieri 7, I-56126 Pisa, Italy}

	\begin{abstract}
		By decoupling the geometric from the dynamical contributions in the scattering processes, we develop a method to compute the scattering matrix of electrons in a one-dimensional coherent conductor connected to two electrodes. In particular, we demonstrate that, in the high-energy regime, the transmission matrix converges to the Berry operator of the system. We showcase the method through several examples featuring different in-plane magnetic field profiles. Notably, our results reveal the possibility of achieving near-perfect spin-flip transmission, highlighting potential applications in spintronics.
	\end{abstract}

	\maketitle

	\section{Introduction}
	Understanding the scattering properties of electrons in mesoscopic systems is essential for advancing spin-dependent transport technologies~\cite{spin-tech-1,spin-tech-2,spin-tech-3}. Such properties are encoded in the scattering matrix, a central object of quantum transport theory~\cite{smatrix-1,smatrix-2,smatrix-3,qtrans-1,qtrans-2,qtrans-3} that relates incoming to outgoing states in the presence of inhomogeneities and magnetic textures. Beyond conductance quantization, the scattering matrix captures subtle quantum interference and geometric effects, which are key ingredients in spintronics and topological transport~\cite{recent-spintronics,das-sarma-spintronics}.
	
	Recent years have seen a surge of interest in transport phenomena arising from spatially varying magnetic fields and textures. Early proposals such as the Datta–Das spin transistor~\cite{datta-das} demonstrated the control of electron spin via engineered precession. Charge and spin pumping through domain walls and magnetic textures has been widely investigated~\cite{textures-pumping}, and the scattering of electrons on skyrmion configurations revealed striking consequences for spin transport~\cite{skyrmion-scatt}. 
	Nontrivial spin dynamics appear through Landau–Zener and Rabi oscillations in spin-dependent conductance~\cite{EPL-LZ-Rabi}, and in mesoscopic spin transport where adiabaticity and nonadiabatic corrections can be tuned by disorder and geometry~\cite{meso-adiabatic,dollinger-thesis}. On the theoretical side, Kato’s seminal work on adiabatic evolution~\cite{Kato1950} laid the foundation for modern treatments, and more recently Wilson-loop~\cite{Wloop1}
	method have been employed to characterize geometric structure in spin and band systems~\cite{wilson-loop}. These advances demonstrate that both dynamical and geometric aspects of scattering must be treated on equal footing to capture the physics of realistic devices.
	
	In this work, we present a novel method—building on Ref.~\cite{Cusumano2020}—for computing the spin-resolved scattering matrix of a coherent conductor, within the well-known Landauer-B\"{u}ttiker formalism~\cite{shot-noise}. The key innovation lies in separating the geometric Berry operator~\cite{note-berry-op} of the system from the dynamical evolution in the scattering dynamics. Differently from Ref.~\cite{Cusumano2020}, which assumes the particle energy to be infinite, our method is exact
	\begin{figure}[H]
		\centering
		\includegraphics[width=\linewidth]{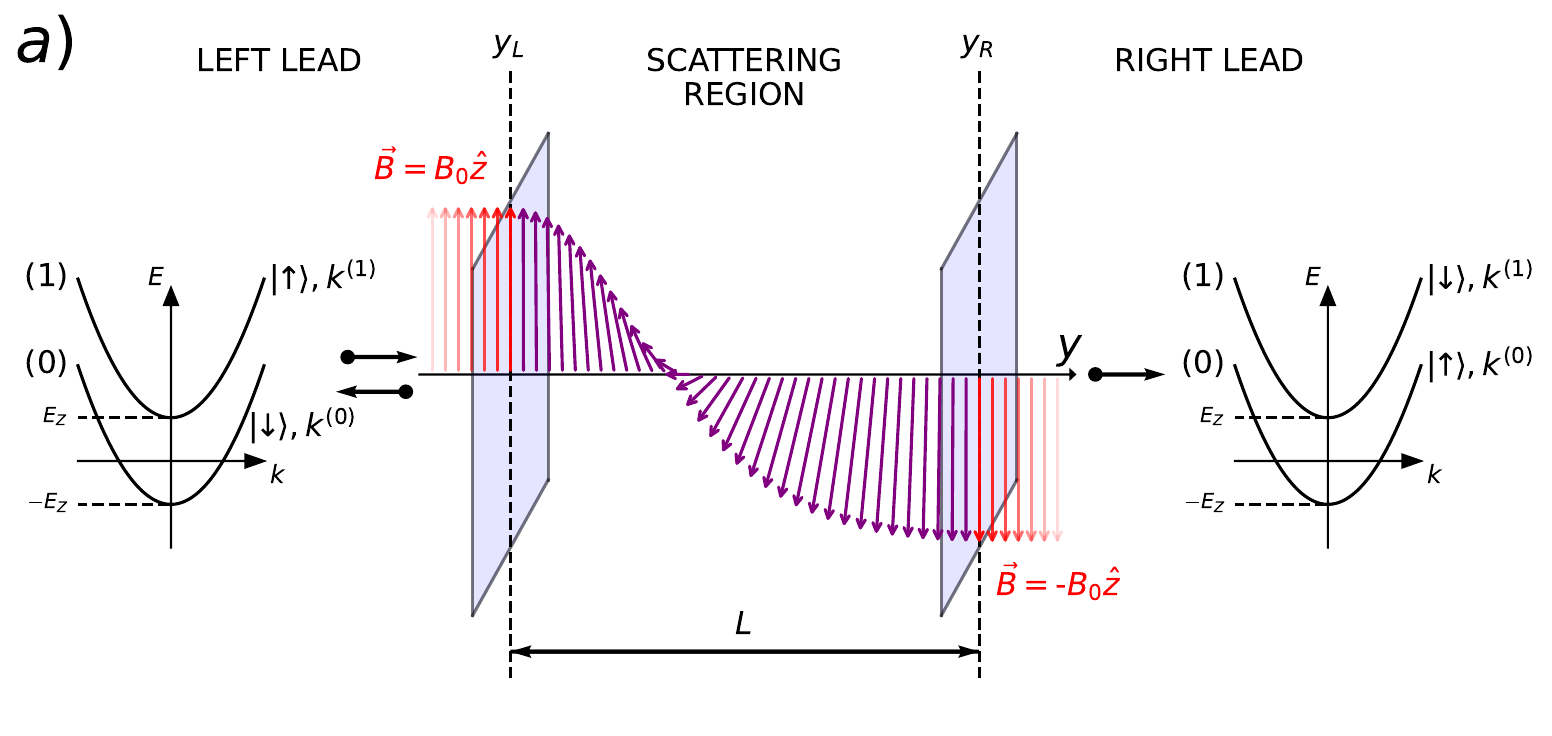}
		\includegraphics[width=\linewidth]{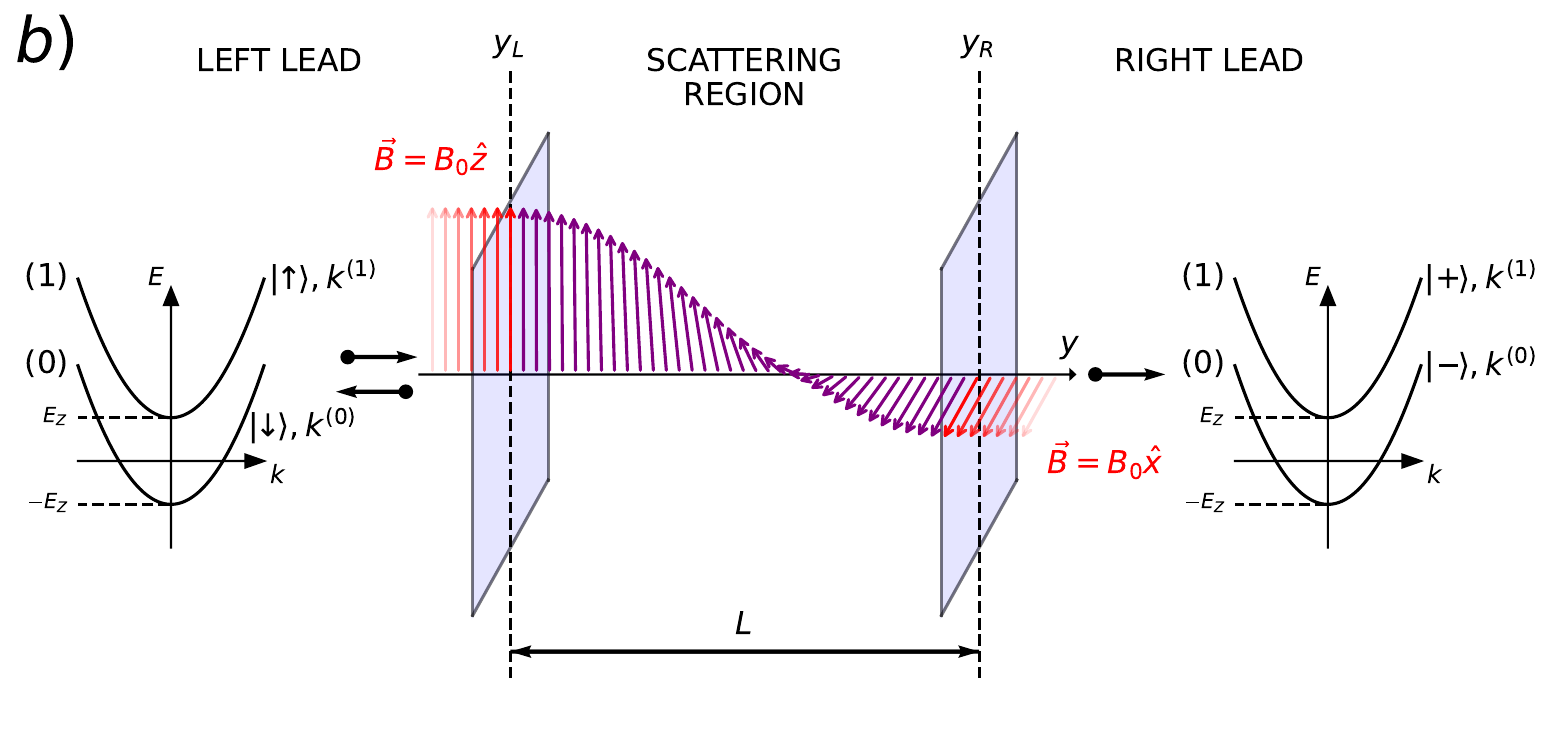}
		\caption{
			Schematic of the model: electrons propagates on a 1D wire parametrized by the longitudinal coordinate $y$, under the action of an external, static magnetic field.  
			The wire is composed by three distinct regions: the left lead ($y \leq y_{\rm L}$), the right lead  ($y_{\rm R} \leq y$) where
			the magnetic field is uniform, assuming constant values  $\bm{B}_{\rm L}$ and  $\bm{B}_{\rm R}$, respectively, and the scattering region ($y_{\rm L}<y<y_{\rm R}$)  where instead
			the magnetic field vector $\bm{B}(y)$ (represented by purple vectors) can vary.
			Panel \textsl{a)}: The plot shows the $y$-varying magnetic field $\bm{B}(y)$ of the scheme I of Sec.~\ref{sec:expI},
			with  components given in \eqref{eq:bzI} setting $q_1=0$ and $q_2=0$. In this specific case, the magnetic field has only a positive $z$ component in the left lead and a negative $z$ components in the right lead.
			Panel \textsl{b)}: The plot shows the $y$-varying magnetic field $\bm{B}(y)$ of the scheme II of Sec.~\ref{sec:expII}, with components given by Eq.~\eqref{eq:bzII} setting $q_1=0$ and $q_2=0$. In this  case, the magnetic field has only a positive $z$ component in the left lead and a negative $x$ components in the right lead. 
			To enhance clarity, we have aligned the directions $\bm{n}_1$ and $\bm{n}_3$, which define the magnetic field orientation along the  wire as described in Eq.~(\ref{defBsca}), with the $x$- and $z$-axis, respectively.
			Dispersion curves for the leads are shown on the sides of the plots.}
		\label{fig:ex1}
	\end{figure}
	\noindent and applicable to arbitrary energies.
	
	As a concrete setup, we consider a 1D wire and calculate the scattering amplitudes which describe the spin-dependent transport in the presence of a spatially varying Zeeman field, see Fig.~\ref{fig:ex1}. This is a typical situation addressed in  {\it spintronics}, the research field which concerns the investigation of spin-dependent transport~\cite{geometricring, adiabaticdisorder, polygonspintronics} and its applications (see the review papers in Refs.~\cite{spintronics-fundamentals-applications,hiroata-spintronics}). In particular, we demonstrate that the  reflection and transmission matrices of the setup can be calculated by solving a differential equation which contains both geometric and dynamical contributions.
	
	We analyze three distinct special regimes, for which the scattering amplitudes admit a closed formula.
	
	To better elucidate these facts we consider two different examples where the vector representing the spatially varying magnetic field lies in a plane. In the first one, the magnetic field in the leads points in opposite directions (see panel \textsl{a)} of Fig.~\ref{fig:ex1}), while in the second it points in 
	orthogonal directions (see panel \textsl{b)} of Fig.~\ref{fig:ex1}). In both cases we consider different profiles of variation of magnetic fields characterized by two parameters we plot the spin-dependent transmission and probabilities as functions of energy $E$ of the impinging electron.
	
	
	We begin by demonstrating that suitable magnetic field profiles can lead to qualitatively different transmission behaviors. For energies $E$ below the Zeeman splitting in the leads, one finds cases where nearly perfect spin-flip transmission occurs (first example), and cases where a spin-down electron is transmitted as a balanced superposition of spin-down and spin-up states (second example). The transmission probabilities depend sensitively on the specific shape of the magnetic field profile, as determined by the two control parameters.
	
	
	The paper is organized as follows:
	in Sec.~\ref{sec:model} we introduce the model;
	in Sec.~\ref{sec:sc} the scattering problem is solved  by decoupling the geometric contribution from the rest; Sec.~\ref{sec:special} focuses on special cases that allow for a fully analytical treatment i.e. the infinite energy limit (Sec.~\ref{large:app}), the infinitesimal scattering region limit (Sec.~\ref{sub:low}), and the piecewise constant regime (Sec.~\ref{sec:piece}); 
	Sec.~\ref{sec:spinholo} is dedicated to compute explicitly the Berry operator for control induced by varying a Zeeman field; in Sec.~\ref{sec:res}  numerical examples are presented to illustrate the results.
	The conclusions are summarized in Sec.~\ref{sec:conclusions}, and a series of appendices provide detailed technical derivations.
	
	\section{Physical model}
	\label{sec:model}
	We consider a minimal theoretical setup where electron dynamics are described within the
	effective-mass approximation~\cite{Datta1997}. In this widely used framework, the kinetic
	term takes a quadratic form, corresponding to a parabolic dispersion relation for the
	propagating modes~\cite{Ando1982}. This approximation is standard in mesoscopic
	transport theory and provides a reliable description over a broad range of energies~\cite{Chiang1979,Kane1957}.
	The Hamiltonian describing the propagation of electrons in a 1D wire  under the action of a position-dependent  magnetic field $\bm{B}(y)$
	writes 
	\begin{eqnarray}
		\hat{\mathcal{H}}&:=&\frac{\hat{p}^2_{y}}{2m}- \bm{\mu}_s\cdot  \bm{B}(y)\;,   \label{eq:H}
	\end{eqnarray}
	with $p_y:= -(i/\hbar) \partial_y$ denoting the electron momentum corresponding to the operator $\hat{p}_y$ along the longitudinal axis $y$ of the device and 
	$\bm{\mu}_s$ the electronic spin magnetic momentum. Indicating with 
	$g(>0)$  the Land\'e factor, $\mu_B(>0)$ the Bohr magneton, and $\hat{\bm{\sigma}}:= (\hat{\sigma}_x, \hat{\sigma}_y,\hat{\sigma}_z)$ the vector of Pauli operators, the second contribution of $\hat{\mathcal{H}}$ can be expressed as 
	\begin{eqnarray}
		\hat{h}_y&:=&- \bm{\mu}_s\cdot  \bm{B}(y)= 
		g \mu_B {B}(y) \hat{{\sigma}}_{\bm{n}(y)}
		\;, 
		\label{eq:zeeman}
	\end{eqnarray}
	where $B(y):=|\bm{B}(y)|$ and $\bm{n}(y):=\bm{B}(y)/B(y)$  indicate respectively the magnitude and the orientation of the magnetic field at position $y$, and  $\hat{{\sigma}}_{\bm{n}(y)} := \bm{n}(y)  \cdot \hat{\bm{\sigma}}$ is the spin operator component along 
	$\bm{n}(y)$. 
	For fixed $y$, the eigenvectors of $\hat{h}_y$ identify the 
	Zeeman spin-eigenstates, 
	\begin{equation}
		\hat{h}_y \ket*{\phi^{(\ell)}_{{\bm n}_y}}={{E}_{{\rm Z}}}_y^{(\ell)} \ket*{\phi^{(\ell)}_{{\bm n}_y}}, \quad  {{E}_{{\rm Z}}}_y^{(\ell)}:= -(-1)^{\ell}  g \mu_B |\bm{B}({y})|\;, 
		\label{eq:Zeemanterm}
	\end{equation}
	with $\ell=\{0,1\}$. In particular, we will refer to $\ket*{\phi^{(0)}_{{\bm n}(y)}}$ as the (local) \textit{lower-energy} Zeeman spin-eigenstate, while to $\ket*{\phi^{(1)}_{{\bm n}(y)}}$ as the (local) \textit{higher-energy} Zeeman spin-eigenstate (recall that $g\mu_B$ is positive). 
	
	In our analysis, we partition the wire into three distinct regions: the left lead ($y \leq y_{\rm L}$), the right lead  ($y_{\rm R} \leq y$), and the scattering region ($y_{\rm L}<y<y_{\rm R}$), see Fig.~\ref{fig:ex1}. In the left and right leads, the magnetic field is uniform, assuming constant values  $\bm{B}_{\rm L}:= \abs{\bm{B}_{\rm L}}  \bm{n}_{\rm L}$ and  $\bm{B}_{\rm R}:=\abs{\bm{B}_{\rm R}}  \bm{n}_{\rm R}$, respectively. 
	We assume the magnitudes of these fields are equal, such that $\abs{\bm{B}_{\rm L}}=\abs{\bm{B}_{\rm R}}=B_0$
	while allowing their orientations, denoted as $\bm{n}_{\rm L}$ and $\bm{n}_{\rm R}$,
	to differ.
	In contrast, the magnetic field in the scattering region is not constant  providing a somehow smooth transition between $\bm{B}_{\rm L}$ and  $\bm{B}_{\rm R}$ (more precisely we shall require $\bm{B}(y)$ be differentiable everywhere with  first derivative that is continuous   at the interface between the leads and the scattering region).
	In virtue of these assumptions
	in the leads the system exhibits spin-resolved energy bands
	\begin{eqnarray} 
		E^{(\ell)}(k) := \frac{\hbar^2 k^2}{2m}+{{E}_{{\rm Z}}}^{(\ell)}\qquad \ell \in\{0,1\}\;,
		\label{eq:energybands}
	\end{eqnarray} 
	where 
	${{E}_{{\rm Z}}}^{(\ell)}:= -(-1)^{\ell} g \mu_B B_0$ are the eigenvalues~(\ref{eq:Zeemanterm})  evaluated outside the scattering region, whose gap 
	\begin{eqnarray} E_{\rm Z}:=\frac{{{E}_{{\rm Z}}}^{(1)}-{{E}_{{\rm Z}}}^{(0)}}{2}=
		g \mu_B B_0\;,\end{eqnarray} 
	defines the fundamental energy scale of the problem.
	Electrons with sufficiently high energy $E$ can occupy these bands. Their redistribution across the scattering region, as well as the current flowing through the device,
	is governed by the transmission and reflection scattering matrices $t_E$ and $r_E$ of the model. Computing them accounts in solving the energy-resolved Schrödinger equation of the setup, i.e.  
	\begin{equation}
		( \hat{\mathcal{H}}-E)\ket*{\psi^{(E)}_y}=    -\frac{\hbar^2}{2m} \partial^2_y\ket*{ \psi^{(E)}_y}+(\hat{h}_y-E)\ket*{\psi^{(E)}_y}
		=0\;,
		\label{eq:xrepresentation}
	\end{equation}
	with $\ket*{ \psi^{(E)}_y}$  the spinor wave-function  associated with energy eigenvalue $E$. 
	In the leads the solution of Eq.~(\ref{eq:xrepresentation}) can be expressed as 
	\begin{equation}
		\ket*{\psi^{(E)}_{y}}=\left\{ \begin{array}{l}\sum_{\ell\in \{0,1\}}\left(A_E^{(\ell)} e^{i k_E^{(\ell)}y}+R_E^{(\ell)}e^{-i k_E^{(\ell)}y}\right) \ket*{\phi^{(\ell)}_{\bm{n}_{\rm L}}}, 
			\\ \qquad  \qquad \qquad \qquad  \qquad \qquad  \qquad  \forall y \leq y_{\rm L} \\\\
			\sum_{\ell\in \{0,1\}} T_E^{(\ell)} e^{i k_E^{(\ell)}y}\ket*{\phi^{(\ell)}_{\bm{n}_{\rm R}}}\;, \qquad  \forall y \geq y_{\rm R}\;, \end{array} \right. 
		\\ 
		\label{eq:asymptR}
	\end{equation}
	with $\ket*{\phi^{(\ell)}_{\bm{n}_{{\rm L}/{\rm R}}}}$ the  Zeeman eigenvectors~(\ref{eq:Zeemanterm}) evaluated for $y=y_{\rm L},y_{\rm R}$ respectively,
	and 
	\begin{eqnarray} \label{wavek} 
		k_E^{(\ell)}:=\sqrt{ 2m (E- {{E}_{{\rm Z}}}^{(\ell)})}/\hbar\;, \end{eqnarray}  determined by the dispersion relation~(\ref{eq:energybands}) by imposing the constraint $E^{(\ell)}(k)=E$. 
	We emphasize that for $E \leq {{E}_{{\rm Z}}}^{(\ell)}$ 
	the coefficients $A_E^{(\ell)}$, $R_E^{(\ell)}$ and $T_E^{(\ell)}$ 
	describe evanescent waves. These waves decay exponentially with distance from the scattering region and, as a result, do not carry any current.
	In contrast for 
	$E > {{E}_{{\rm Z}}}^{(\ell)}$,  $A_E^{(\ell)}$, $R_E^{(\ell)}$, and $T_E^{(\ell)}$  characterize the asymptotic behavior of  $\ket*{\psi^{(E)}_{y}}$ far away 
	from the scattering region. 
	Specifically in this regime $A_E^{(\ell)}$ represents right-going electrons injected into the $\ell$-th 
	energy band~(\ref{eq:energybands}) of the left lead,  $R_E^{(\ell)}$ 
	denotes the associated amplitude of left-going  reflected electrons, and $T_E^{(\ell)}$  the amplitude of right-going transmitted electrons.
	The functional dependence among these terms, determined by  explicitly solving 
	Eq.~(\ref{eq:xrepresentation}),  define the transmission $t_E$  and reflection $r_E$ matrix of the model.
	
	Consider for instance the  {\it two-channel} regime where 
	$E$ exceeds the biggest Zeeman eigenvalues in the leads, i.e. 
	\begin{eqnarray} E>{{E}^{(1)}_{{\rm Z}}}={{E}_{{\rm Z}}}\;. \label{2channels}\end{eqnarray} In this regime,  both the energy bands of the system can be populated, and 
	$t_E$ and $r_E$ correspond to
	$2\times 2$  matrices. The elements of these matrices~\cite{Datta1997} are  defined by the following relations:
	\begin{equation}\label{def:tandr} 
		[t_E]_{\ell,\ell'}:=\tau_E^{(\ell,\ell')}\sqrt{\tfrac{k_E^{(\ell)}}{k_E^{(\ell')}}} \;, \quad 
		[r_E]_{\ell,\ell'}:= \rho_E^{(\ell,\ell')}\sqrt{\tfrac{k_E^{(\ell)}}{k_E^{(\ell')}}} \;, 
	\end{equation} 
	where $\tau_E^{(\ell,\ell')}$ and $\rho_E^{(\ell,\ell')}$) are, respectively, the values of amplitudes $T_E^{(\ell)}$ and  
	$R_E^{(\ell)}$ appearing in Eq.~(\ref{eq:asymptR})
	that emerge when we take
	$A_E^{(\ell)}=\delta_{\ell,\ell'}$.
	In particular, $[t_E]_{\ell,0}$ (resp. $[r_E]_{\ell,0}$) represents the probability amplitude for an electron starting in the \textit{lower-energy} eigenstate in the left lead and arriving in the \textit{lower-energy} right (left) lead  eigenstate if $\ell=0$, or in the \textit{higher-energy} right (left) lead eigenstate if $\ell=1$. Contrarily, $[t_E]_{\ell,1}$  (resp. $[r_E]_{\ell,1}$) represents the probability amplitude for an electron starting in the \textit{higher-energy} left lead eigenstate and arriving in the \textit{lower-energy} right (left) lead eigenstate if $\ell=0$, or in the \textit{higher-energy} right (left) lead  eigenstate if $\ell=1$. 
	With these definitions, current conservation is ensured by the unitarity relation
	\begin{equation}
		r_E^{\dagger} r_E + t_E^{\dagger} t_E=\openone\;.
		\label{eq:currentconservation}
	\end{equation}
	For the sake of completeness, we derive this result in Appx.~\ref{appx:rewriting}.
	
	In the {\it single-channel} regime the injection energy $E$ is larger than the lowest eigenvalue in the leads but smaller than their highest eigenvalue, i.e.
	\begin{eqnarray}\label{oneband}
		E^{(1)}_{\rm Z}=E_{\rm Z} > E>  - E_{\rm Z}=E^{(0)}_{\rm Z}\;. \end{eqnarray} 
	Under this condition only the lowest band supports electrons far from the scattering region. In this case $t_E$ and $r_E$ can still be defined as in~(\ref{def:tandr}), but the only elements 
	associated with the low-energy eigenstates  contribute to the current. Specifically,
	the term $[t_E]_{0,0}$ which represents the probability amplitude for an electron starting in the \textit{lower-energy} eigenstate in the left lead and arriving in the \textit{lower-energy} right lead eigenstate, and the element  $[r_E]_{0,0}$, which represents
	the reflection amplitude for the same eigenstate.
	In this case, the current conservation identity~(\ref{eq:currentconservation}) simplifies to:
	\begin{equation}
		|[r_E]_{0,0}|^2 + |[t_E]_{0,0}|^2=1\;.
		\label{eq:currentconservationsingle}
	\end{equation}
	
	\section{Solving the scattering problem}
	\label{sec:sc}
	
	To compute the values of the  matrices $t_E$ and $r_E$ we need to explicitly  find a spinor
	wave-function $\ket*{\psi^{(E)}_y}$ which solves
	Eq.~(\ref{eq:xrepresentation}) for all $y$, and used it to identifying the coefficients $A_E^{(\ell)}$,
	$R_E^{(\ell)}$, and $T_E^{(\ell)}$ that enters in Eq.~(\ref{eq:asymptR}). 
	For this purpose, following Ref.~\cite{Cusumano2020}, we expand
	$\ket*{\psi^{(E)}_y}$ w.r.t. the local eigenstates~(\ref{eq:Zeemanterm}) of the spin operator $\hat{h}_y$, i.e. \begin{equation}
		\ket*{\psi^{(E)}_y}=\sum_{\ell=0,1} C^{(\ell)}_{y} \ket*{\phi^{(\ell)}_{{\bm n}_y}}\;,
		\label{eq:decompose}
	\end{equation}
	with $C^{(\ell)}_y:= \langle{\phi^{(\ell)}_{{\bm n}_y}}\ket*{\psi^{(E)}_y}$ being (coordinate-dependent) complex probabilities amplitudes. With this choice Eq.~\eqref{eq:xrepresentation} assumes the form 
	\begin{equation}
		(\partial_y+K_y)^2\bm{C}_y+\frac{2m}{\hbar^2}(E\mathbb{1}-\Omega_y)\bm{C}_y=0\;,
		\label{eq:equivSCH}
	\end{equation}
	where $\bm{C}_y:= (C^{(0)}_y,C^{(1)}_y)^T$ is the column vector of coefficients $C^{(\ell)}_y$,
	$\Omega_y$ is a $2\times 2$ diagonal matrix whose elements are the instantaneous eigenvalues in Eq.~\eqref{eq:Zeemanterm}, 
	\begin{eqnarray}
		[\Omega_y]_{\ell,\ell'}:=E^{(\ell)}_{Z_y} \delta_{\ell,\ell'}\;, 
	\end{eqnarray}  and finally $K_y$ is the $2\times 2$ Berry  matrix~\cite{adiabatic-nonabelian} of the process, whose elements are defined as
	\begin{equation}
		[K_y]_{\ell,\ell'}:=\braket*{\phi^{(\ell)}_{{\bm n}_y}}{\partial_y \phi^{(\ell')}_{{\bm n}_y}}.
		\label{eq:berrymat}
	\end{equation}
	By construction $K_{y}$ is a skew-matrix 
	which assumes zero value in the leads and varies continuously at the interface with the 
	scattering region, i.e. 
	\begin{eqnarray}K_{y}&=& -K^\dag_{y} \;,   \label{eq:skew} \\ 
		K_{y}&=&0\;, \qquad \forall y\notin\;  ] y_{\rm L},y_{\rm R} [ \;.
		\label{eq:conditionK}
	\end{eqnarray} 
	We stress that the second request is less demanding than it might initially appear. This is due to the fact that, at large distances from the scattering region, the magnetic field becomes constant, taking the values $\bm{B}_{\rm L}$ and $\bm{B}_{\rm R}$.
	Consequently, we can enforce condition (\ref{eq:conditionK}) by simply redefining the boundaries of the scattering region as 
	\begin{eqnarray} y_{\rm L} \rightarrow y_{\rm L}^{-}:=  y_{\rm L}-\epsilon\;,  \qquad 
		y_{\rm R} \rightarrow y_{\rm R}^{+}:= y_{\rm R}+\epsilon\;,
	\end{eqnarray}  with $\epsilon >0$ is an
	infinitesimal  positive shift.
	This regularization allows us to accommodate configurations in which $\bm{B}({y})$ exhibits an explicit discontinuity at the borders of the scattering region.
	The matrix $K_{y}$ defines the Berry operator of the model via the following unitary transformation
	\begin{equation}\label{def:holo} 
		\mathcal{U}_{y_1 \to y_2}:=\overleftarrow{\exp}\left\{-\int_{y_1}^{y_2}dy K_{y}\right\}\;,
	\end{equation}
	where $y_1 \leq y_2$, and   $\overleftarrow{\exp}[\cdots]$ indicates path ordering of the operators product. In this sense, $\mathcal{U}_{y_1 \to y_2}$ plays the role of a Wilson line (the open-loop analog of a Wilson loop~\cite{Wloop2}) in gauge theory, representing the gauge-invariant holonomy of the Berry connection $K_y$ along the path from $y_1$ to $y_2$. As discussed in Sec.~\ref{subs:der22}, setting $\bm{D}_{y}:=\partial_y \bm{C}_{y}$ the first derivative of the vector $\bm{C}_{y}$, the continuity 
	constraint~(\ref{eq:conditionK}) allows us to formally integrate  
	Eq.~\eqref{eq:equivSCH} producing the identity 
	\begin{equation}
		\begin{bmatrix}
			\bm{C}_{y_{\rm R}} \\
			\bm{D}_{y_{\rm R}}
		\end{bmatrix}=\Gamma^{(E)}_{y_{\rm L} \rightarrow y_{\rm R}}  \begin{bmatrix}
			\bm{C}_{y_{\rm L}} \\
			\bm{D}_{y_{\rm L}}
		\end{bmatrix}
		\label{eq:system}\;, 
	\end{equation}
	with $\Gamma^{(E)}_{y_{\rm L} \rightarrow y_{\rm R}}$ the generalized transfer matrix 
	of the process. Factorizing out the Berry operator contribution, the matrix \(\Gamma^{(E)}_{y_{\rm L} \to y_{\rm R}}\) can be expressed as 
	\begin{eqnarray} 
		\Gamma^{(E)}_{y_{\rm L} \rightarrow y_{\rm R}}& :=&\begin{bmatrix}
			\mathcal{U}_{y_{\rm L} \rightarrow y_{\rm R}}& 0 \\
			0 & \mathcal{U}_{y_{\rm L} \rightarrow y_{\rm R}} 
		\end{bmatrix}
		\tilde{\Gamma}^{(E)}_{y_{\rm L} \rightarrow y_{\rm R}} 
		\label{defGAMMA} \;, 
	\end{eqnarray} 
	with the energy dependent term 
	\begin{eqnarray} \label{defDE} \tilde{\Gamma}^{(E)}_{y_{\rm L} \rightarrow y_{\rm R}} := 
		\overleftarrow{{\rm \exp}}\left\{\int_{y_{\rm L}}^{y_{\rm R}} dy \tilde{\mathcal{M}}^{(E)}_{y}\right\} \;, 
	\end{eqnarray} 
	generated by the $4\times 4$ block matrix 
	\begin{equation}
		\tilde{\mathcal{M}}^{(E)}_y:=
		\begin{bmatrix}
			0 & \mathbb{1} \\
			\frac{2m}{\hbar^2}(\tilde{\Omega}_y-E \mathbb{1}) & 0
			\label{eq:mx}
		\end{bmatrix}\;, \quad
		\tilde{\Omega}_y:=\mathcal{U}^{\dagger}_{y_{\rm L} \rightarrow y} \Omega_y \mathcal{U}_{y_{\rm L} \rightarrow y}\;. 
	\end{equation} 
	The matrix $\tilde{\mathcal{M}}_y^{\dag(E)}$ 
	plays a crucial role in the scattering problem. Specifically, as demonstrated in Appx.~\ref{appx:rewriting}, it guarantees current conservation (\ref{eq:currentconservation}) through the symmetry relation:
	\begin{eqnarray} \label{currentconsM} 
		\tilde{\mathcal{M}}_y^{\dag(E)}   \begin{bmatrix} 0&\mathbb{1}\\-\mathbb{1}&0\end{bmatrix} + 
		\begin{bmatrix} 0&\mathbb{1}\\-\mathbb{1}&0\end{bmatrix} \tilde{\mathcal{M}}_y^{(E)} = 0\;.
	\end{eqnarray} 
	Recalling 
	Eq.~\eqref{eq:asymptR} we can write
	\begin{eqnarray}
		\begin{bmatrix}
			\bm{C}_{y_{\rm L}} \\
			\bm{D}_{y_{\rm L}}
		\end{bmatrix}&=&  
		\begin{pmatrix}
			{\scriptstyle    A_E^{(0)} e^{i k_E^{(0)} y_{\rm L}}+R_E^{(0)}e^{-i k_E^{(0)} y_{\rm L}} }\\
			{\scriptstyle   A_E^{(1)} e^{i k_E^{(1)} y_{\rm L}}+R_E^{(1)}e^{-i k_E^{(1)} y_{\rm L}} }\\
			{\scriptstyle    i k_E^{(0)}( A_E^{(0)} e^{i k_E^{(0)} y_{\rm L}}-R_E^{(0)}e^{-i k_E^{(0)} y_{\rm L}}) }\\
			{\scriptstyle  i k_E^{(1)}(A_E^{(1)} e^{i k_E^{(1)} y_{\rm L}}-R_E^{(1)}e^{-i k_E^{(1)} y_{\rm L}})}
		\end{pmatrix},\nonumber \\
		\begin{bmatrix}
			\bm{C}_{y_{\rm R}} \\
			\bm{D}_{y_{\rm R}}
		\end{bmatrix}&=&  
		\begin{pmatrix}
			{\scriptstyle  T_E^{(0)}e^{i k_E^{(0)} y_{\rm R}} }\\
			{\scriptstyle    T_E^{(1)}e^{i k_E^{(1)} y_{\rm R}} }\\
			{\scriptstyle  i k_E^{(0)} T_E^{(0)}e^{i k_E^{(0)} y_{\rm R}} }\\
			{\scriptstyle  i k_E^{(1)}T_E^{(1)}e^{i k_E^{(1)} y_{\rm R}} }
		\end{pmatrix}\;,\label{defCED} 
	\end{eqnarray}
	which replaced in Eq.~\eqref{eq:system},  leads to a set of 4 linear identities that link the amplitudes $T_E^{(\ell)}$ and $R_E^{(\ell)}$ to 
	the input amplitudes $A_E^{(\ell)}$. 
	In conjunction with Eq.~(\ref{def:tandr}) this results in two linear 
	equations for $t_E$ and $r_E$, i.e. 
	\begin{equation}\label{fondamentale} \left\{ \begin{array}{l} 
			W^{-1} t_E W  
			= \; F^{\dagger}_{\rm R} \mathcal{U}_{y_{\rm L} \to y_{\rm R}} \Big(
			X_{0,0} +  i X_{0,1} V \Big) F_{\rm L} \\
			\qquad \qquad   +  F^{\dagger}_{\rm R} \mathcal{U}_{y_{\rm L} \to y_{\rm R}} \Big(
			X_{0,0} -  i X_{0,1} V \Big) F^\dag_{\rm L}  W^{-1} r_E W\;, 
			\\ \\
			i V W^{-1} t_E W  
			= \; F^{\dagger}_{\rm R} \mathcal{U}_{y_{\rm L} \to y_{\rm R}} \Big(
			X_{1,0} +  i X_{1,1} V \Big) F_{\rm L}  \\
			\qquad  \qquad +  F^{\dagger}_{\rm R} \mathcal{U}_{y_{\rm L} \to y_{\rm R}} \Big(
			X_{1,0} -  i X_{1,1} V \Big) F^\dag_{\rm L}  W^{-1} r_E W\;,
		\end{array} \right.
	\end{equation} 
	where
	\begin{equation} \label{defW} 
		W:=\begin{pmatrix}
			1 & 0\\
			0 & \sqrt{\tfrac{k_E^{(1)}}{k_E^{(0)}}}
		\end{pmatrix}\;,  \;\;\;	V:=\begin{pmatrix}
			k_E^{(0)} & 0\\
			0 &  k_E^{(1)} 
		\end{pmatrix}\;
	\end{equation}
	and
	\begin{equation}\label{defV} 
		F_{{\rm L}}:=\begin{pmatrix}
			e^{ik_E^{(0)}y_{{\rm L}}} & 0\\
			0 &  e^{ik_E^{(1)}y_{{\rm L}}} 
		\end{pmatrix}\;, \;\;\; F_{{\rm R}}:=\begin{pmatrix}
			e^{ik_E^{(0)}y_{{\rm R}}} & 0\\
			0 &  e^{ik_E^{(1)}y_{{\rm R}}} 
		\end{pmatrix}\;,
	\end{equation}
	are $2\times 2$ diagonal matrices defined also in Eqs.~(\ref{defW}) and (\ref{defV}) of 
	Appx.~\ref{app:pre}. Moreover, \(X_{i,j}\) are the $2\times 2$ blocks of $\tilde{\Gamma}^{(E)}_{y_{\rm L} \rightarrow y_{\rm R}}$, namely
	\begin{equation}
		\tilde{\Gamma}^{(E)}_{y_{\rm L} \rightarrow y_{\rm R}} = \overleftarrow{{\rm \exp}}\left\{\int_{y_{\rm L}}^{y_{\rm R}} dy \, \tilde{\mathcal{M}}^{(E)}_y\right\} = 
		\begin{bmatrix}
			X_{0,0} & X_{0,1} \\
			X_{1,0} & X_{1,1}
		\end{bmatrix}, \label{main-questaqui-2} 
	\end{equation}
	defined also in Eq.~\eqref{questaqui-2}. At odds with Ref.~\cite{Cusumano2020}, the scattering amplitudes are now calculated exactly at any given energy $E$.
	
	With the exception of the Berry operator $\mathcal{U}_{y_{\rm L} \to y_{\rm R}}$ and of the block matrices $X_{i,j}$, all quantities entering into Eq.~\eqref{fondamentale} admit a closed form solution. Indeed, the main challenge lies in evaluating $\mathcal{U}_{y_{\rm L} \to y_{\rm R}}$ and $X_{i,j}$, since both are defined as solutions of path-ordered exponentials that, in general, do not admit a closed form solution. As will be illustrated in the following sections through specific examples and special regimes, both $\mathcal{U}_{y_{\rm L} \to y_{\rm R}}$ and $X_{i,j}$ can admit a closed form; in such cases, the reflection and transmission matrices themselves admit a fully closed form. Most importantly, the system of Eq.~\eqref{main-questaqui-2} highlights the separation between the dynamical and geometric contributions to the scattering problem. On the one hand, the matrices $W$, $F_{\rm L}$, $F_{\rm R}$, and $V$ are purely dynamical, as they depend solely on the wave vectors of the asymptotic states $k^{(0/1)}_E$, while the Berry operator $\mathcal{U}_{y_{\rm L} \to y_{\rm R}}$ is purely geometric. On the other hand, the block matrices $X_{i,j}$ of Eq.~\eqref{main-questaqui-2} generally cannot be separated into dynamical and geometric components. However, we anticipate that in special regimes the $X_{i,j}$ matrices simplify and become fully dynamical.
	
	\subsection{Derivation of Eq.~(\ref{eq:system})} \label{subs:der22} 
	Define the vectors 
	\begin{eqnarray} \label{defCtilde} 
		\tilde{\bm{C}}_y&: =& \mathcal{U}^\dag_{y_{\rm L} \to y}  \bm{C}_y\;, \\
		\tilde{\bm{D}}_y&:=&\partial_y \tilde{\bm{C}}_y =\mathcal{U}^{\dagger}_{y_{\rm L} \rightarrow y} \left( \bm{D}_{y}+ K_{y} {\bm{C}}_{y}\right)\;,\label{def:dtilde}
	\end{eqnarray} 
	where in the second identity of Eq.~(\ref{def:dtilde}) we used 
	\begin{equation} 
		\partial_y \mathcal{U}_{y_{\rm L} \to y} = - K_y \; \mathcal{U}_{y_{\rm L} \to y}  \Longrightarrow 
		\partial_y \mathcal{U}^\dag_{y_{\rm L} \to y} =  \mathcal{U}^{\dag}_{y_{\rm L} \to y}\; K_y \;.
	\end{equation} 
	This allows one to convert Eq.~(\ref{eq:equivSCH}) in  
	\begin{equation}
		\partial^2_y \tilde{\bm{C}}_y+\frac{2m}{\hbar^2}(E-\tilde{\Omega}_y)\tilde{\bm{C}}_y=0\;,
		\label{eq:presystem}
	\end{equation}
	and hence in the first order differential equation 
	\begin{equation}
		\label{mat}
		\partial_y \begin{bmatrix}
			\tilde{\bm{C}}_y \\
			\tilde{\bm{D}}_y
		\end{bmatrix} =
		\tilde{\mathcal{M}}^{(E)}_y  \begin{bmatrix}
			\tilde{\bm{C}}_y \\
			\tilde{\bm{D}}_y
		\end{bmatrix},
	\end{equation}
	with $\tilde{\mathcal{M}}^{(E)}_y$ and  $\tilde{\Omega}_y$ the matrices defined in  Eq.~(\ref{eq:mx}).
	It is worth noticing that the current conservation condition~(\ref{currentconservation1})
	expressed in terms of $\tilde{\bm{C}}_y$ and $\tilde{\bm{D}}_y$, becomes
	\begin{eqnarray} \label{currentconservation} 
		\tilde{\bm{C}}^\dag_y \cdot \tilde{\bm{D}}_y-\tilde{\bm{D}}^\dag_y \cdot \tilde{\bm{C}}_y 
		=\mbox{const.} 
	\end{eqnarray} 
	Alternatively we can write this by saying that the first derivative of 
	$\tilde{\bm{C}}^\dag_y \cdot \tilde{\bm{D}}_y-\tilde{\bm{D}}^\dag_y \cdot \tilde{\bm{C}}_y$ w.r.t.
	$y$ must be zero, i.e.
	\begin{eqnarray} \label{currentconservation11} 
		\partial_y \left( \tilde{\bm{C}}^\dag_y \cdot \tilde{\bm{D}}_y-\tilde{\bm{D}}^\dag_y \cdot \tilde{\bm{C}}_y \right)=0\;.
	\end{eqnarray}
	Invoking~(\ref{mat})  we can finally notice that~\eqref{currentconservation11} is guaranteed
	by the symmetry~(\ref{currentconsM}) of the matrix $\tilde{\mathcal{M}}^{(E)}$. Indeed we can write
	\begin{widetext}
		\begin{equation} \nonumber 
			\partial_y \left( \tilde{\bm{C}}^\dag_y \cdot \tilde{\bm{D}}_y-\tilde{\bm{D}}^\dag_y \cdot \tilde{\bm{C}}_y \right) 
			= \partial_y \left( \begin{bmatrix}
				\tilde{\bm{C}}^\dag_y ,
				\tilde{\bm{D}}^\dag_y\end{bmatrix}\begin{bmatrix} 0&\mathbb{1} \\-\mathbb{1} &0\end{bmatrix}
			\begin{bmatrix}
				\tilde{\bm{C}}_{y_{\rm R}} \\
				\tilde{\bm{D}}_{y_{\rm R}} 
			\end{bmatrix}\right)\nonumber
			= \begin{bmatrix}
				\tilde{\bm{C}}^\dag_y ,
				\tilde{\bm{D}}^\dag_y\end{bmatrix}
			\left( \tilde{\mathcal{M}}_y^{\dag(E)}    \begin{bmatrix} 0&\mathbb{1} \\-\mathbb{1} &0\end{bmatrix} + 
			\begin{bmatrix} 0&\mathbb{1} \\-\mathbb{1} &0\end{bmatrix} \tilde{\mathcal{M}}_y^{(E)} \right)
			\begin{bmatrix}
				\tilde{\bm{C}}_{y_{\rm R}} \\
				\tilde{\bm{D}}_{y_{\rm R}} 
			\end{bmatrix} =0\;. \label{currentconservation12} 
		\end{equation} 
	\end{widetext}
	
	The derivation of Eq.~\eqref{eq:system} finally follows by integrating  the system~(\ref{mat}) over the scattering region, from $y_{\rm L}$ to $y_{\rm R}$. This leads to
	\begin{equation}
		\begin{bmatrix}
			\tilde{\bm{C}}_{y_{\rm R}} \\
			\tilde{\bm{D}}_{y_{\rm R}}
		\end{bmatrix} = \overleftarrow{\exp}\left\{\int_{y_{\rm L}}^{y_{\rm R}} dy \tilde{\mathcal{M}}^{(E)}_y\right\} \begin{bmatrix}
			\tilde{\bm{C}}_{y_{\rm L}} \\
			\tilde{\bm{D}}_{y_{\rm L}}
		\end{bmatrix}\;,
		\label{eq:systemtilde2}
	\end{equation}
	which corresponds to Eq.~\eqref{eq:equivSCH} thanks to the identities 
	\begin{eqnarray} \tilde{\bm{C}}_{y_{\rm L}}&=&\bm{C}_{y_{\rm L}}\;, \qquad  \tilde{\bm{C}}_{y_{\rm R}}= \mathcal{U}^{\dagger}_{y_{\rm L} \rightarrow y_{\rm R}} \bm{C}_{y_{\rm R}} \;, \label{first} \\
		\tilde{\bm{D}}_{y_{\rm L}}&=& \bm{D}_{y_{\rm L}} \;, \label{second}
		\qquad \tilde{\bm{D}}_{y_{\rm R}}= \mathcal{U}^{\dagger}_{y_{\rm L} \rightarrow y_{\rm R}} \bm{D}_{y_{\rm R}} \;,\end{eqnarray} 
	which follows from~(\ref{defCtilde}) and the continuity condition~(\ref{eq:conditionK}).
	
	\section{Special regimes} \label{sec:special} 
	The analysis allows for a fully analytical treatment in at least three distinct regimes. The first is the asymptotic limit where $E$ is significantly larger than any other energy scale in the model. The second occurs when the variation of the magnetic field within the scattering region is minimal. The third regime arises when the magnetic field in the scattering region is piecewise constant.

	\subsection{Large energy limit} \label{large:app} 
	Let us first  consider the case where the particle energy is 
	much larger than the Zeeman energy splitting in the wire, i.e.
	\begin{eqnarray} 
		E\gg \Delta E_{\rm Z}^{(\max)} &:=& \max_{y} ({E_{\rm Z}}^{(1)}_y-{E_{\rm Z}}^{(0)}_y) \nonumber \\
		&=&  2g \mu_B  \max_{y} |\bm{B}({y})|\;.  \label{zeemansmall} 
	\end{eqnarray} 
	In this regime one has that 
	\begin{eqnarray}\label{condk} 
		k_E^{(0)}\simeq k_E^{(1)}\simeq k_E:= \sqrt{ 2m E}/\hbar\;,
		\label{eq:klhigh}
	\end{eqnarray}
	and we can  disregard the spatial modulation of $\tilde{\mathcal{M}}^{(E)}_y$ induced by the instantaneous Zeeman eigen-energy term $2m\tilde{\Omega}_y/\hbar^2$,  
	\begin{equation}
		\tilde{\mathcal{M}}^{(E)}_y\Big|_{E\gg \Delta E_{\rm Z}^{(\max)} } \simeq \mathcal{M}_E:= 
		\begin{bmatrix}
			0 & \mathbb{1} \\
			-k^2_E \mathbb{1}& 0
		\end{bmatrix}\;.
	\end{equation}
	Accordingly the path-ordered exponent associated with such operator  can be explicitly computed obtaining 
	\begin{eqnarray}\label{approxBIGE} 
		\tilde{\Gamma}^{(E)}_{y_{\rm L} \rightarrow y_{\rm R}}\Big|_{E\gg \Delta E_{\rm Z}^{(\max)} }   &\simeq& \exp\{\mathcal{M}_E \, L  \} \\
		& = &\begin{bmatrix}
			{\scriptstyle \cos(k_E L) \mathbb{1}} &   {\scriptstyle k_E^{-1}\sin(k_E L)\mathbb{1} }\\
			{\scriptstyle - k_E\sin(k_E L)\mathbb{1}} & {\scriptstyle \cos(k_E L)\mathbb{1}}
		\end{bmatrix}\nonumber ,
	\end{eqnarray}
	where 
	$L:= {y_{\rm R}}-{y_{\rm L}}$ represents the length of the scattering region. It is worth noticing that in this special regime the above matrix becomes fully dynamical, depending explicitly on the wave vectors $k_E$ of the asymptotic states. By replacing the above identity  in Eq.~(\ref{eq:system}) and solving for $T_E^{(\ell)}$ and $R_E^{(\ell)}$, this leads finally to 
	\begin{equation}
		t_E\Big|_{E\gg \Delta E_{\rm Z}^{(\max)} } \simeq\mathcal{U}_{y_{\rm L} \rightarrow y_{\rm R}}, \qquad r_E\Big|_{E\gg \Delta E_{\rm Z}^{(\max)} }\simeq 0\;, 
		\label{eq:tequalsholo}
	\end{equation}
	i.e.~the transmission matrix is equal to the Berry operator, with no reflection (see Appx.~\ref{appx:Einf}
		for calculation breakdown). We conclude by noting that higher-order corrections can be computed by tracking the expansion of 
	$\tilde{\Gamma}^{(E)}_{y_{\rm L} \rightarrow y_{\rm R}}$ in terms of the difference $\tilde{\mathcal{M}}^{(E)}_y - \mathcal{M}^{(E)}$. For example, in Appx.~\ref{appendixH}, we use this approach to determine the first-order correction to the reflection matrix.
	
	Equation~(\ref{eq:tequalsholo}) has a significant implication when one considers that, in the large-energy limit, the scattering process should have a negligible effect on the spin of the particle. Physically, this can be understood through an analogy with dynamical quenches: a high-energy particle propagating from left to right along the wire encounters a sudden, abrupt change in the external magnetic field. This rapid variation leaves the spin invariant and induces no reflections.
	If one adopts this ansatz, as detailed in Appx..~\ref{app:ansatz}, Eq.~(\ref{eq:tequalsholo}) leads to the following key identity
	\begin{eqnarray}\label{IIMPP} 
		[\mathcal{U}_{y_{\rm L} \rightarrow y_{\rm R}}]_{\ell',\ell}  =\bra*{\phi^{(\ell')}_{\bm{n}_{\rm R}}}
		\ket*{\phi^{(\ell)}_{\bm{n}_{\rm L}}} \;,
	\end{eqnarray}
	In Sec.~\ref{sec:spinholo}, we will provide a proof of Eq.~(\ref{IIMPP})
	for the special case where 
	the variation of the magnetic field along the 1D wire is constrained on a  fixed plane. Additionally, we will present an argument that extends the result to the general case.
	The identity~(\ref{IIMPP}) is a remarkable result: it reveals that the Berry operator $\mathcal{U}_{y_{\rm L} \rightarrow y_{\rm R}}$ can be determined analytically and depends solely on the magnetic field's components at the boundaries, hence reducing to a purely topological form. Crucially, it is independent of the magnetic field's behavior in the scattering region or the length $L=y_{\rm R} -y_{\rm L}$. Notably, this result predicts that if the magnetic fields at the left and right leads are identical (i.e. $\bm{n}_{\rm R}=\bm{n}_{\rm L}$), the Berry operator transformation reduces to the identity~\cite{windings}.

	\subsection{Infinitesimal scattering region limit} \label{sub:low}

	When the variation of the magnetic field inside the scattering region is sufficiently small 
	the path-ordered exponent in Eq.~(\ref{eq:system}) can be approximated by a regular exponential
	\begin{eqnarray}\label{approxslow} 
		\tilde{\Gamma}^{(E)}_{y_{\rm L} \rightarrow y_{\rm R}}\simeq 
		{\exp}\left\{\int_{y_{\rm L}}^{y_{\rm R}} dy \tilde{\mathcal{M}}^{(E)}_y\right\}=\mathcal{D}_{\rm L}(Q_E)\;,
	\end{eqnarray}
	where $Q_E$ is the Hermitian matrix 
	\begin{eqnarray} 
		Q_E :=  \frac{2m}{\hbar^2}\left(E \mathbb{1}-\int_{y_{\rm L}}^{y_{\rm R}} dy \tilde{\Omega}_y /L\right) \;,  
	\end{eqnarray} and
	\begin{equation} 
		\label{defDQ} 
		\mathcal{D}_{\rm L}(Q_E): = \begin{bmatrix}
			{\scriptstyle \cos(\sqrt{Q_E}L)} & {\scriptstyle \frac{1}{\sqrt{Q_E}} \sin(\sqrt{Q_E}L)} \\
			{\scriptstyle  - \sqrt{Q_E} \sin(\sqrt{Q_E}L) }&  {\scriptstyle \cos(\sqrt{Q_E}L)}
		\end{bmatrix}\;,
	\end{equation}
	(see  Appx..~\ref{appx:infinitesimalscattering} for details).
	Observe in particular, that for infinitesimally short scattering regions (i.e. $L\rightarrow 0$), 
	the matrix
	$\mathcal{D}_{\rm L}(Q_E)$, and hence $\tilde{\Gamma}^{(E)}_{y_{\rm L} \rightarrow y_{\rm R}}$,
	reduces to the identity. However, this behavior does not necessarily extend to the Berry operator. Specifically, as long as the boundary conditions at the edges of the scattering region are preserved in the $L\rightarrow 0$ limit,
	Eq.~(\ref{IIMPP}) remains applicable and can assign a non-trivial value to
	$\mathcal{U}_{y_{\rm L} \rightarrow y_{\rm R}}$. 
	These special conditions are satisfied when the magnetic field $\bm{B}(y)$ 
	undergoes an abrupt transition from $\bm{B}_{\rm L}$ to $\bm{B}_{\rm R}$ at a single point 
	$y_{\rm L}=y_{\rm R}=y_{\star}$  of the 1D wire.   Under such circumstances, 
	leveraging the preceding analysis, we can hence write  \begin{equation} 
		\Gamma^{(E)}_{y_{\rm L} \rightarrow y_{\rm R} }\Big|_{L=0}  =\begin{bmatrix}
			\mathcal{U}_{y_{\star}}& 0 \\
			0 & \mathcal{U}_{y_{\star}} 
		\end{bmatrix}
		\label{defGAMMAL0} \;,
	\end{equation}
	with $\mathcal{U}_{y_{\star}} :=\mathcal{U}_{y_{\star}^{-} \rightarrow y_{\star}^{+}}$ 
	the Berry operator acquired when crossing the discontinuity point. In this case, notice that the matrix \(\Gamma^{(E)}_{y_{\rm L} \to y_{\rm R}}\) matrix becomes fully dynamical, depending explicitly on the wave vectors $k^{(0/1)}_E$ of the asymptotic states.
	Replaced into Eq.~(\ref{fondamentale}) this 
	implies
	\begin{eqnarray} 
		r_E\Big|_{L=0} 
		&=& \tfrac{W\mathcal{U}^\dag_{y_{\star}} W^{-2} \mathcal{U}_{y_{\star}} W-\mathbb{1}}{ W\mathcal{U}^\dag_{y_{\star}} W^{-2} \mathcal{U}_{y_{\star}} W +\mathbb{1} }\;, \\
		t_E\Big|_{L=0}  \nonumber 
		&=& 2  W^{-1} \mathcal{U}_{y_{\star}} W
		\tfrac{\mathbb{1} }{ W\mathcal{U}^\dag_{y_{\star}} W^{-2} \mathcal{U}_{y_{\star}} W +\mathbb{1} } \;,
		\label{rL0} 
	\end{eqnarray} 
	which represents the reflection and transmission matrices through the discontinuity (as shown in 
	Appx..~\ref{appx:infinitesimalscattering}   for $E\rightarrow \infty$ these solutions behave as predicted by Eq.~(\ref{eq:tequalsholo})).

	\subsection{Piecewise constant field regime} \label{sec:piece} 
	When the  magnetic field is a piecewise constant function  the operator $\Gamma^{(E)}_{y_{\rm L} \rightarrow y_{\rm R}}$
	can be expressed as an ordered product of simpler terms. This approach is structurally analogous to lattice gauge formulations of Wilson loops~\cite{Wloop3,Wloop4}, where continuous line integrals are replaced by ordered products of link variables. Specifically, let us assume that there exists a collection of ordered points \begin{eqnarray}\label{partition}
		y_0= y_{\rm L} < y_1 < y_2 < \dots < y_{N-1} < y_N = y_{\rm R}\;, \end{eqnarray}   
	that  identify  a collection of $N$ non-overlapping intervals 
	$I_{j,j+1}:= ] y_j, y_{j+1}[$ 
	over which the magnetic field (and hence
	${\Omega}_y$), take constant values.
	Under this condition  we can write 
	\begin{eqnarray}&&\label{decomp}  \Gamma^{(E)}_{y_{\rm L} \rightarrow y_{\rm R}} =
		\begin{bmatrix}
			\mathcal{U}_{N} & 0 \\
			0 & \mathcal{U}_{N}  
		\end{bmatrix}
		\mathcal{D}_{L_{N-1}}(Q^{(N-1)}_E) 
		\cdots  \\ &&  \quad \cdots
		\begin{bmatrix}
			\mathcal{U}_{2} & 0 \\
			0 & \mathcal{U}_{2}  
		\end{bmatrix}
		\mathcal{D}_{L_{1}}(Q^{(1)}_E) 
		\begin{bmatrix}
			\mathcal{U}_{1} & 0 \\
			0 & \mathcal{U}_{1}  
		\end{bmatrix}
		\mathcal{D}_{L_{0}}(Q^{(0)}_E)  \begin{bmatrix}
			\mathcal{U}_{0} & 0 \\
			0 & \mathcal{U}_{0}  
		\end{bmatrix},\nonumber 
	\end{eqnarray}
	where for all $j\in \{0,1,\cdots, N\}$, 
	$\mathcal{U}_{{j}}:=\mathcal{U}_{y^{-}_{j} \rightarrow y^{+}_{j}}$ represents the Berry operator transformation
	one gets when crossing the discontinuity point $y_{j}$,
	$\mathcal{D}_{L_j}(\cdots)$ is the matrix function defined in Eq.~(\ref{defDQ}), and  finally 
	\begin{eqnarray} 
		L_j&:=&  y_{j+1}-y_j\;,\nonumber  \\
		Q^{(j)}_E &:=& \label{defQjE}
		\frac{2m }{\hbar^2}(E \mathbb{1}-{\Omega}_j)\;, 
	\end{eqnarray} 
	with  ${\Omega}_j$  being the  value  of ${\Omega}_y$ in the  interval $I_{j,j+1}$.
	The derivation of Eq.~(\ref{decomp}) is provided in 
	Appx..~\ref{app:scatteringpiecewise}. 
	Notably, by factoring out the total Berry operator contribution as defined in~(\ref{defGAMMA}), the right-hand side of Eq.~(\ref{decomp}) can be equivalently rewritten in the form:
	\begin{eqnarray}  \Gamma^{(E)}_{y_{\rm L} \rightarrow y_{\rm R}} &=&
		\begin{bmatrix}\nonumber 
			\mathcal{U}_{y_{\rm L} \rightarrow y_{\rm R}}& 0 \\
			0 & \mathcal{U}_{y_{\rm L} \rightarrow y_{\rm R}} 
		\end{bmatrix}
		\tilde{\mathcal{D}}_{L_{N-1}}(Q^{(N-1)}_E)   
		\cdots  \\ && \cdots 
		\tilde{\mathcal{D}}_{L_{1}}(Q^{(1)}_E) 
		\tilde{\mathcal{D}}_{L_{0}}(Q^{(0)}_E) \;,\label{decom1} 
	\end{eqnarray}
	where  now
	\begin{eqnarray}\label{defDtildepc} &&
		\tilde{\mathcal{D}}_{L_j}(Q^{(j)}_E) 
		:= \\
		&& \nonumber \qquad  \begin{bmatrix}
			\mathcal{U}^\dag_{y_{\rm L} \rightarrow y_{j}} & 0 \\
			0 & \mathcal{U}^\dag_{y_{\rm L} \rightarrow y_{j}}
		\end{bmatrix}
		\mathcal{D}_{L_j}(Q^{(j)}_E) 
		\begin{bmatrix}
			\mathcal{U}_{y_{\rm L} \rightarrow y_{j}} & 0 \\
			0 & \mathcal{U}_{y_{\rm L} \rightarrow y_{j}}
		\end{bmatrix}\;,
	\end{eqnarray}
	with 
	\begin{eqnarray} 
		\mathcal{U}_{y_{\rm L} \rightarrow y_{j}}= \mathcal{U}_{j}\cdots \mathcal{U}_{1}\mathcal{U}_{0}\;, 
	\end{eqnarray} 
	the Berry operator accumulated moving from  $y_{\rm L}^{-}$ to $y_{j}^{+}$. 
	As a particular application  let us consider the special case  of a piecewise constant 
	scattering region where 
	\begin{eqnarray}
		{\bm B}(y) = 0 \qquad \forall y\in] y_{\rm L}, y_{\rm R}[\;. 
	\end{eqnarray} 
	Since in this case $Q^{(0)}_E =
	\frac{2m }{\hbar^2}E \mathbb{1}$, Eq.~(\ref{decomp}) reduces to 
	\begin{equation} 
		\Gamma^{(E)}_{y_{\rm L} \rightarrow y_{\rm R}} =         
		\begin{bmatrix}
			{\scriptstyle \cos(k_E L) \mathcal{U}_{y_{\rm L} \rightarrow y_{\rm R}}}&  
			{\scriptstyle k_E^{-1}\sin(k_E L)\mathcal{U}_{y_{\rm L} \rightarrow y_{\rm R}}} \\
			{\scriptstyle - k_E\sin(k_E L)\mathcal{U}_{y_{\rm L} \rightarrow y_{\rm R}}}&
			{\scriptstyle \cos(k_E L)\mathcal{U}_{y_{\rm L} \rightarrow y_{\rm R}}}
		\end{bmatrix} \;. 
	\end{equation} 
	
	We finally observe that  the identity (\ref{decomp}), together with Eq.~(\ref{approxslow}), allows us to 
	derive approximate solutions for $\Gamma^{(E)}_{y_{\rm L} \rightarrow y_{\rm R}}$
	for arbitrary choices of ${\bm B}_y$.  Specifically, a finite-length scattering region can be divided into $N$ equally spaced segments $[y_j,y_{j+1}[$ (with $j\in\{0,1,...,N-1\}$, $y_0=y_{\rm L}$ and $y_N=y_{\rm R}$) of length $\Delta y:=L/N$. Provided that  $\Delta y$ is sufficiently small to ensure the validity of  Eq.~\eqref{approxslow}  across all segments, Eq.~\eqref{decomp} can then
	be employed as a well-defined numerical procedure for solving the scattering problem.
	
	\section{Berry operator} \label{sec:spinholo}
	To compute the Berry operator transformation~(\ref{def:holo}) 
	we begin by considering the special case where the variation of the magnetic field along the 1D wire is restricted to a fixed plane. Under this assumption, the 3D vector $\bm{n}(y)$, which specifies the orientation of the magnetic field $\bm{B}(y)$, can be expressed in terms of a single polar angle $\theta_y$ as:
	\begin{eqnarray}\label{defn} 
		\bm{n}(y):=\sin{(\theta_y)} \bm{n}_1 + \cos{(\theta_y)}\bm{n}_3\;, \end{eqnarray} 
	with $\bm{n}_1$ and $\bm{n}_3$ are orthonormal vectors.
	The  Zeeman eigenvectors~(\ref{eq:Zeemanterm})  can hence be expressed as
	\begin{equation}\left\{ \begin{array}{l}   
			\ket*{\phi^{(0)}_{{\bm n}(y)}}=\cos{(\tfrac{\theta_y}{2})}\ket{\downarrow}-\sin{(\tfrac{\theta_y}{2})}\ket{\uparrow} = e^{-i (\theta_y/2)\hat{\sigma}_{{\bm{n}_2}}} \ket{\downarrow}\;, \\
			\ket*{\phi^{(1)}_{{\bm n}(y)}}=\cos{(\tfrac{\theta_y}{2})}\ket{\uparrow}+\sin{(\tfrac{\theta_y}{2})}\ket{\downarrow} = 
			e^{-i (\theta_y/2)\hat{\sigma}_{{\bm{n}_2}}} \ket{\uparrow} \label{eigenv1} \;,
		\end{array} \right. 
	\end{equation}
	with $\ket{\downarrow}:= \ket*{\phi^{(0)}_{{\bm n}_3}}$ and $\ket{\uparrow}:= \ket*{\phi^{(1)}_{{\bm n}_3}}$ the eigenvectors of the Pauli operator $\hat{\sigma}_{{\bm{n}_3}}:= {\bm{n}_3} \cdot
	\hat{\sigma}$ associated to the eigenvalues $-1$ and $1$ respectively, and $\hat{\sigma}_{{\bm{n}_2}}: = {\bm{n}_2} \cdot
	\hat{\bm{\sigma}}$
	the Pauli operator associated to the vector $\bm{n}_2:= \bm{n}_3\times\bm{n}_1$ (expressed in matrix form in the 
	representation for which $\ket*{\uparrow}:=\icol{1\\0}$ and
	$\ket*{\downarrow}:=\icol{0\\1}$, $\sigma_{{\bm{n}_2}}$ is just the matrix 
	$\begin{pmatrix}
		0 & -i \\
		i & 0 
	\end{pmatrix}$). 
	From Eq.~(\ref{eigenv1}) it follows 
	\begin{eqnarray}\left\{ \begin{array}{l}  
			\ket*{\partial_y \phi^{(0)}_{{\bm n}(y)}}=-(\partial_y \theta_y) \ket*{\phi^{(1)}_{{\bm n}(y)}}/2\;, \\
			\ket*{\partial_y \phi^{(1)}_{{\bm n}(y)}}= (\partial_y \theta_y) \ket*{\phi^{(0)}_{{\bm n}(y)}}/2\;,\end{array} \right.
	\end{eqnarray}
	so that 
	the Berry matrix~(\ref{eq:berrymat}) becomes 
	\begin{equation}
		K_y = (\partial_y \theta_y)  \begin{pmatrix}
			0 & 1/2 \\
			-1/2 & 0 
		\end{pmatrix}
		= i (\partial_y \theta_y) \sigma_{{\bm{n}_2}}/2\;.
		\label{kx}
	\end{equation}
	Now note that Eq.~(\ref{kx}) implies that the commutator $\left[K_y,K_{y'}\right]=0$ for all $y$ and $y'$.
	Since the ordered exponential simplifies to a standard exponential, the Berry operator admits a closed formula, resulting in the following expression
	\begin{eqnarray} 
		\mathcal{U}_{y_{\rm L}\rightarrow y}&=& \exp[ -(i/2) \int_{\theta_{y_{\rm L}}}^{\theta_{y}} dy\; 
		(\partial_y \theta_y) \sigma_{{\bm{n}_2}}]\nonumber \\
		&=& \exp[- i(\tfrac{{\theta_{y}} -{\theta_{y_{\rm L}}}}{2}) \sigma_{{\bm{n}_2}}
		]\label{eq:holo-x}\;,  
	\end{eqnarray}
	which in particular, for $y=  y_{\rm R}$  yields 
	\begin{equation}
		\mathcal{U}_{y_{\rm L}\rightarrow  y_{\rm R}}=  
		\exp[- i (\tfrac{\alpha}{2})\sigma_{{\bm{n}_2}}
		] 
		= \begin{pmatrix}
			\cos(\alpha/2) &  -\sin(\alpha/2)  \\
			\sin(\alpha/2)  &  \cos(\alpha/2)  
		\end{pmatrix} \;, 
		\label{eq:holo}
	\end{equation}
	with  $\alpha:=\theta_{y_{\rm R}}-\theta_{y_{\rm L}}$.
	Using   Eq.~(\ref{eigenv1}) one can easily verify that the matrix elements of 
	Eq.~(\ref{eq:holo}) 
	matches exactly with  the prediction of Eq.~(\ref{IIMPP}). 
	We can hence conclude that such an identity can be explicitly proved for the special cases where the $\bm{n}(y)$ evolve an a 2D plane as in~(\ref{defn}).
	Extending  this result for arbitrary trajectories of $\bm{n}(y)$ 
	can be achieved by employing the same strategy outlined at the end of
	Sec.~\ref{sub:low} for the numerical evaluation of 
	$\overleftarrow{{\exp}}\left\{\int_{y_{\rm L}}^{y_{\rm R}} dy \tilde{\mathcal{M}}^{(E)}_y\right\}$. 
	Specifically this time we decompose the scattering region into $N$ equally spaced segments $[y_j,y_{j+1}[$ (with $j\in\{0,1,...,N-1\}$, $y_0=y_{\rm L}$ and $y_N=y_{\rm R}$) of the length $\Delta y=L/N$ so that 
	\begin{eqnarray}\mathcal{U}_{y_{\rm L}\rightarrow y_{\rm R}} = 
		\mathcal{U}_{y_{N-1}\rightarrow y_{N}} \cdots 
		\mathcal{U}_{y_{2}\rightarrow y_{3}} 
		\mathcal{U}_{y_{1}\rightarrow y_{2}} 
		\mathcal{U}_{y_{0}\rightarrow y_{1}}\;. \label{decdd}
	\end{eqnarray} 
	Now, assume that the length  $\Delta y=L/N$ 
	is sufficiently small so that, within each interval, the evolution of
	$\bm{n}(y)$ can be approximated as occurring on a 2D plane. More precisely, for each
	$j$, we assume that there exists
	a couple of orthonormal axis $\bm{n}_1^{(j)}$ and $\bm{n}_3^{(j)}$, along with a function $\theta_y^{(j)}$ such that 
	\begin{equation}\label{defnj} 
		\bm{n}(y)\simeq \sin{(\theta^{(j)}_y)} \bm{n}^{(j)}_1 + \cos{(\theta^{(j)}_y)}\bm{n}^{(j)}_3\;, 
		\; \forall y\in [y_j,y_{j+1}[\;.\end{equation} 
	By invoking  (\ref{eq:holo}) we can hence evaluate the individual terms 
	in the product (\ref{decdd}) via (\ref{IIMPP}). Specifically  for all $j$ we have
	$[\mathcal{U}_{y_{j}\rightarrow y_{j+1}}]_{\ell',\ell} = 
	\bra*{\phi^{(\ell')}_{\bm{n}_{j+1}}}
	\ket*{\phi^{(\ell)}_{\bm{n}_{j}}}$. 
	Note then that this property implies that
	\begin{eqnarray}&&[\mathcal{U}_{y_{j+1}\rightarrow y_{j+2}} \mathcal{U}_{y_{j}\rightarrow y_{j+1}}]_{\ell',\ell}\nonumber \\
		&&\qquad \qquad = \sum_{\ell''=\{0,1\}} [\mathcal{U}_{y_{j+1}\rightarrow y_{j+2}}]_{\ell'\ell''} [ \mathcal{U}_{y_{j}\rightarrow y_{j+1}}]_{\ell'',\ell}\nonumber \\
		&&\qquad \qquad=\sum_{\ell''=\{0,1\}}\bra*{\phi^{(\ell')}_{\bm{n}_{j+2}}}
		\ket*{\phi^{(\ell'')}_{\bm{n}_{j+1}}}
		\bra*{\phi^{(\ell'')}_{\bm{n}_{j+1}}}
		\ket*{\phi^{(\ell)}_{\bm{n}_{j}}} \nonumber \\
		&&\qquad \qquad= \bra*{\phi^{(\ell')}_{\bm{n}_{j+2}}}
		\ket*{\phi^{(\ell)}_{\bm{n}_{j}}} \;. 
	\end{eqnarray} 
	which  used in~(\ref{decdd}) allows us to conclude that 
	\begin{eqnarray}[\mathcal{U}_{y_{\rm L}\rightarrow y_{\rm R}}]_{\ell',\ell} = 
		\bra*{\phi^{(\ell')}_{\bm{n}_{N}}}
		\ket*{\phi^{(\ell)}_{\bm{n}_{0}}}=\bra*{\phi^{(\ell')}_{\bm{n}_{\rm R}}}
		\ket*{\phi^{(\ell)}_{\bm{n}_{\rm L}}}\;, 
		\label{decdd1}
	\end{eqnarray} 
	hence proving (\ref{IIMPP}).

	\section{Examples}
	\label{sec:res}
	In this section we apply the method developed in Sec.~\ref{sec:sc} to 
	two different examples under the assumption that along the 1D wire 
	the Zeeman field varies continuously  in a fixed 2D plane determined by the directions 
	$\bm{n}_1$ and $\bm{n}_3$ as indicated in Eq.~(\ref{defn}), i.e.
	\begin{eqnarray} \label{defBsca}
		\bm{B}(y) = B_1(y) \bm{n}_1 + B_3(y) \bm{n}_3\;. 
	\end{eqnarray}
	
	\subsection{Scheme I}
	\label{sec:expI}
	Consider the scenario depicted in  panel \textsl{a)} of Fig.~\ref{fig:ex1},
	where the Zeeman field in the two leads has the same magnitude, is directed along the some axis $\bm{n}_3$, but with opposite orientations, i.e. 
	\begin{eqnarray} 
		\label{BFI} \bm{B}_{\rm L}=B_0\,{\bm n}_3\;, \qquad \bm{B}_{\rm R}=-B_0\,{\bm n}_3\;,\end{eqnarray}
	corresponding to set ${\bm n}_{\rm L}= {\bm n}_3$ and ${\bm n}_{\rm R}= -{\bm n}_3$.
	Adopting the notation of Eq.~(\ref{eigenv1}), this in particular implies 
	\begin{eqnarray} \ket*{\phi^{(0)}_{y_{\rm L}}}&=& \ket*{\phi^{(0)}_{{\bm n}_3}}=\ket{\downarrow}=\ket*{\phi^{(1)}_{-{\bm n}_3}} =\ket*{\phi^{(1)}_{y_{\rm R}}}\;, \\
		\ket*{\phi^{(1)}_{y_{\rm L}}}&=&\ket*{\phi^{(1)}_{{\bm n}_3}}=\ket{\uparrow}= -
		\ket*{\phi^{(0)}_{-{\bm n}_3}}=-\ket*{\phi^{(0)}_{y_{\rm R}}}\;.\label{autostati} 
	\end{eqnarray} 
	To interpolate among the values~(\ref{BFI})
	we define the components of the field~(\ref{defBsca}) in the scattering region 
	to have the form 
	\begin{eqnarray}\left\{ \begin{array}{l}
			{B_1(y)}:= \beta_0(y)B_0\left[\sin{\left( (1+2q_2)\tfrac{\pi(y-y_{\rm L})}{y_{\rm R}-y_{\rm L}}\right)}\right]^{(2+2q_1)}\;, 
			\\ 
			{B_3(y)}:=B_0 \left[\cos{\left((1+2q_2)\tfrac{\pi(y-y_{\rm L})}{y_{\rm R}-y_{\rm L}}\right)}\right]^{(1+2q_1)}\;. \end{array} 
		\right.
		\label{eq:bzI}
	\end{eqnarray}
	where $q_1$ and $q_2$ are non-negative integer numbers. In particular $q_1$ accounts for the sharpness of the magnetic field components: if $q_1$ increases, $B_1(y)$ will have support in an increasingly smaller region around the middle point of the scattering region while $B_3(y)$ will nullify in the whole scattering domain, connecting with the correct boundary conditions to the leads. Indeed, in the limiting case of $q_1\rightarrow \infty$, this scheme reduced to the first example in Appx..~\ref{appx:magneticwall} (magnetic wall model) with $\bm{n}_{\rm L}=\bm{n}_3$ and $\bm{n}_{\rm R}=-\bm{n}_3$. Starting from the left lead, the integer $q_2$ accounts for complete winding (in the $\bm{n}_1$-$\bm{n}_3$ plane) of the magnetic field before ending with the correct boundary condition on the right lead, for which we need to introduce the phase function $\beta_0(y)$ (with values $\pm 1$). In the limiting case where $q_2 \rightarrow \infty$, with magnetic field components oscillating very rapidly and having zero average, we expect the system to reduce to the magnetic wall model, illustrated in the first example of Appx.~\ref{appx:magneticwall} with $\bm{n}_{\rm L}=\bm{n}_3$ and $\bm{n}_{\rm R}=-\bm{n}_3$. Specifically, while $|\bm{B}(y)| >0$ for all $y$, both $\partial_y B_1(y)$ and $\partial_y B_3(y)$ are zero for $y\rightarrow y_{\rm R/L}$. Consequently, we have $K_{y_{\rm L}}=K_{y_{\rm{R}}}=0$, indicating  that~(\ref{eq:system}) is  valid for use. This allows us to apply the procedure outlined in Appx.~\ref{appendixB} to calculate the transmission matrix $t_E$. Furthermore since $\lim_{y \rightarrow y_{\rm R}} \theta_y=\pi$ and $\theta_{y_{\rm L}}=0$ we obtain $ \alpha = \pi$, and thus
	\begin{equation}
		\mathcal{U}_{y_{\rm L} \to y_{\rm R}} =- i \sigma_{{\bm{n}_2}}=
		\begin{pmatrix}
			0 & -1 \\
			1 & 0 
		\end{pmatrix}\;.
		\label{eq:isigmay}
	\end{equation}
	Alternatively, this result can be derived using~(\ref{IIMPP}) and (\ref{autostati}). 
	We next compute the probability $P_E^{(\downarrow\mapsto \uparrow)}$ for an electron injected in the left lead in the lower-energy state $\ket*{\phi^{(0)}_{y_{\rm L}}}$ (corresponding to the spin down state $\ket{\downarrow}$)
	to be transmitted in the lower-energy state in the right lead  $\ket*{\phi^{(0)}_{y_{\rm L}}}$ (corresponding to the spin up state $\ket{\uparrow}$), i.e.~
	\begin{eqnarray} P_E^{(\downarrow\mapsto \uparrow)}:=|[t_E]_{0,0}|^2\;.\end{eqnarray} 
	This, along with the corresponding reflection probability $|[r_{E}]_{0,0}|^2$, is the only scattering process which is allowed when we are in the single-channel regime~(\ref{oneband}).
	Indeed, for $E<E_{\rm Z}$,  no higher-energy states are available for transport, as $E$ is below the bottom of the higher-energy band.
	In contrast, in the two-channel regime~(\ref{2channels}), where $E>E_{\rm Z}$, we can also define two more transmission probabilities. Specifically, $P_E^{(\uparrow\mapsto \downarrow)}=|[t_{E}]_{1,1}|^2$ represents the probability that an electron injected in the left lead in the higher-energy state (corresponding to the spin up) to be transmitted in the higher-energy state in the right lead (corresponding to the spin down). Similarly,  $P_E^{(\downarrow\mapsto \downarrow)}=|[t_{E}]_{1,0}|^2$ is the probability that an electron injected in the left lead in the low-energy state
	(spin-down) is transmitted into the higher-energy state in the right lead (spin down). Finally, we have the  $P_E^{(\uparrow\mapsto \uparrow)}=|[ t_E]_{0,1}|^2$, which corresponds to an electron injected in the left lead in the lower-energy state (spin-up) being transmitted into the lower-energy state in the right lead (spin-up).
	In Fig.~\ref{fig:PT_I} we plot the transmission probabilities 
	defined above
	as functions of $E$ (in units of the Zeeman energy splitting $E_{\rm Z}$) for a wire of length $L=3 L_{\rm Z}$, where $L_{\rm Z}:=\sqrt{\hbar^2/(2m E_{\rm Z}})$ is the Zeeman length. The various panels refer to different values of the parameters $q_1$ and $q_2$. In the main panels, the energy ranges between $-E_{\rm Z}$ and $5E_{\rm Z}$, covering both the 
	single- and two-channel regime, while in the insets we focus only on the two-channel regime, considering  energies within the range $[E_{\rm Z},10E_{\rm Z}]$. 
	The green curve represents the probability $P_E^{(\downarrow\mapsto \uparrow)}$, the red curve represents $P_E^{(\uparrow\mapsto \downarrow)}$, and the orange curve represents $P_E^{(\downarrow\mapsto \downarrow)}$. 
	Interestingly,  $P_E^{(\downarrow\mapsto \downarrow)}$ is always equal to $P_E^{(\uparrow\mapsto \uparrow)}$ (as we shall see in the following when $E$ is sufficiently larger, such behaviour follows directly from Eqs.~(\ref{eq:holo}) and (\ref{eq:tequalsholo})).
	Let us first discuss panel \textsl{a)}, relative to $q_1=0$ and $q_2=0$.
	The numerical results show that, in the single-channel regime, the only transmission probability that is different from zero is $P_E^{(\downarrow\mapsto \uparrow)}$, as expected. This probability 
	increases monotonically from zero at $E=-E_{\rm Z}$ (bottom of the lower band), rapidly approaching 1 (i.e. a perfect transition from the spin-down state to the spin-up state) as the energy nears $E_{\rm Z}$.
	For $E>E_{\rm Z}$ the green curve decreases monotonically, while the red curve, after a rapid increase, starts decrease and eventually joins the green curve at $E\simeq 2E_{\rm Z}$. Beyond this point, 
	both $P_E^{(\downarrow\mapsto \uparrow)}$ and $P_E^{(\uparrow\mapsto \downarrow)}$ decrease to zero for sufficiently large energies (see inset). 
	The orange curve, representing $P_E^{(\downarrow\mapsto \downarrow)}$ and
	$P_E^{(\uparrow\mapsto \uparrow)}$, also starts 
	from zero for $E>E_{\rm Z}$ monotonously increases and eventually reaching 1 for large values of $E$ (see inset).
	The behavior of the transmission probabilities for $E\gg E_{\rm Z}$ is as expected, since, according to Eq.~\eqref{eq:tequalsholo}, the transmission matrix is completely determined by the Berry operator in Eq.~(\ref{eq:isigmay}). As a result,  we have $P_E^{(\downarrow\mapsto \uparrow)}=P_E^{(\uparrow\mapsto \downarrow)}=0$ and $P_E^{(\uparrow\mapsto \uparrow)}=P_E^{(\downarrow\mapsto \downarrow)}=1$. In this regime, the reflection amplitudes are suppressed, and the particle retains the same spin value of the incoming field as the incoming state.
	Notably, $P_E^{(\downarrow\mapsto \uparrow)}$ and $P_E^{(\uparrow\mapsto \downarrow)}$ become virtually equal for an energy as small as $E\simeq 2E_{\rm Z}$.
	Notably, even at 
	$E \simeq 6E_{\rm Z}$, both $P_E^{(\downarrow\mapsto \downarrow)}$ and 
	$P_E^{(\uparrow\mapsto \uparrow)}$  reach a value of $0.9$, indicating that 
	$90\%$ inversion of the input energy band population has been achieved. For example, electrons entering the left lead in the low-energy band will exit the right lead in the opposite band with  $90\%$ probability. It is important to highlight that, due to energy conservation, this effect is accompanied by a partial conversion of the kinetic energy component into spinor energy. This conversion results in a net reduction in momentum, given by\begin{eqnarray} \label{wavekD} 
		\Delta k_E:= k_E^{(1)}-k_E^{(0)}&=&-\sqrt{ \tfrac{2m E}{\hbar^2}} \left(\sqrt{1+ \tfrac{{E}_{{\rm Z}}}{E}
		} - 
		\sqrt{1-  \tfrac{{E}_{{\rm Z}}}{E}}\right)\nonumber \\
		&\simeq&-\sqrt{ \tfrac{2m E^2_{\rm Z}}{\hbar^2E}} \;. \end{eqnarray} 
	The opposite effect occurs for electrons entering the left lead in the high-energy band: with a $90\%$ probability, they will emerge in the low-energy band on the right lead, accompanied by an increase in their kinetic energy. The net gain in momentum in this case is  $|\Delta k_E|$.
	
	By increasing the value of $q_1$, for fixed $q_2=0$, the transmission probabilities are modified as shown in panels \textsl{b)} and \textsl{c)}.
	The main effect of increasing $q_1$ is that the probability  $P_E^{(\downarrow\mapsto \downarrow)}$, see panel \textsl{c)}, reaches more quickly the value 1 for $E>E_{\rm Z}$, i.e.~
	the transmission matrix is completely determined by the Berry operator.
	A similar effect is observed when increasing the value of $q_2$, for a fixed $q_1=0$ [panels \textsl{d)} and \textsl{e)}].
	A notable additional effect of larger values of $q_2$ is the significant suppression of the probability $P_E^{(\downarrow\mapsto \uparrow)}$, even for energies $E<E_{\rm Z}$. We quantify how much the transmission matrix aligns to the Berry operator~\eqref{eq:isigmay} as a function of the injection energy $E$, using the Hilbert-Schmidt distance $\|A-B\|_{\rm HS}:=\sqrt{\mbox{Tr}[(A-B)(A-B)^{\dagger}]}$. In the main panel of Fig.~\ref{fig:distI}\textsl{a)} we plot the distance $\|t_E(L)- \mathcal{U}_{y_{\rm L} \to y_{\rm R}}\|_{\rm HS}$ for $q_1=0,1,10$, with $q_2=0$. Moreover we plot the Hilbert-Schmidt distance $\|t^{\text{I}}_E(L)-\mathcal{U}_{y_{\rm L} \to y_{\rm R}}\|_{\rm HS}$ for the \textit{magnetic wall} configuration described in Appx.~\ref{appx:magneticwall} with $\bm{n}_{\rm L}=\bm{n}_3$ and $\bm{n}_{\rm R}=-\bm{n}_3$, see the red dashed curve in the main panel of Fig.~\ref{fig:distI}\textsl{b)}. Then we quantify the contribution of the back scattering in the process by plotting the norm $\|r_E(L)\|_{\rm HS}$ as a function og the injection energy in the inset of Fig.~\ref{fig:distI}\textsl{a)} (the analytical behavior of $\|r^{\text{I}}_E(L)\|_{\rm HS}$ is plotted with red dashed line). Similarly, in Fig~\ref{fig:distI}\textsl{a)} we quantify both $\|t_E(L)-\mathcal{U}_{y_{\rm L} \to y_{\rm R}}\|_{\rm HS}$ (main panel) and $\|r^{\text{I}}_E(L)\|_{\rm HS}$ (inset) for $q_2=0,1,10$, with $q_1=0$, as a function of the injection energy $E$.
	\subsection{Scheme II}
	\label{sec:expII}
	In this second example, we assume that the Zeeman field in the left lead and the right lead 
	is directed along orthogonal axes. Specifically, as shown in panel \textsl{b)} of Fig.~\ref{fig:ex1} we set 
	\begin{eqnarray} \bm{B}_{\rm L}=B_0\,{\bm n}_3\;, \qquad \bm{B}_{\rm R}=B_0\,{\bm n}_1\;.\end{eqnarray}
	which corresponds to choosing  ${\bm n}_{\rm L}= {\bm n}_3$ and ${\bm n}_{\rm R}= {\bm n}_1$. Hence, we have the following spin states:  
	\begin{eqnarray}
		\begin{array}{l}
			\ket*{\phi^{(0)}_{y_{\rm L}}}= \ket*{\phi^{(0)}_{{\bm n}_3}}=\ket*{\downarrow}\;,\\
			\ket*{\phi^{(1)}_{y_{\rm L}}}= \ket*{\phi^{(1)}_{{\bm n}_3}}=\ket*{\uparrow}\;,\\
			\ket*{\phi^{(0)}_{y_{\rm R}}}=\ket*{\phi^{(0)}_{{\bm n}_1}}=\ket*{-}:=({\ket*{\downarrow}-\ket*{\uparrow}})/{\sqrt{2}}\;, \\
			\ket*{\phi^{(1)}_{y_{\rm R}}}=\ket*{\phi^{(1)}_{{\bm n}_1}}= \ket*{+}  := ({\ket*{\downarrow}+\ket*{\uparrow}})/{\sqrt{2}}\;.
	\end{array} \end{eqnarray}  
	For the sake of definiteness, we also assume that the magnetic field components vary in the scattering region as follows:
	\begin{equation} \left\{ \begin{array}{l}
			{B_1(y)}={B_0} \beta_1(y)\sin^{ (2+2q_1)}{\left( (1+4q_2)\tfrac{\pi(y-y_{\rm L})}{2(y_{\rm R}-y_{\rm L})}\right)}\;,\\
			{B_3(y)}={B_0} \beta_2(y)\cos^{(2+2q_1)}{\left((1+4q_2)\tfrac{\pi(y-y_{\rm L})}{2(y_{\rm R}-y_{\rm L})}\right)}\;,
			\label{eq:bzII}\end{array} \right.
	\end{equation}
	where $q_1$ and $q_2$ are again non-negative integer numbers. In this scheme if $q_1$ increases, both $B_1(y)$ and $B_3(y)$ will nullify in the whole scattering region.
	Indeed, in the limiting case of $q_1\to \infty$, with $q_2=0$, this scheme reduced to the second example of the model in Appx.~\ref{appx:magneticwall} with $\bm{n}_{\rm L}=\bm{n}_3$ and $\bm{n}_{\rm R}=\bm{n}_1$. As for the first scheme, starting from the left lead, the integer $q_2$ accounts for complete winding of the magnetic field.
	In order to end with the correct boundary conditions on the right lead, we introduce the two phase function $\beta_1(y)$ and $\beta_2(y)$ (with values $\pm 1$). In the limiting case of $q_2\rightarrow \infty$, we expect to find the same results of the second example of the model described in Appx.~\ref{appx:magneticwall} with $\bm{n}_{\rm L}=\bm{n}_3$ and $\bm{n}_{\rm R}=\bm{n}_1$.
	In this case, the rotation angle is $\alpha={\pi}/{2}$, so that  the Berry operator becomes
	\begin{equation}
		\mathcal{U}_{y_{\rm L} \to y_{\rm R}} = 
		\exp[- i (\tfrac{\pi}{2})\sigma_{{\bm{n}_2}}
		]  = \frac{1}{\sqrt{2}}\begin{pmatrix}
			1 & -1 \\
			1 & 1 
		\end{pmatrix}.
		\label{eq:holomix}
	\end{equation}
	We now apply the procedure outlined in Secs.~\ref{sec:sc} to compute the transmission matrix $t_E$.
	First, we analyze the probability $P_E^{(\downarrow\mapsto -)}=|[t_{E}]_{0,0}|^2$ 
	which represents the probability for an electron injected in the low-energy state of the left lead  $\ket*{\downarrow}$ to be transmitted in the low-energy state of the right lead $\ket*{-}$. This, with the corresponding reflection probability $|[r_E]_{0,0}|^2$, are the only scattering processes occurring in the single-channel regime. 
	In the main panel of Fig.~\ref{fig:PT_II}\textsl{a)}, for example, we plot $P_E^{(\downarrow\mapsto -)}$ (green curve) as a function of energy for the case $q_1=0$ and $q_2=0$.
	We show that for an injection energy $E$ sufficiently close to $E_{\rm Z}$, the electron is fully transmitted, i.e. the probability transmission $P_E^{(\downarrow\mapsto -)}$ reaches 1. 
	On the other hand, for $E\geq E_{\rm Z}$, 
	we can also define the probability $P_E^{(\uparrow\mapsto -)}=|[t_{E}]_{0,1}|^2$ (orange curve) for an electron injected in the high-energy state of the left lead $\ket*{\uparrow}$ to be transmitted in the low-energy state of the right lead $\ket*{-}$, 
	and analogously the transmission probabilities $P_E^{(\downarrow\mapsto +)}=|[t_E]_{1,0}|^2$ and $P_E^{(\uparrow\mapsto +)}=|[t_E]_{1,1}|^2$ (red curve).
	As demonstrated in Appx.~\ref{appx:equalprob}, we find that
	$P_E^{(\uparrow\mapsto -)}=\abs{[t_{E}]_{0,1}}^2$ is equal to $P_E^{(\downarrow\mapsto +)}=\abs{[t_E]_{1,0}}^2$.
	The plot shows that $P_E^{(\downarrow\mapsto -)}$ (green curve) monotonically decreases for $E> E_{\rm Z}$, while $P_E^{(\uparrow\mapsto +)}$ (red curve) rapidly increases, joining the green curve, and thereafter decreasing with it.
	On the other hand, $P_E^{(\uparrow\mapsto -)}$ monotonically increases from zero for increasing energy.
	The specific behavior of the various probabilities depends on the details of the scattering region, such as the actual spatial dependence of the Zeeman field $\bm{B}(y)$ and the length $L$.
	This is shown in panels \textsl{b)}-\textsl{e)} of Fig.~\ref{fig:PT_II}, where the transmission probabilities are plotted for different values of $q_1$ and $q_2$.
	Panels \textsl{b)} and \textsl{c)} show what happens for increasing $q_1$ with a fixed value of $q_2=0$, namely orange and green curves converge more quickly to 1/2 for $E>E_{\rm Z}$.
	On the other hand, panels \textsl{d)} and \textsl{e)} presents the behavior of the transmission probabilities for a fixed value of $q_1=0$ and $q_2=1$ and $q_2=10$, respectively. In the latter case all curves collapse on the value 1/2 at $E\gtrsim E_{\rm Z}$. We emphasize that the asymptotic behavior ($E\gg E_{\rm Z}$) of the curves in Fig.~\ref{fig:PT_II} does not depend on the details of $\bm{B}(y)$ but only on its boundary condition. At odds with respect to Fig.~\ref{fig:PT_I}, here all probabilities tend to the value $1/2$ (denoted by the horizontal black solid line) for large values of energy. This behaviour is actually expected since for $E\gg E_{\rm Z}$ the transmission matrix reduces to the Berry operator [Eq.~\eqref{eq:tequalsholo} in Sec.~\ref{sec:sc}]. In the latter, given in Eq.~\eqref{eq:holomix}, all matrix elements have an absolute squared value equal to $1/2$.
	
	As for the Scheme I,  we quantify how much the transmission matrix aligns to the Berry operator~\eqref{eq:holomix} as a function of the injection energy $E$, using the Hilbert-Schmidt distances. In the main panel of Fig.~\ref{fig:distII}\textsl{a)} we plot the distance $\|t_E(L)-\mathcal{U}_{y_{\rm L} \to y_{\rm R}}\|_{\rm HS}$ for $q_1=0,1,10$, with $q_2=0$. In the main panel of Fig.~\ref{fig:distII}\textsl{b)} we plot the Hilbert-Schmidt distance $\|t^{\text{II}}_E(L)-\mathcal{U}_{y_{\rm L} \to y_{\rm R}}\|_{\rm HS}$ (see the red dashed curve), for the \textit{magnetic wall} configuration described in Appx.~\ref{appx:magneticwall} with $\bm{n}_{\rm L}=\bm{n}_3$ and $\bm{n}_{\rm R}=\bm{n}_1$. Then we quantify the contribution of the back scattering in the process by plotting the distance $\|r_E(L)\|_{\rm HS}$ as a function of the injection energy in the inset of Fig.~\ref{fig:distII}\textsl{a)} (the analytical behavior of $\|r^{\text{II}}_E(L)\|_{\rm HS}$ is plotted with red dashed line). Similarly, in Fig~\ref{fig:distII}\textsl{b)} we quantify both $\|t_E(L)-\mathcal{U}_{y_{\rm L} \to y_{\rm R}}\|_{\rm HS}$ (main panel) and $\|r^{\text{II}}_E(L)\|_{\rm HS}$ (inset) for $q_2=0,1,10$, with $q_1=0$, as a function of the injection energy $E$.
	
	We now analyze the scattering process in terms of the incoming and outgoing states, highlighting the role of the Berry operator in the entanglement between the spin degree of freedom and the electron momentum (see, e.g.,~\cite{spinmomentumentanglement}). Consider an electron injected from the left lead in the low-energy band. For large, but not excessively large energy values (i.e., 
	$E$ moderately larger than $E_{\rm Z}$), the transmission matrix closely approximates the Berry operator described in  Eq.~\eqref{eq:holomix}. Consequently, the outgoing state in the right lead exhibits strong correlations (entanglement) between the spin and kinetic components. Specifically, the state can be expressed as \begin{equation}
		|{\psi^{(E)}_y}\rangle\simeq \frac{e^{i k_E^{(0)}y}\ket*{-} + e^{i \gamma}e^{i k_E^{(1)}y}\ket*{+}}{\sqrt{2}},
	\end{equation}
	where  $\gamma:=e^{i(k_E^{(1)} - k_E^{(0)})L}$ represents a dynamic phase arising from the free evolution along the scattering region.
	Similarly, if the incoming state is in the high-energy band, the corresponding outgoing state in the right lead takes the form
	\begin{equation}
		|{\psi^{(E)}_y}\rangle\simeq \frac{e^{i k_E^{(0)}y}\ket*{-} - e^{i \gamma}e^{i k_E^{(1)}y}\ket*{+}}{\sqrt{2}}.
	\end{equation}
	
	\section{Conclusions}\label{sec:conclusions} 
	In summary, we have proposed a method to describe the coherent transport of electrons through a 1D wire subject to a magnetic field. We have derived a closed-form differential equation, which contains both geometric and dynamical contributions, that governs (spin-resolved) electron transport for a generic spatially varying magnetic field which interacts with the electrons spin via the Zeeman coupling.
	We have shown that electrons are fully transmitted when the injection energy is much larger than the Zeeman splitting energy. In this regime, the quantum state of the transmitted electrons undergoes a topological transformation, i.e. the Berry operator of the system. Indeed, such Berry operator depends solely on the values of the magnetic field at the nanowire boundaries, regardless of the magnetic field’s behavior within the wire. We have then analyzed two specific examples. In the first one, when the injection energy is below the Zeeman energy, we have demonstrated that perfect spin-flip occurs in the transmission through the wire.
	In the second example, for injection energies below the Zeeman splitting energy, we observed a balanced spin-mixing. Moreover, for an infinite injection energy (or an infinitesimal wire length), the quantum state of the transmitted electron emerges with a different wavevector in the first example. In the second example, the quantum state becomes entangled between its spin and wavevector degrees of freedom.
	
	We mention that we have checked our results in Figs.~\ref{fig:PT_I} and~\ref{fig:PT_II} using an independent numerical method which uses a wavefunction matching technique (KWANT-toolkit~\cite{kwant}). We mention that the numerical probabilities plotted in all the figures coincide with the results obtained with KWANT-toolkit up to a relative error of the order of \(10^{-5}\).
	
	To conclude, our approach can be extended to realistic three dimensional nanowires (where multiple transverse modes are present). As long as the number of modes available for transport is restricted to two by choosing the injection energy of electrons within a reduced range of values, we have obtained similar results to the ones reported in Figs.~\ref{fig:PT_I} and~\ref{fig:PT_II}.
	We mention that a possible implementation of a Zeeman field having the required behavior can be obtained by bending a nanowire in a uniform magnetic field, or by depositing a nanowire on a substrate containing a magnetic material with a suitable pattern of non-collinear magnetic moments. Examples are Heusler compounds~\cite{Manna2018}, low-dimensional systems which lack structural inversion symmetry, such as a single atomic layer of manganese on a tungsten substrate~\cite{Bode2007}, and other materials, such as insulating multiferroics~\cite{Khomskii2009} and van der Waals magnets~\cite{Du2023}.
	
	\begin{acknowledgments}
		F.T. acknowledges funding from MUR-PRIN 2022 - Grant No. 2022B9P8LN
		- (PE3)-Project NEThEQS “Non-equilibrium coherent
		thermal effects in quantum systems” in PNRR Mission 4 -
		Component 2 - Investment 1.1 “Fondo per il Programma
		Nazionale di Ricerca e Progetti di Rilevante Interesse
		Nazionale (PRIN)” funded by the European Union - Next
		Generation EU and from the Royal Society through the International Exchanges between the UK and Italy (Grants No. IEC R2 192166).
		V.G. acknowledges financial support by MUR (Ministero dell’Istruzione, dell’Università e della Ricerca) through the following projects: PNRR MUR project PE0000023-NQSTI, PRIN 2017 Taming complexity via Quantum Strategies: a Hybrid Integrated Photonic approach (QUSHIP) Id. 2017SRN-BRK.
	\end{acknowledgments}
	\newpage
	
	\appendix
	\begin{widetext} 
		\section{Current conservation and preliminary observations} 
		\label{appx:rewriting}
		In this section we  review some basic properties of the model clarifying how the current operator associated with the eigenvector 
		$|\psi^{(E)}_y\rangle$ is expressed in terms of the amplitudes $\bm{C}_y$ introduced in Eq.~(\ref{eq:decompose}).

		\subsection{Preliminaries}\label{app:pre}

		We start by using the transmission and reflection matrices $t_E$ and $r_E$ defined in Eq.~(\ref{def:tandr}) to
		link the coefficients $\bm{C}_{y_{\rm L}}$ and 
		$\bm{C}_{y_{\rm R}}$ appearing in Eq.~(\ref{eq:system}), in terms of the amplitudes
		$R_E^{(\ell)}$, $T_E^{(\ell)}$, and $A_E^{(\ell)}$ appearing
		in Eq.~(\ref{eq:asymptR}).
		
		From  Eq.~(\ref{def:tandr}) we can write 
		\begin{eqnarray}\left\{ \begin{array}{l} 
				T_E^{(\ell)} =\sum_{\ell'=0}^{1} \tau_E^{(\ell,\ell')} A_E^{(\ell')} = 
				\sum_{\ell'=0}^{1} [t_E]_{\ell,\ell'} \sqrt{\tfrac{k_E^{(\ell')}}{k_E^{(\ell)}}} A_E^{(\ell')} \;, \\
				R_E^{(\ell)}=\sum_{\ell'=0}^{1} \rho_E^{(\ell,\ell')} A_E^{(\ell')} = 
				\sum_{\ell'=0}^{1} [r_E]_{\ell,\ell'} \sqrt{\tfrac{k_E^{(\ell')}}{k_E^{(\ell)}}} A_E^{(\ell')} \;,\end{array} 
			\right.
		\end{eqnarray} 
		which, introducing $\bm{T}:=(T_E^{(0)},T_E^{(1)})^T$, 
		$\bm{R}:=(R_E^{(0)},R_E^{(1)})^T$, $\bm{A}:=(A_E^{(0)},A_E^{(1)})^T$, and the matrices
		\begin{equation} \label{defW} 
			W:=\begin{pmatrix}
				1 & 0\\
				0 & \sqrt{\tfrac{k_E^{(1)}}{k_E^{(0)}}}
			\end{pmatrix}\;, \qquad \qquad    W^{-1}:=\begin{pmatrix}
				1 & 0\\
				0 & \sqrt{\tfrac{k_E^{(0)}}{k_E^{(1)}}}
			\end{pmatrix}\;,
		\end{equation}
		can be conveniently expressed in the following vectorial form 
		\begin{eqnarray} \left\{ \begin{array}{l} 
				\bm{T} = W^{-1} t_E W \bm{A}\;, \\ \\
				\bm{R} = W^{-1} r_E W \bm{A} \;. \end{array} 
			\right. \label{identityT} 
		\end{eqnarray} 
		Accordingly, setting 
		\begin{equation}\label{defV} 
			V:=\begin{pmatrix}
				k_E^{(0)} & 0\\
				0 &  k_E^{(1)} 
			\end{pmatrix}\;, \qquad\qquad   F_{{\rm L}/{\rm R}}:=\begin{pmatrix}
				e^{ik_E^{(0)}y_{{\rm L}/{\rm R}}} & 0\\
				0 &  e^{ik_E^{(1)}y_{{\rm L}/{\rm R}}} 
			\end{pmatrix}\;,
		\end{equation}
		we can rewrite Eq.~(\ref{defCED}) in the form 
		\begin{eqnarray}
			\begin{bmatrix} \label{DEFCL} 
				\bm{C}_{y_{\rm L}} \\
				\bm{D}_{y_{\rm L}}
			\end{bmatrix}&=& \begin{bmatrix}
				F_{\rm L}  \bm{A}+ F^\dag_{\rm L}  \bm{R}  \\
				i V (F_{\rm L} \bm{A}- F^\dag_{\rm L}\bm{R} )
			\end{bmatrix}=
			\begin{bmatrix}
				(F_{\rm L} + F^\dag_{\rm L}W^{-1} r_E W) \bm{A}  \\
				i V (F_{\rm L} - F^\dag_{\rm L}W^{-1} r_E W)\bm{A} 
			\end{bmatrix}\;,
			\nonumber \\  \nonumber \\
			\begin{bmatrix} 
				\bm{C}_{y_{\rm R}} \\
				\bm{D}_{y_{\rm R}}
			\end{bmatrix}&=& \begin{bmatrix}
				F_{\rm R}  \bm{T}  \\
				i V F_{\rm R} \bm{T}
			\end{bmatrix}
			=
			\begin{bmatrix}
				F_{\rm R} W^{-1} t_E W  \bm{A} \\
				i V F_{\rm R} W^{-1} t_E W  \bm{A}
			\end{bmatrix}\;.
		\end{eqnarray}

		\subsection{Current conservation} 
		The current operator associated with the spinor eigenstate  $|\psi^{(E)}_y\rangle$ is
		given by
		\begin{eqnarray} 
			\hat{J}^{(E)}_y:= - \frac{i\hbar}{2m}\left(  |\partial_y\psi^{(E)}_y\rangle \langle \psi^{(E)}_y| -  |\psi^{(E)}_y\rangle \langle \partial_y \psi^{(E)}_y|\right) \;,
		\end{eqnarray} 
		where for easy of notation we defined $|\partial_y \psi^{(E)}_y\rangle:=\partial_y |\psi^{(E)}_y\rangle$.
		Since $|\psi^{(E)}_y\rangle$ is an eigenvector of $\hat{\cal H}$ 
		we can write
		\begin{eqnarray} 
			\partial_y \hat{J}^{(E)}_y &=&- \frac{i\hbar}{2m}\left(  |\partial_y^2\psi^{(E)}_y\rangle \langle \psi^{(E)}_y| -  |\psi^{(E)}_y\rangle \langle \partial_y^2 \psi^{(E)}_y|\right) \nonumber \\
			&=&  \frac{i}{\hbar} \left[ \hat{{\cal H}} -  \hat{h}_y,  |\psi^{(E)}_y\rangle \langle \psi^{(E)}_y| \right] 
			\nonumber \\
			&=& - \frac{i}{\hbar} \left[ \hat{h}_y, |\psi^{(E)}_y\rangle \langle \psi^{(E)}_y|\right]\;, 
		\end{eqnarray}  
		which represents the continuity equation of the model. 
		Taking the trace of the above identity we arrive to relation $\partial_y \mbox{Tr}[ \hat{J}^{(E)}_y] = 0$ 
		which establishes current conservation along the 1D wire, i.e. 
		\begin{equation} 
			\mbox{Tr}[ \hat{J}^{(E)}_y] = 
			-  \tfrac{i\hbar}{2m} \left( \langle \psi^{(E)}_y|\partial_y \psi^{(E)}_y\rangle - 
			\langle\partial_y  \psi^{(E)}_y|\psi^{(E)}_y\rangle\right) \nonumber \\
			= \mbox{const} 
			\;. 
		\end{equation}  
		Expanding the w.r.t. to the decomposition~(\ref{eq:decompose}) we can rewrite such constraint in the form 
		\begin{eqnarray} 
			\sum_{\ell\in\{0,1\}}\left[ \left( C_y^{(\ell)}\right)^* D_y^{(\ell)} - 
			C_y^{(\ell)}\left( D_y^{(\ell)}\right)^* \right]  + 2\sum_{\ell,\ell'\in\{0,1\}} \left( C_y^{(\ell)}\right)^*  [K_y]_{\ell,\ell'} C_y^{(\ell')} =\mbox{const.} 
		\end{eqnarray} 
		where we used the fact that the Berry matrix  $K_y$ of the process defined in Eq.~(\ref{eq:berrymat}) is skew-symmetric. 
		Expressed in a more compact form we can rewrite the above identity as
		\begin{eqnarray} \label{currentconservation1} 
			\bm{C}^\dag_y \cdot \bm{D}_y-\bm{D}^\dag_y \cdot \bm{C}_y + 2 \bm{C}^\dag_y \cdot K_y 
			\cdot \bm{C}_y 
			=\mbox{const.} 
		\end{eqnarray} 
		Applying Eq.~(\ref{currentconservation1})
		at the borders of the scattering region we obtain the identity 
		\begin{equation} 
			\bm{C}^\dag_{y_{\rm L}}  \cdot \bm{D}_{y_{\rm L}}-\bm{D}^\dag_{y_{\rm L}} \cdot \bm{C}_{y_{\rm L}} = \bm{C}^\dag_{y_{\rm R}}  \cdot \bm{D}_{y_{\rm R}}-\bm{D}^\dag_{y_{\rm R}} \cdot \bm{C}_{y_{\rm R}}\;,  \label{fin} 
		\end{equation} 
		where we used the continuity condition~(\ref{eq:conditionK}). 
		Invoking~(\ref{DEFCL}) we can then write
		\begin{eqnarray}
			\bm{C}^\dag_{y_{\rm L}}  \cdot \bm{D}_{y_{\rm L}}-\bm{D}^\dag_{y_{\rm L}} \cdot \bm{C}_{y_{\rm L}}
			&=&   i \bm{A}^\dag \left[ F^\dag_{\rm L} + W^\dag r^\dag_E \left(W^{-1}\right)^\dag F_{\rm L}\right] V\left[ F_{\rm L} - F^\dag_{\rm L}W^{-1} r_E W\right]\bm{A}\nonumber \\
			&&+
			i \bm{A}^\dag \left[ F^\dag_{\rm L} - W^\dag r^\dag_E \left(W^{-1}\right)^\dag F_{\rm L}\right] V^\dag\left[ F_{\rm L} + F^\dag_{\rm L}W^{-1} r_E W\right]\bm{A} \nonumber \\
			&=&
			i \bm{A}^\dag \left[ F^\dag_{\rm L} (V+ V^\dag) F_{\rm L} - W^\dag r^\dag_E \left(W^{-1}\right)^\dag F_{\rm L} (V+ V^\dag) F^\dag_{\rm L}W^{-1} r_E W\right] \bm{A} \;, 
		\end{eqnarray} 
		and 
		\begin{eqnarray}
			\bm{C}^\dag_{y_{\rm R}}  \cdot \bm{D}_{y_{\rm R}}-\bm{D}^\dag_{y_{\rm R}} \cdot \bm{C}_{y_{\rm R}}
			&=&   i \bm{A}^\dag  \left[ W^\dag t^\dag_E \left(W^{-1}\right)^\dag F^\dag_{\rm R}  (V + V^\dag) 
			F_{\rm R} W^{-1} t_E W\right] \bm{A} \;.
		\end{eqnarray} 
		Accordingly Eq.~(\ref{fin}) can be expressed as
		\begin{equation} 
			i \bm{A}^\dag \left[ F^\dag_{\rm L} (V+ V^\dag) F_{\rm L} - W^\dag r^\dag_E \left(W^{-1}\right)^\dag F_{\rm L} (V+ V^\dag) F^\dag_{\rm L}W^{-1} r_E W-W^\dag t^\dag_E \left(W^{-1}\right)^\dag F^\dag_{\rm R}  (V + V^\dag) 
			F_{\rm R} W^{-1} t_E W
			\right]
			\bm{A} =0\;, 
		\end{equation}
		which, since it must hold for all choices of  the input vector $\bm{A}$, 
		can be recast in the operator identity 
		\begin{equation} \label{questaqui-1} 
			W^{\dag} r^\dag_E \left[  \left(W^{-1}\right)^\dag F_{\rm L} (V+ V^\dag) F^\dag_{\rm L}W^{-1}
			\right]  r_E  W +W^{\dag} t^\dag_E \left[ \left(W^{-1}\right)^\dag F^\dag_{\rm R}  (V + V^\dag) 
			F_{\rm R} W^{-1}\right] t_E W
			=  F^\dag_{\rm L} (V+ V^\dag) F_{\rm L}  \;.
		\end{equation}
		Consider first the two-channel scenario. Here  $k_E(0)$ and $k_E(1)$ are both real so that 
		$V$ and $W$ are Hermitian and $F_{\rm L,\rm R}$ are unitary matrices. Invoking the fact that  all commute  and using the  identity
		\begin{eqnarray} 
			W^{-1} VW^{-1} = k_E^{(0)}  \mathbb{1}\;,\label{IDENTITY0}
		\end{eqnarray}
		equation~\eqref{questaqui-1} can be expressed as 
		\begin{equation} 
			k_E^{(0)}  W ( r^\dag_E  r_E +   t^\dag_E  t_E ) W=V \quad 
			\Longleftrightarrow \quad  k_E^{(0)}  (r^\dag_E  r_E +   t^\dag_E  t_E) =W^{-1} V W^{-1}  = k_E^{(0)} \mathbb{1}\;, 
		\end{equation} 
		that corresponds to (\ref{eq:currentconservation}) in the main text. 
		In the single-channel scenario 
		only $k_E(0)$ is real, while $k_E^{(1)}$ is an imaginary quantity so neither $V$ or $W$ are
		Hermitian and $F_{\rm L,\rm R}$ are no longer unitary matrices.
		Since all these operators still commutes it follows that 
		\begin{eqnarray} 
			\left(W^{-1}\right)^\dag F_{\rm L} (V+ V^\dag) F^\dag_{\rm L}W^{-1} =
			\left(W^{-1}\right)^\dag F^\dag_{\rm R}  (V + V^\dag) 
			F_{\rm R} W^{-1}= 
			\begin{pmatrix}
				2 k_E^{(0)} & 0\\
				0 &  0
			\end{pmatrix} = 2 k_E^{(0)} \Pi_0 \;,  
		\end{eqnarray} 
		with $\Pi_0:= \begin{pmatrix}
			1 & 0\\
			0 &  0
		\end{pmatrix}$ the projector on the low-energy  band of the model. 
		Therefore 
		defining 
		\begin{eqnarray} 
			\bar{r}_E := \Pi_0{r}_E \Pi_0 =\begin{pmatrix}
				[r_E]_{00} & 0\\
				0 &  0
			\end{pmatrix}\;,  \qquad 
			\bar{t}_E := \Pi_0{t}_E \Pi_0=\begin{pmatrix}
				[t_E]_{00} & 0\\
				0 &  0
			\end{pmatrix}\;,\end{eqnarray}  the restrictions of $r_E$ and 
		$t_E$ to the lowest energy level of the model,  
		Eq.~\eqref{questaqui-1} can be written as 
		\begin{equation} 
			\bar{r}^\dag_E  \bar{r}_E +   \bar{t}^\dag_E  \bar{t}_E=\Pi_0 \qquad \Longleftrightarrow \qquad 
			|[r_E]_{00}|^2 + |[t_E]_{00}|^2 =1 \;,
		\end{equation}
		as anticipated in the main text. 
		We conclude by reminding that, 
		according to the scattering (Landauer-B\"uttiker) approach~\cite{shot-noise}, the current flowing in the 1D wire can be expressed in terms of the transmission matrix $t_E$, via the formula
		\begin{equation}
			\mathcal{I}=\frac{e}{2 \pi \hbar} \int dE \, G_E\left( f_{\rm L}(E)-f_{\rm R}(E)\right),
		\end{equation}
		where $f_{{\rm L}/{\rm R}}(E)$ are the Fermi function in the left and right leads and 
		$G_E= \mbox{Tr}\left[ t_E^{\dagger}t_E \right]$ in the two-channel scenario, and 
		$G_E= \mbox{Tr}\left[ \bar{t}_E^{\dagger}\bar{t}_E \right]= |[t_E]_{00}|^2$ in the single-channel scenario. 
		
		\section{Transmission and reflection matrices} \label{appendixB} 
		In this section, we show how to solve the system in Eq.~\eqref{eq:system} for the matrices $r_E$ and $t_E$. 
		Before going deeper, we decompose in blocks the ordered exponential in Eq.~\eqref{eq:system} by writing 
		\begin{equation}
			\tilde{\Gamma}^{(E)}_{y_{\rm L} \rightarrow y_{\rm R}} = \overleftarrow{{\rm \exp}}\left\{\int_{y_{\rm L}}^{y_{\rm R}} dy \, \tilde{\mathcal{M}}^{(E)}_y\right\} = 
			\begin{bmatrix}
				X_{0,0} & X_{0,1} \\
				X_{1,0} & X_{1,1}
			\end{bmatrix}, \label{questaqui-2} 
		\end{equation}
		where each $X_{i,j}$ is a two-by-two matrix. By doing so, Eq.~\eqref{eq:system} can be equivalently written as
		\begin{equation}
			\begin{bmatrix}
				F_{\rm R} W^{-1} t_E W  \bm{A} \\
				i V F_{\rm R} W^{-1} t_E W \bm{A} 
			\end{bmatrix}
			=    \begin{bmatrix}
				\mathcal{U}_{y_{\rm L} \to y_{\rm R}} & 0 \\
				0 & \mathcal{U}_{y_{\rm L} \to y_{\rm R}}
			\end{bmatrix}
			\begin{bmatrix}
				X_{0,0} & X_{0,1} \\
				X_{1,0} & X_{1,1}
			\end{bmatrix}
			\begin{bmatrix} 
				( F_{\rm L} +  F^\dag_{\rm L} W^{-1} r_E W) \bm{A} \\
				i V  ( F_{\rm L} - F^\dag_{\rm L} W^{-1} r_E W) \bm{A}
			\end{bmatrix}\;.
			\label{eqa} 
		\end{equation}
		Exploiting the fact that the above expression must hold for all input vectors $\bm A$ we can finally translate it into the system~(\ref{fondamentale}) which we report here for completeness 
		\begin{eqnarray}\left\{ \begin{array}{rl} 
				W^{-1} t_E W  
				=& \; F^{\dagger}_{\rm R} \mathcal{U}_{y_{\rm L} \to y_{\rm R}} \Big(
				X_{0,0} +  i X_{0,1} V \Big)F_{\rm L}   +  F^{\dagger}_{\rm R} \mathcal{U}_{y_{\rm L} \to y_{\rm R}} \Big(
				X_{0,0} -  i X_{0,1} V \Big)F^\dag_{\rm L}  W^{-1} r_E W\;, 
				\\ \\
				i V W^{-1} t_E W  
				=& \; F^{\dagger}_{\rm R} \mathcal{U}_{y_{\rm L} \to y_{\rm R}} \Big(
				X_{1,0} +  i X_{1,1} V \Big) F_{\rm L}  +  F^{\dagger}_{\rm R} \mathcal{U}_{y_{\rm L} \to y_{\rm R}} \Big(
				X_{1,0} -  i X_{1,1} V \Big) F^\dag_{\rm L}  W^{-1} r_E W\;. 
			\end{array} \right.
			\label{nuovaeq}  
		\end{eqnarray}
		These can be equivalently expressed as 
		\begin{equation}
			\left\{ \begin{array}{l}   \Big[ \mathcal{U}_{y_{\rm L} \to y_{\rm R}} (X_{1,1}V + i X_{1,0})  +V \mathcal{U}_{y_{\rm L} \to y_{\rm R}} (X_{0,0} - i X_{0,1} V)\Big]  W^{-1} r_E W =\\
				\qquad \qquad \qquad\qquad \qquad \qquad   \mathcal{U}_{y_{\rm L} \to y_{\rm R}} (X_{1,1}V - i X_{1,0})  -V \mathcal{U}_{y_{\rm L} \to y_{\rm R}} (X_{0,0} + i X_{0,1} V)\;, \\\\
				t_E  =  WF^{\dagger}_{\rm R} \mathcal{U}_{y_{\rm L} \to y_{\rm R}} \Big[
				X_{0,0}+i X_{0,1} V\Big] W^{-1} \label{eq:tsolfin1} 
				+ WF^{\dagger}_{\rm R} \mathcal{U}_{y_{\rm L} \to y_{\rm R}} \Big[
				X_{0,0}-i X_{0,1} V\Big] W^{-1} r_E
				\;,\end{array} \right.
		\end{equation}
		which we wrote  assuming without loss of generality $y_{\rm L}=0$, 
		so that $F_{\rm L}=\mathbb{1}$.
		By matrix inversion the first equation leads to the following solution for $r_E$
		\begin{equation}
			\label{eq:rsystem}
			r_E = W \Big[ \mathcal{U}_{y_{\rm L} \to y_{\rm R}} (X_{1,1}V + i X_{1,0})  +V \mathcal{U}_{y_{\rm L} \to y_{\rm R}} (X_{0,0} - i X_{0,1} V)\Big]^{-1} \Big[ \mathcal{U}_{y_{\rm L} \to y_{\rm R}} (X_{1,1}V - i X_{1,0})  -V \mathcal{U}_{y_{\rm L} \to y_{\rm R}} (X_{0,0} + i X_{0,1} V)\Big]W^{-1} \;, 
		\end{equation}
		which inserted in the second equation gives $t_E$. The explicit evolution of this function
		requires us to 
		to compute~\eqref{questaqui-2}, which can be done numerically in the case of interested.

		\section{Scattering problem with large injection energy}
		\label{appx:Einf}
		In this appendix we show how the system in Eq.~\eqref{eq:system} can be solved analytically in the 
		limiting case~(\ref{zeemansmall}) obtaining (\ref{eq:tequalsholo}).
		Indeed when the injection energy is much larger than the
		Zeeman gap, we can invoke the approximation Eq.~(\ref{condk}) 
		from which it follows that the matrices $W$, $V$, $F_{\rm R}$ are all proportional to the
		identity, i.e. 
		\begin{eqnarray} W\simeq  \mathbb{1}\;, \qquad \quad 
			V\simeq k_E \mathbb{1}\;, \qquad \quad  \label{wcond} 
			F_{{\rm R}}\simeq e^{ik_E y_{\rm R}}\mathbb{1}\;. \end{eqnarray} 
		Similarly from Eq.~(\ref{approxBIGE}) it also follows that 
		\begin{eqnarray} X_{0,0} \simeq X_{1,1} \simeq \cos(k_E L)\mathbb{1}\;, \qquad \quad 
			X_{0,1} \simeq  k_E^{-1}\sin(k_E L)\mathbb{1}\;, \qquad\quad 
			X_{1,0} \simeq  - k_E\sin(k_E L)\mathbb{1}\;.
		\end{eqnarray} 
		Accordingly we can conclude that in the limit~(\ref{zeemansmall}) the following identities hold
		\begin{eqnarray}\nonumber
			\mathcal{U}_{y_{\rm L} \to y_{\rm R}} (X_{1,1}V - i X_{1,0}) 
			- V \mathcal{U}_{y_{\rm L} \to y_{\rm R}} (X_{0,0} + i X_{0,1} V)
			&\simeq& \mathcal{U}_{y_{\rm L} \to y_{\rm R}} ( k_EX_{1,1} - i X_{1,0} -  k_EX_{0,0} - ik_E^2 X_{0,1} )\nonumber \\
			&=& \mathcal{U}_{y_{\rm L} \to y_{\rm R}} ( k_E(X_{1,1} - X_{0,0}) -i (X_{1,0} + k_E^2 X_{0,1}) )=0\;, \nonumber \\
			\nonumber \\
			\mathcal{U}_{y_{\rm L} \to y_{\rm R}} (X_{1,1}V + i X_{1,0})  +V \mathcal{U}_{y_{\rm L} \to y_{\rm R}} (X_{0,0} - i X_{0,1} V) &\simeq& \mathcal{U}_{y_{\rm L} \to y_{\rm R}} ( k_EX_{1,1} + i X_{1,0} + k_EX_{0,0} - ik_E^2 X_{0,1} )\nonumber \\
			&=& \mathcal{U}_{y_{\rm L} \to y_{\rm R}} ( k_E(X_{1,1} + X_{0,0}) +i (X_{1,0} - k_E^2 X_{0,1}) ) \nonumber\\
			&=& 
			2k_E ( \cos(k_E L)-i\sin(k_E L))\mathcal{U}_{y_{\rm L} \to y_{\rm R}}\nonumber \\
			&=&
			2k_E e^{-i k_E L}  \mathcal{U}_{y_{\rm L} \to y_{\rm R}}\;, 
		\end{eqnarray} 
		which replaced into~(\ref{eq:tsolfin1}) gives $ r_E\simeq 0$ and
		\begin{eqnarray}
			t_E  &\simeq &  WF^{\dagger}_{\rm R} \mathcal{U}_{y_{\rm L} \to y_{\rm R}} \Big[
			X_{0,0}+i X_{0,1} V\Big] W^{-1} 
			\simeq e^{-ik_E y_{\rm R}}\mathcal{U}_{y_{\rm L} \to y_{\rm R}}(\cos(k_E L)+i\sin(k_E L))
			=
			\mathcal{U}_{y_{\rm L} \to y_{\rm R}} \;,\label{eq:tsolfin1EINF} 
		\end{eqnarray}
		(recall that  Eqs.~(\ref{eq:tsolfin1}) assumed
		$y_{\rm L}=0$ so that $y_{\rm R}=L$, which we used in the last identity to simplify the exponential function).

		\subsection{Formal derivation of the quench ansatz} \label{app:ansatz} 
		In this section we clarify how the quench ansatz leads to the identity (\ref{IIMPP}).
		As a prerequisite for this derivation 
		recall that given $O$  a $3\times 3$ real, orthogonal matrix describing a rotation in the 3D cartesian space, its unitary representation $\hat{\mathcal{U}}(O)$  on the Hilbert space induces the following transformation on the Pauli operators 
		\begin{eqnarray}\hat{\mathcal{U}}^\dag(O){\hat{\bm \sigma}} \hat{\mathcal{U}}(O)&=& O {\hat{\bm \sigma}}\;, \end{eqnarray} 
		which, given $\bm n$ a 3D real vector, and ${\hat{\bm \sigma}}_{\bm n} = {\bm n}\cdot \hat{\bm \sigma}
		$ implies \begin{eqnarray}
			\hat{\mathcal{U}}^\dag(O){\hat{\bm \sigma}}_{\bm n} \hat{\mathcal{U}}(O)&=& {\hat{\bm \sigma}}_{\bm n'}\;, \qquad 
			{\bm n'} = O^{-1} {\bm n}  \;, 
		\end{eqnarray} 
		Accordingly the eigenvectors $\{ \ket*{\phi^{(\ell)}_{{\bm n}}}\}_{\ell=0,1}$ of ${\hat{\bm \sigma}}_{\bm n}$ can be related to the eigenvectors   
		$\{ \ket*{\phi^{(\ell)}_{{\bm n}'}}\}_{\ell=0,1}$ of ${\hat{\bm \sigma}}_{\bm n'}$ via the identity
		\begin{eqnarray}\label{IMO} 
			\hat{\mathcal{U}}(O) \ket*{\phi^{(\ell)}_{{\bm n}'}} = \ket*{\phi^{(\ell)}_{{\bm n}}}\;. 
		\end{eqnarray} 
		
		\subsubsection{The quench ansatz} 
		
		In this large energy regime, one might be tempted to conclude that the scattering process has a negligible impact on the particle's propagation. However, it is important to note that in the present scenario, due to the differing orientations of the magnetic field in the left and right leads, the condition~(\ref{zeemansmall}) does not necessarily imply that the scattering region can be treated as a minor perturbation. A better ansatz can be obtained by 
		drawing an analogy with dynamical quenches, suggesting that a particle propagating from left to right in the wire "experiences" a sudden, abrupt change in the external magnetic field
		that leaves the spin invariant and induces no reflections. To better frame this
		property we find it useful to rewrite the right lead component of the  eigenstate $\ket*{\psi^{(E)}_{y}}$ given in Eq.~(\ref{eq:asymptR}) in terms of 
		the local Zeeman spinors of the left lead. This can be done by using Eq.~(\ref{IMO}) to write  
		\begin{eqnarray} \label{identityrot} 
			\ket*{\phi^{(\ell)}_{{\bm n}_{\rm R}}} = \hat{\mathcal{U}}(O) \ket*{\phi^{(\ell)}_{{\bm n}_{\rm L}}}\;, \qquad \forall \ell\in\{0,1\}\;,
		\end{eqnarray}  
		with $O$ being a $3\times 3$ real orthogonal matrix which connects $\bm{n}_{\rm L}$ and $\bm{n}_{\rm R}$, i.e. $\bm{n}_{\rm R}=  O \; \bm{n}_{\rm L}$. 
		Observe next that for large $E$, invoking~(\ref{condk}) the solution 
		Eq.~(\ref{eq:asymptR}) can be expressed as 
		\begin{equation}
			\ket*{\psi^{(E)}_{y}}\simeq \left\{ \begin{array}{ll}e^{i k_Ey} \sum_{\ell\in \{0,1\}} A_E^{(\ell)} \ket*{\phi^{(\ell)}_{\bm{n}_{\rm L}}}+e^{-i k_Ey}  \sum_{\ell\in \{0,1\}}R_E^{(\ell)}\ket*{\phi^{(\ell)}_{\bm{n}_{\rm L}}}\;,&\qquad  \forall y \leq y_{\rm L} \\\\
				e^{i k_Ey} \sum_{\ell\in \{0,1\}} T_E^{(\ell)} \ket*{\phi^{(\ell)}_{\bm{n}_{\rm R}}}=e^{i k_Ey}  \sum_{\ell\in \{0,1\}} T_E^{(\ell)} \hat{\mathcal{U}}(O) \ket*{\phi^{(\ell)}_{\bm{n}_{\rm L}}} \;,    &\qquad \forall y \geq y_{\rm R}\;.\end{array} \right. 
			\label{eq:asymptRRas}
		\end{equation}
		The quench ansatz requires that for very large $E$ the reflection term disappears, and that the spinor component of the input coincides with the transmitted one, i.e. 
		\begin{eqnarray}
			\sum_{\ell\in \{0,1\}} A_E^{(\ell)} \ket*{\phi^{(\ell)}_{\bm{n}_{\rm L}}}&=&
			\sum_{\ell\in \{0,1\}} T_E^{(\ell)}\hat{\mathcal{U}}(O) \ket*{\phi^{(\ell)}_{\bm{n}_{\rm L}}} = \sum_{\ell,\ell' \in \{0,1\}}  [W^{-1} t_E W]_{\ell,\ell'} 
			A_E^{(\ell')}\hat{\mathcal{U}}(O)  \ket*{\phi^{(\ell)}_{\bm{n}_{\rm L}}}  \nonumber \\
			&\simeq&   \sum_{\ell,\ell' \in \{0,1\}}  [\mathcal{U}_{y_{\rm L} \rightarrow y_{\rm R}}]_{\ell,\ell'} 
			A_E^{(\ell')}\hat{\mathcal{U}}(O)  \ket*{\phi^{(\ell)}_{\bm{n}_{\rm L}}} \;,
			\label{eq:asymptRR1}
		\end{eqnarray}
		where we wrote $T_E^{(\ell)}$
		in terms of the input amplitudes $A_E^{(\ell)}$ via
		the identity (\ref{identityT}) of App.~\ref{app:pre} and used Eqs.~(\ref{eq:tequalsholo}) and (\ref{wcond}). 
		Projecting on  $\ket*{\phi^{(\ell)}_{\bm{n}_{\rm L}}}$ we obtain
		\begin{equation}
			\sum_{\ell'',\ell' \in \{0,1\}}  \bra*{\phi^{(\ell)}_{\bm{n}_{\rm L}}}
			\hat{\mathcal{U}}(O)  \ket*{\phi^{(\ell'')}_{\bm{n}_{\rm L}}} [\mathcal{U}_{y_{\rm L} \rightarrow y_{\rm R}}]_{\ell'',\ell'} 
			A_{E}^{(\ell')}=  A_E^{(\ell)}\;,
			\label{eq:asymptRR1ss} 
		\end{equation}
		which, since it  must hold for all input coefficients $A_E^{(\ell)}$, translates into the following matrix identity, 
		\begin{eqnarray}
			\sum_{\ell'' \in \{0,1\}}  \bra*{\phi^{(\ell)}_{\bm{n}_{\rm L}}}
			\hat{\mathcal{U}}(O)  \ket*{\phi^{(\ell'')}_{\bm{n}_{\rm L}}} [\mathcal{U}_{y_{\rm L} \rightarrow y_{\rm R}}]_{\ell'',\ell'} 
			=  \delta_{\ell,\ell'}  \quad &\Longrightarrow& \quad  \hat{\mathcal{U}}(O)  
			\sum_{\ell'' \in \{0,1\}}
			[\mathcal{U}_{y_{\rm L} \rightarrow y_{\rm R}}]_{\ell'',\ell'}  \ket*{\phi^{(\ell'')}_{\bm{n}_{\rm L}}} \bra*{\phi^{(\ell')}_{\bm{n}_{\rm L}}}
			=\mathbb{1} \nonumber \\\nonumber
			\quad &\Longrightarrow& \quad  
			\sum_{\ell'' \in \{0,1\}}
			[\mathcal{U}_{y_{\rm L} \rightarrow y_{\rm R}}]_{\ell'',\ell'}  \ket*{\phi^{(\ell'')}_{\bm{n}_{\rm L}}} \bra*{\phi^{(\ell')}_{\bm{n}_{\rm L}}}=\hat{\mathcal{U}}^\dag(O)\;, 
			\label{eq:asymptRR1ss1}  \end{eqnarray}
		so that 
		\begin{eqnarray} 
			[\mathcal{U}_{y_{\rm L} \rightarrow y_{\rm R}}]_{\ell'',\ell'}  =\bra*{\phi^{(\ell'')}_{\bm{n}_{\rm L}}}
			\hat{\mathcal{U}}^\dag(O) \ket*{\phi^{(\ell')}_{\bm{n}_{\rm L}}}=\bra*{\phi^{(\ell'')}_{\bm{n}_{\rm R}}} \ket*{\phi^{(\ell')}_{\bm{n}_{\rm L}}}
			\;,
		\end{eqnarray} 
		which corresponds to Eq.~(\ref{IIMPP}).
		
		\subsection{Leading order}\label{appendixH} 
		Here we would like to express Eq.~\eqref{defDE} by evidencing the solution in the high energy regime, namely~\eqref{approxBIGE}. By doing so we can rewrite~\eqref{defDE} according to 
		\begin{equation}
			\tilde{\Gamma}^{(E)}_{y_{\rm L} \rightarrow y_{\rm R}}=e^{\mathcal{M}_E L} \; \overleftarrow{\exp}\left[\int_{y_{\rm L}}^{y_{\rm R}} dy e^{-\mathcal{M}_E y} \Delta_{y} e^{\mathcal{M}_E y}\right]
			\label{eq:develop}\;, 
		\end{equation}
		formally defining \(\Delta_y:=\tilde{\mathcal{M}}^{(E)}_{y}-\mathcal{M}_E=\begin{bmatrix}
			0 & 0 \\
			-\frac{2m}{\hbar^2}\tilde{\Omega}_y & 0
		\end{bmatrix}\).
		At the first order in the series expansion of the ordered exponential in~\eqref{eq:develop} we have
		\begin{equation}
			\tilde{\Gamma}^{(E)}_{y_{\rm L} \rightarrow y_{\rm R}}\simeq e^{\mathcal{M}_E L}\left(\mathbb{1}+ \int_{y_{L}}^{y_{\rm R}} dy \; e^{-\mathcal{M}_E y} \Delta_{y} e^{\mathcal{M}_E y}
			\right)
			\equiv
			\begin{bmatrix}
				X^{[0]}_{00} & 	X^{[0]}_{01} \\
				X^{[0]}_{10} & 	X^{[0]}_{11}
			\end{bmatrix}
			+
			\begin{bmatrix}
				X^{[1]}_{00} & 	X^{[1]}_{01} \\
				X^{[1]}_{10} & 	X^{[1]}_{11}
			\end{bmatrix}\;, 
		\end{equation}
		with 
		\begin{align}
			\begin{bmatrix}
				X^{[0]}_{00} & 	X^{[0]}_{01} \\
				X^{[0]}_{10} & 	X^{[0]}_{11}
			\end{bmatrix}\equiv e^{\mathcal{M}_E L}=\begin{bmatrix}
				{\scriptstyle \cos(k_E L) \mathbb{1}} &   {\scriptstyle k_E^{-1}\sin(k_E L)\mathbb{1} }\\
				{\scriptstyle - k_E\sin(k_E L)\mathbb{1}} & {\scriptstyle \cos(k_E L)\mathbb{1}}
			\end{bmatrix}, \qquad
			\begin{bmatrix}
				X^{[1]}_{00} & 	X^{[1]}_{01} \\
				X^{[1]}_{10} & 	X^{[1]}_{11}
			\end{bmatrix}
			=
			e^{\mathcal{M}_E L} \int_{y_{L}}^{y_{\rm R}} dy \; e^{-\mathcal{M}_E y} \Delta_{y} e^{\mathcal{M}_E y}\;.
		\end{align}
		Then we can write that \(t_E=t^{[0]}_E+t^{[1]}_E=\mathcal{U}_{y_{\rm L} \to y_{\rm R}} + t^{[1]}_E\) and \(r_E=r^{[0]}_E+r^{[1]}_E=r^{[1]}_E\), with \(t^{[1]}_E\) and \(r^{[1]}_E\) to be determined. Similarly we have that
		\begin{equation}
			V=V^{[0]} + V^{[1]},\qquad V^{[0]}=k_E \mathbb{1},\qquad V^{[1]}=v^{[1]}_E\sigma_z,\qquad v^{[1]}_E=\frac{k_E^2-(k^{(1)}_E)^2}{2k_E},
		\end{equation}
		\begin{equation}
			W=W^{[0]} + W^{[1]},\qquad W^{[0]}=\mathbb{1},\qquad W^{[1]}=\begin{pmatrix}
				0&0\\
				0& w^{[1]}_E
			\end{pmatrix},\qquad w^{[1]}_E=-\frac{v_E^{[1]}}{k_E}\;, 
		\end{equation}
		and
		\begin{equation}
			F_{\rm R}=F^{[0]}_{\rm R}+F^{[1]}_{\rm R},\qquad F^{[0]}_{\rm R}=e^{ik_EL}\mathbb{1},\qquad F^{[1]}_{\rm R}=iLv^{[1]}_Ee^{ik_EL}\hat{\sigma}_z\;. 
		\end{equation}
		By looking at~\eqref{eq:rsystem}, the first order for the reflection matrix results to be
		\begin{eqnarray}
			r^{[1]}_E &=& \frac{e^{i2k_EL}}{2k_E}\left[v^{[1]}_E \left(\tilde{\sigma}_z(y_{\rm R}) + \hat{\sigma}_z
			\right) +i\frac{2m}{\hbar^2} \int_{y_{\rm L}}^{y_{\rm R}}dy	\tilde{\Omega}_y\right] \\
			&=& \frac{e^{i2k_EL}}{2}\left[1-\left(\tfrac{k^{(1)}_E}{k_E}\right)^2\right]\left[\left(\tilde{\sigma}_z(y_{\rm R}) + \hat{\sigma}_z
			\right)-k_E\int_{y_{\rm L}}^{y_{\rm R}} dy  \tilde{\sigma}_z(y)\right]\;, \end{eqnarray}
		with  $\tilde{\sigma}_z(y):=\mathcal{U}^{\dagger}_{y_{\rm L} \to y} \hat{\sigma}_z \mathcal{U}_{y_{\rm L} \to y}$.
		In a similar way, we can evaluate the first order transmission matrix \(t^{[1]}_E\) by looking at Eq.~\eqref{eq:tsolfin1}.

		\section{Low-varying field approximation}
		\label{appx:infinitesimalscattering}
		The derivation of the identities (\ref{approxBIGE}) and (\ref{approxslow}) follow by observing that, for $n$ integer,  the 
		block matrix of the form
		\begin{equation}
			{\mathcal{M}} = 
			\begin{bmatrix}
				0 & \mathbb{1} \\ A   & 0
				\label{eq:mx3}
			\end{bmatrix}\;, 
		\end{equation} 
		fulfills the identities 
		\begin{equation}
			{\mathcal{M}}^{2n} = \begin{bmatrix}
				A^{n} & 0 \\   0 & A^{n}
			\end{bmatrix}\;,  \qquad {\mathcal{M}}^{2n+1} = \begin{bmatrix}
				0 &A^{n}  \\    A^{n+1}&0
			\end{bmatrix}\;.
		\end{equation} 
		Accordingly for $x$ real, we can write 
		\begin{eqnarray} 
			{{\rm \exp}}\left\{x{\mathcal{M}} \right\} = \sum_{k=0}^\infty \frac{x^k \mathcal{M}^{k}}{k!} 
			= \begin{bmatrix}
				\sum_{n=0}^\infty  \frac{x^{2n}A^{n}}{(2n)!}& \sum_{n=0}^\infty \frac{x^{2n+1}A^{n}}{(2n+1)!} \\   \sum_{n=0}^\infty \frac{x^{2n+1}A^{n+1}}{(2n+1)!} &\sum_{n=0}^\infty \frac{x^{2n}A^{n}}{(2n)!}
			\end{bmatrix}\;.
		\end{eqnarray}  
		In case $A$ is positive semidefinite one has 
		\begin{eqnarray} 
			A = \left(\sqrt{A}\right)^2\;, 
		\end{eqnarray}  
		so that 
		\begin{eqnarray} 
			\sum_{n=0}^\infty  \frac{x^{2n}A^{n}}{(2n)!}
			&=&   \sum_{n=0}^\infty  \frac{\left(x\sqrt{A}\right)^{2n}}{(2n)!}= \cos\left(x\sqrt{A}\right)\;,\\
			\sum_{n=0}^\infty \frac{x^{2n+1}A^{n}}{(2n+1)!} 
			&=&  \frac{1}{\sqrt{A}} \sum_{n=0}^\infty  \frac{\left(x\sqrt{A}\right)^{2n+1}}{(2n+1)!}=
			\frac{1}{\sqrt{A}}   \sin\left(x\sqrt{A}\right)\;,
			\\
			\sum_{n=0}^\infty \frac{x^{2n+1}A^{n+1}}{(2n+1)!} 
			&=&{\sqrt{A}} \sum_{n=0}^\infty  \frac{\left(x\sqrt{A}\right)^{2n+1}}{(2n+1)!}=
			{\sqrt{A}}   \sin\left(x\sqrt{A}\right)\;,
		\end{eqnarray}  
		Accordingly we can write 
		\begin{eqnarray} 
			{{\rm \exp}}\left\{x{\mathcal{M}} \right\} = {\cal D}_x(A):= 
			\begin{bmatrix}
				\cos(x\sqrt{A}) & \frac{1}{\sqrt{A}} \sin(x\sqrt{A}) \\
				- \sqrt{A} \sin(x\sqrt{A}) & \cos(x\sqrt{A})
			\end{bmatrix} \label{equationD} \;.
		\end{eqnarray}  
		A similar expression applies also when $A$ is Hermitian but not necessarily positive semidefinite. 
		In such case it is useful to write 
		\begin{eqnarray}
			A =A_{+} - A_{-} \;, 
		\end{eqnarray} 
		with $A_{+} ,A_{-}\geq0$,  $A_{+} A_{-}=0$  representing the positive and negative parts of $A$. 
		Accordingly we have
		\begin{eqnarray} 
			\sqrt{A}= \sqrt{A_+} + i \sqrt{ A_{-}}\;, \qquad \qquad 
			\frac{1}{A} =  \frac{1}{\sqrt{A_+}}+ \frac{-i}{ \sqrt{A_-}} \;, \end{eqnarray} 
		and
		the cosine and sine operators appearing in Eq.~(\ref{equationD}) can be computed as
		\begin{eqnarray} 
			\cos(x\sqrt{A})&=& \cos(x\sqrt{A_+}) \oplus \cosh(x\sqrt{A_-})\;, \\ 
			\sin(x\sqrt{A})&=& \sin(x\sqrt{A_+})\oplus   i \sinh(x\sqrt{A_-})=\sin(x\sqrt{A_+})+   i \sinh(x\sqrt{A_-})\;,
		\end{eqnarray} 
		where  the symbol $\oplus$  has been introduced to indicate that the sum must be done
		by projecting the  functions $f({A_\pm})$ on the support of ${A_\pm}$. 
		Therefore we can now write 
		\begin{eqnarray} 
			\tfrac{1}{\sqrt{A}}\sin(x\sqrt{A})&=& \tfrac{1}{\sqrt{A_+}}\sin(x\sqrt{A_+})+
			\tfrac{1}{\sqrt{A_-}}\ \sinh(x\sqrt{A_-})\;, \\
			{\sqrt{A}}\sin(x\sqrt{A})&=& {\sqrt{A_+}}\sin(x\sqrt{A_+}) - 
			{\sqrt{A_-}} \sinh(x\sqrt{A_-})\;.
		\end{eqnarray} 
		Equations  (\ref{approxslow}) and  (\ref{approxBIGE})  follow finally by setting $x=L$ and taking the operator $A$  in Eq.~(\ref{equationD}) equal to $Q_E$ and $-k_E^2 \mathbb{1}$ respectively.

		\subsection{Infinitesimal scattering regions} 
		
		In the limit $L\rightarrow 0$ 
		from Eq.~(\ref{approxslow}) it follows that 
		\begin{eqnarray}\label{approxslow1} 
			\tilde{\Gamma}^{(E)}_{y_{\rm L} \rightarrow y_{\rm R}}
			= \mathbb{1} \qquad  \Longrightarrow \qquad X_{0,0} = X_{1,1} =  \mathbb{1} \;, X_{1,0} = X_{0,1} =  0\;. \end{eqnarray}
		Accordingly the identities~(\ref{nuovaeq}) become
		\begin{eqnarray}
			\left\{ \begin{array}{rl} 
				W^{-1} t_E W  
				=& \;  \mathcal{U}_{y_{\rm L} \to y_{\rm R}} ( \mathbb{1} +    W^{-1} r_E W)\;, 
				\\ \\
				W^{-1} t_E W  
				=& \; V^{-1} \mathcal{U}_{y_{\rm L} \to y_{\rm R}} 
				V ( \mathbb{1} -    W^{-1} r_E W)\;.
			\end{array} 
			\right.
		\end{eqnarray}
		Introducing the matrix 
		\begin{eqnarray} 
			G:=  \mathcal{U}^\dag_{y_{\rm L} \to y_{\rm R}} V^{-1} \mathcal{U}_{y_{\rm L} \to y_{\rm R}} V\;, 
		\end{eqnarray} 
		and reorganizing the various terms this leads to the following recursive expression for 
		the reflection matrix 
		\begin{eqnarray}
			(W^{-1} r_E W) = (G-\mathbb{1} ) - G (W^{-1} r_E W)\end{eqnarray}
		whose solution can be expressed as 
		\begin{eqnarray} 
			(W^{-1} r_E W) = \sum_{k=0}^\infty (- G)^k (G-\mathbb{1} ) = \frac{G-\mathbb{1}}{G+\mathbb{1} }
			=  \frac{\mathcal{U}^\dag_{y_{\rm L} \to y_{\rm R}} V^{-1} \mathcal{U}_{y_{\rm L} \to y_{\rm R}} V-\mathbb{1}}{\mathcal{U}^\dag_{y_{\rm L} \to y_{\rm R}} V^{-1} \mathcal{U}_{y_{\rm L} \to y_{\rm R}} V+\mathbb{1} }
		\end{eqnarray} 
		which can be turned into the second expression of Eq.~(\ref{rL0}), i.e. 
		\begin{eqnarray} 
			r_E \label{rel0} 
			=  \frac{k_E^{(0)}(W\mathcal{U}^\dag_{y_{\rm L} \to y_{\rm R}} V^{-1} \mathcal{U}_{y_{\rm L} \to y_{\rm R}} W)-\mathbb{1}}{k_E^{(0)} ( W\mathcal{U}^\dag_{y_{\rm L} \to y_{\rm R}} V^{-1} \mathcal{U}_{y_{\rm L} \to y_{\rm R}} W) +\mathbb{1} }=
			\frac{W\mathcal{U}^\dag_{y_{\rm L} \to y_{\rm R}} W^{-2} \mathcal{U}_{y_{\rm L} \to y_{\rm R}} W-\mathbb{1}}{W\mathcal{U}^\dag_{y_{\rm L} \to y_{\rm R}} W^{-2}  \mathcal{U}_{y_{\rm L} \to y_{\rm R}} W +\mathbb{1} }\;,
		\end{eqnarray} 
		with the help of Eq.~(\ref{IDENTITY0}). Observe that by construction the solution is always 
		self-adjoint, i.e. $r_E^\dag = r_E$. 
		Replacing this into the first of Eq.~(\ref{first}) we can then arrive at 
		\begin{eqnarray} 
			t_E 
			&=& W\mathcal{U}^\dag_{y_{\rm L} \to y_{\rm R}} W^{-1} 
			\frac{2k_E^{(0)}(W\mathcal{U}_{y_{\rm L} \to y_{\rm R}} V^{-1} \mathcal{U}_{y_{\rm L} \to y_{\rm R}} W)}{k_E^{(0)} ( W\mathcal{U}^\dag_{y_{\rm L} \to y_{\rm R}} V^{-1} \mathcal{U}_{y_{\rm L} \to y_{\rm R}} W) +\mathbb{1} } \nonumber \\
			&=& 2 k_E^{(0)}
			WV^{-1} \mathcal{U}_{y_{\rm L} \to y_{\rm R}} W
			\frac{\mathbb{1} }{k_E^{(0)} ( W\mathcal{U}^\dag_{y_{\rm L} \to y_{\rm R}} V^{-1} \mathcal{U}_{y_{\rm L} \to y_{\rm R}} W) +\mathbb{1} } \nonumber \\
			&=& 2  W^{-1} \mathcal{U}_{y_{\rm L} \to y_{\rm R}} W
			\frac{\mathbb{1} }{k_E^{(0)} ( W\mathcal{U}^\dag_{y_{\rm L} \to y_{\rm R}} V^{-1} \mathcal{U}_{y_{\rm L} \to y_{\rm R}} W) +\mathbb{1} }\nonumber \\
			&=&2  W^{-1} \mathcal{U}_{y_{\rm L} \to y_{\rm R}} W
			\frac{\mathbb{1} }{W\mathcal{U}^\dag_{y_{\rm L} \to y_{\rm R}} W^{-2} \mathcal{U}_{y_{\rm L} \to y_{\rm R}} W +\mathbb{1} } \;, \label{teLO} 
		\end{eqnarray} 
		which corresponds to the first identity of  Eq.~(\ref{rL0}).
		It is worth observing  that the solutions described here could have been directly obtained by imposing continuity conditions 
		of the asymptotic functions ~(\ref{eq:asymptR}) at the point $y_{\rm L}$. 
		Notice also that for $E\rightarrow \infty$,  
		$W\rightarrow \mathbb{1}$ and
		Eqs.~(\ref{rel0}) and (\ref{teLO}) behave as predicted by 
		Eq.~(\ref{eq:tequalsholo}). 
		
		\section{Scattering problem for piecewise constant elements}\label{app:scatteringpiecewise}
		As anticipated in the main text, when the magnetic field in the scattering region is piecewise constant, the transfer matrix $\Gamma^{(E)}_{y_{\rm L} \rightarrow y_{\rm R}}$
		can be formally integrated as in Eq.~(\ref{decomp}). To see this, given $\epsilon >0$
		an infinitesimal increment, 
		we find it useful to introduce two extra points $y_j^{\pm} := y_j \pm \epsilon$ for each of the $N+1$ elements of the partition~(\ref{partition}). Observe that the set of points 
		\begin{eqnarray}\label{partitionepsilon} 
			y_0^{-} < y_0^{+} < y^{-}_1 < y^{+}_1< y^{-}_2 < y^{+}_2< \dots <y^{-}_{N-1} < y^{+}_{N-1}< y^{-}_{N} < y^{+}_{N} \;, \end{eqnarray}   
		identifies a new collection of non overlapping intervals 
		\begin{eqnarray} \label{newint} 
			I_{j}^{(\epsilon)} :=] y_j^{-},y_j^{+}[ \;,   \qquad \qquad I_{j,j+1}^{(\epsilon)} :=] y_j^{+},y_{j+1}^{-}[ \;, 
		\end{eqnarray} 
		that provide a covering of the scattering region. In particular, 
		$I_{j}^{(\epsilon)}$ are infinitesimal intervals over which  the Berry matrix 
		$K_y$ experience an abrupt change; on the contrary from the properties of (\ref{partition}), it follows that 
		on $I_{j,j+1}^{(\epsilon)}$ the matrix $\Omega_y$
		assumes constant value $\Omega_j$ and the  Berry matrix $K_y$ is null, i.e.
		\begin{equation} 
			\left\{ \begin{array}{l} 
				K_y = 0 \;, \\
				\Omega_y = \Omega_j \;, 
			\end{array} \right. \qquad \forall y\in I_{j,j+1}^{(\epsilon)}   \qquad 
			\Longrightarrow  \qquad 
			\left\{ \begin{array}{l} 
				\mathcal{U}_{y^{+}_{j} \rightarrow y}= \mathbb{1} \\
				\tilde{\Omega}_y = \Omega_j \;, \\
				\mathcal{M}_{y}^{(E)}= \mathcal{M}_{E}^{(j)} := 
				\begin{bmatrix}
					0 & \mathbb{1} \\ \frac{2m}{\hbar^2}({\Omega}_j-E \mathbb{1})   & 0
				\end{bmatrix}=
				\begin{bmatrix}
					0 & \mathbb{1} \\ -Q^{(j)}_E    & 0
					\label{eq:mxapp}
				\end{bmatrix}\;,
			\end{array} \right. \qquad \forall y\in I_{j,j+1}^{(\epsilon)} \;,
		\end{equation} 
		with $Q^{(j)}_E$ as in Eq.~(\ref{defQjE}). 
		Most importantly each of the intervals~(\ref{newint})
		fulfill the continuity condition~(\ref{eq:conditionK}) so that we can invoke the general formula Eq.~(\ref{eq:system}) to propagate the vector $\begin{bmatrix}
			\bm{C}_{y} \\
			\bm{D}_{y}
		\end{bmatrix}$ across them.
		Specifically, for all $j$ we can write 
		\begin{equation}
			\begin{bmatrix}
				\bm{C}_{y_{j}^{+}} \\
				\bm{D}_{y_{j}^{+}}
			\end{bmatrix}=\Gamma^{(E)}_{y_{j}^{-}\rightarrow y_{j}^{+}}  \begin{bmatrix}
				\bm{C}_{y_{j}^{-}} \\
				\bm{D}_{y_{j}^{-}}
			\end{bmatrix}
			\;, \qquad 
			\begin{bmatrix}
				\bm{C}_{y_{j+1}^{-}} \\
				\bm{D}_{y_{j+1}^{-}}
			\end{bmatrix}=\Gamma^{(E)}_{y_{j}^{+}\rightarrow y_{j+1}^{-}}  \begin{bmatrix}
				\bm{C}_{y_{j}^{+}} \\
				\bm{D}_{y_{j}^{+}}
			\end{bmatrix}
			\label{eq:systemj+}\;, 
		\end{equation}
		which by concatenation gives 
		\begin{equation} \Gamma^{(E)}_{y^{-}_{\rm L} \rightarrow y^{+}_{\rm R}} =  
			\Gamma^{(E)}_{y^{-}_{0} \rightarrow y^{+}_{N}}  =
			\Gamma^{(E)}_{y^{-}_{N} \rightarrow y^{+}_{N}} \Gamma^{(E)}_{y^{+}_{N-1} \rightarrow y^{-}_{N}}
			\cdots 
			\Gamma^{(E)}_{y^{+}_{1} \rightarrow y^{-}_{2}}  \Gamma^{(E)}_{y^{-}_{1} \rightarrow y^{+}_{1}} 
			\Gamma^{(E)}_{y^{+}_{0} \rightarrow y^{-}_{1}}  \Gamma^{(E)}_{y^{-}_{0} \rightarrow y^{+}_{0}}  \;.\end{equation}
		The identity~(\ref{decomp}) follows from this by invoking~(\ref{defGAMMAL0}) to compute 
		the transfer matrix over the infinitesimal intervals $I_{j}^{(\epsilon)}$, i.e. 
		\begin{equation} 
			\Gamma^{(E)}_{y^{-}_{j} \rightarrow y^{+}_{j}}  =\begin{bmatrix}
				\mathcal{U}_{j}& 0 \\
				0 & \mathcal{U}_{j} 
			\end{bmatrix}
			\label{defGAMMAL0j} \;,\qquad \mathcal{U}_{j} :=  \mathcal{U}_{y^{-}_{j} \rightarrow y^{+}_{j}}\;,
		\end{equation}
		and using Eq.~(\ref{eq:mxapp}) to write 
		\begin{equation} 
			\Gamma^{(E)}_{y^{+}_{j} \rightarrow y^{-}_{j+1}}  =
			{{\rm \exp}}\left\{\int_{y^{+}_{j} }^{y^{-}_{j+1} } dy \mathcal{M}_{E}^{(j)} \right\}=
			{{\rm \exp}}\left\{({y^{-}_{j+1} -y^{+}_{j} )} \mathcal{M}_{E}^{(j)} \right\}
			\label{defGAMMAL0jj+1}= \mathcal{D}_{L_j}(Q^{(j)}_E)  \;,
		\end{equation}
		where we used  the identity~(\ref{equationD}) with 
		$A= Q^{(j)}_E$ and $x=y^{-}_{j+1} -y^{+}_{j}=L_j$. 
		
		\section{Property of the scattering matrix}
		\label{appx:equalprob}
		The transmission and reflection matrices $t_E$ and $r_E$ computed for electrons injected from the left lead,  contribute in defining the 
		scattering matrix $S_E$ via the identity, 
		\begin{equation}
			S_{E}= \label{scatter}
			\begin{bmatrix}
				r _{E} & t'_{E} \\
				t_{E} & r'_{E}
			\end{bmatrix}\;, 
		\end{equation}
		with  $t'_{E}$ ($r'_{E}$) being  the transmission (reflection) matrix for electrons injected from the right lead.
		As discussed in Ref.~\cite{Datta1997}, for a given position-dependent magnetic field $\bm{B}(y)$, the matrix
		$S_E$ satisfies the reciprocity relation
		\begin{equation}
			\label{rever}
			S_E({\bm{B}(y)})=[S_E({-\bm{B}(y)})]^{T},
		\end{equation}
		where the superscript $T$ means transpose. 
		From Eq.~\eqref{rever} and (\ref{scatter}) follow straightforwardly that 
		\begin{equation}
			t_{E}({\bm{B}(y)})=[t'_E({-\bm{B}(y)})]^{T}, \qquad \quad  r_{E}({\bm{B}(y)})=[r'_E({-\bm{B}(y)})]^{T}\;. 
			\label{eq:transmissionB}
		\end{equation}
		Now we will show that, under two specific conditions, the scattering problem for electrons that are injected from the left lead with $\bm{B}(y)$ is formally equivalent to the scattering problem for electrons that are injected from the right lead with the reverse field, $-\bm{B}(y)$.
		More precisely, if these two conditions are satisfied, the system in Eq.~\eqref{eq:system} is identical for the two scattering problems, meaning that
		\begin{equation}
			t_E({\bm{B}(y)})=t'_{E}({-\bm{B}(y)}).
			\label{eq:propertyt}
		\end{equation}
		The first condition is that
		$[\Omega_y]_{\bm{B}(y)}=[\Omega_{y'}]_{\pm \bm{B}(y')}$
		where $y\in[y_{\rm L},y_{\rm R}]$ and $y'=:y_{\rm L}+y_{\rm R}-y$ so that $y'\in[y_{\rm R},y_{\rm L}]$.
		In other words, the Zeeman eigenvalues, contained in the matrix $\Omega_y$, must be symmetric with respect to the middle point of the scattering region. The second condition is that $\mathcal{U}_{y_{\rm L} \to y}|_{\bm{B}(y)}=\mathcal{U}_{y_{\rm R} \to y'}|_{-\bm{B}(y')}$, which holds true if $\theta_y|_{\bm{B}(y)}=\theta_{y'}|_{-\bm{B}(y')}$, see Eq.~\eqref{eq:holo}. Combining Eq.~\eqref{eq:transmissionB} with Eq.~\eqref{eq:propertyt}, it is straightforward to find
		\begin{equation}
			t_{E}({\bm{B}(y)})=[t_{E}(\bm{B}(y)]^T  \qquad \Longrightarrow \qquad [t_E]_{0,1}=[t_E]_{1,0}\;. 
			\label{eq:ttT}
		\end{equation}
		Notice that in the Example II  (Sec.\ref{sec:expII}) both conditions are satisfied, leading to the property in Eq.~\eqref{eq:ttT}. On the other hand, in the Example I (Sec.\ref{sec:expI}), the first condition is satisfied, while the second one is not. Indeed, we find $\abs{\theta_y}_{\bm{B}(y)}=\abs{\theta_{y'}}_{-\bm{B}(y')}$, meaning that $\abs{\mathcal{U}_{y_{\rm L} \to y}}_{\bm{B}(y)}=\abs{\mathcal{U}_{y_{\rm R} \to y'}}_{-\bm{B}(y')}$. 
		Although we are not able to prove it, numerically in this case we find 
		$|[t_E]_{0,1}|^2=|[t_E]_{1,0}|^2$.
		
		\section{Magnetic wall}
		\label{appx:magneticwall}
		In this section, we define an analytically solvable scattering problem, and we refer to this configuration as a \textit{magnetic wall}.
		The leads are subjected to two magnetic fields of equal magnitude, $B_0$, pointing in different directions, $\bm{n}_{\rm L}$ and $\bm{n}_{\rm R}$:
		\begin{equation}
			\bm{B}_{\rm L}=B_0 \bm{n}_{\rm L},\quad \bm{B}_{\rm R}=B_0 \bm{n}_{\rm R}.
		\end{equation}
		In the scattering region (length $L$), the magnetic field is zero.
		This configuration represents a limiting case of Scheme I (Sec.~\ref{sec:expI}) and Scheme II (Sec.~\ref{sec:expII}). In both schemes, the magnetic wall configuration is achieved when $q_1\rightarrow \infty$ with $q_2=0$, or when $q_2\rightarrow \infty$ with $q_1=0$. Specifically, Scheme I reduces to a magnetic wall with $\bm{n}_{\rm L}=\bm{n}_3$ and $\bm{n}_{\rm R}=-\bm{n}_3$, while Scheme II reduces to a magnetic wall with $\bm{n}_{\rm L}=\bm{n}_3$ and $\bm{n}_{\rm R}=\bm{n}_1$.
		Analytical solutions exist for the reflection and transmission matrices, $r_E^{\text{I}/\text{II}}(L)$ and $t_E^{\text{I}/\text{II}}(L)$, of the magnetic wall configuration in Schemes I and II. These solutions (valid for any magnetic field direction in the leads) are found by matching the spinor wavefunction and its first $y$-derivative at $y=y_{\rm L}$ and $y=y_{\rm R}$. Due to space constraints, we do not provide the analytical forms for Scheme I. Instead, we present the Hilbert-Schmidt norms $\|t^{\text{I}}_E(L)-\mathcal{U}_{y_{\rm L} \to y_{\rm R}}\|_{\rm HS}$ in Fig.~\ref{fig:distI} (red dashed curve, main plots) and $\|r^{\text{I}}_E(L)\|_{\rm HS}$ (red dashed curve, inset plots). Analogously, Fig.~\ref{fig:distII} displays the distances $\|t^{\text{II}}_E(L)-\mathcal{U}_{y_{\rm L} \to y_{\rm R}}\|_{\rm HS}$ (red dashed curve, main plots) and $\|r^{\text{II}}_E(L)\|_{\rm HS}$ (red dashed curve, inset plots) for Scheme II.
		
	\end{widetext}

	\begin{figure*}[!htbp]
		\centering
		\includegraphics[width=0.3\textwidth]{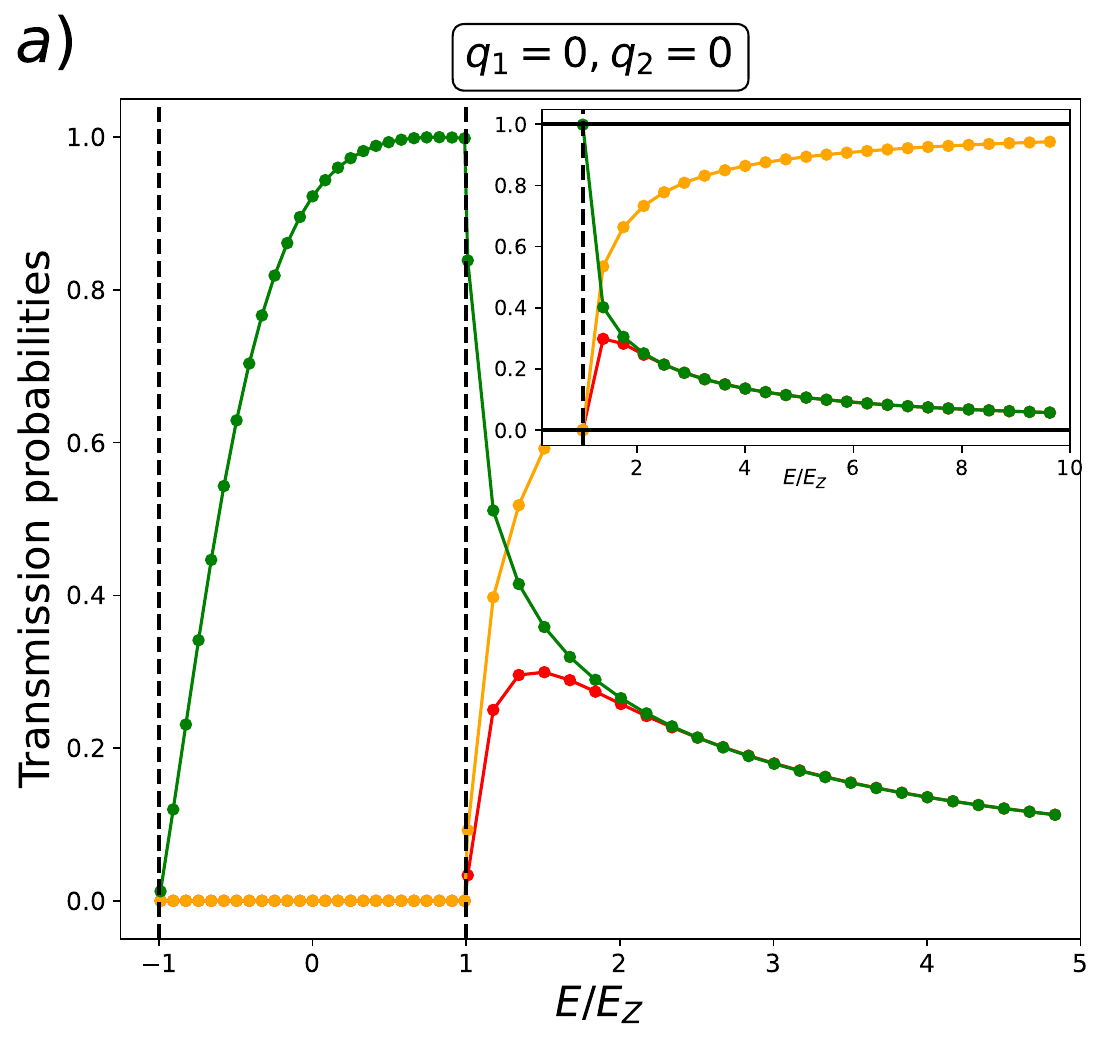}
		\includegraphics[width=0.3\textwidth]{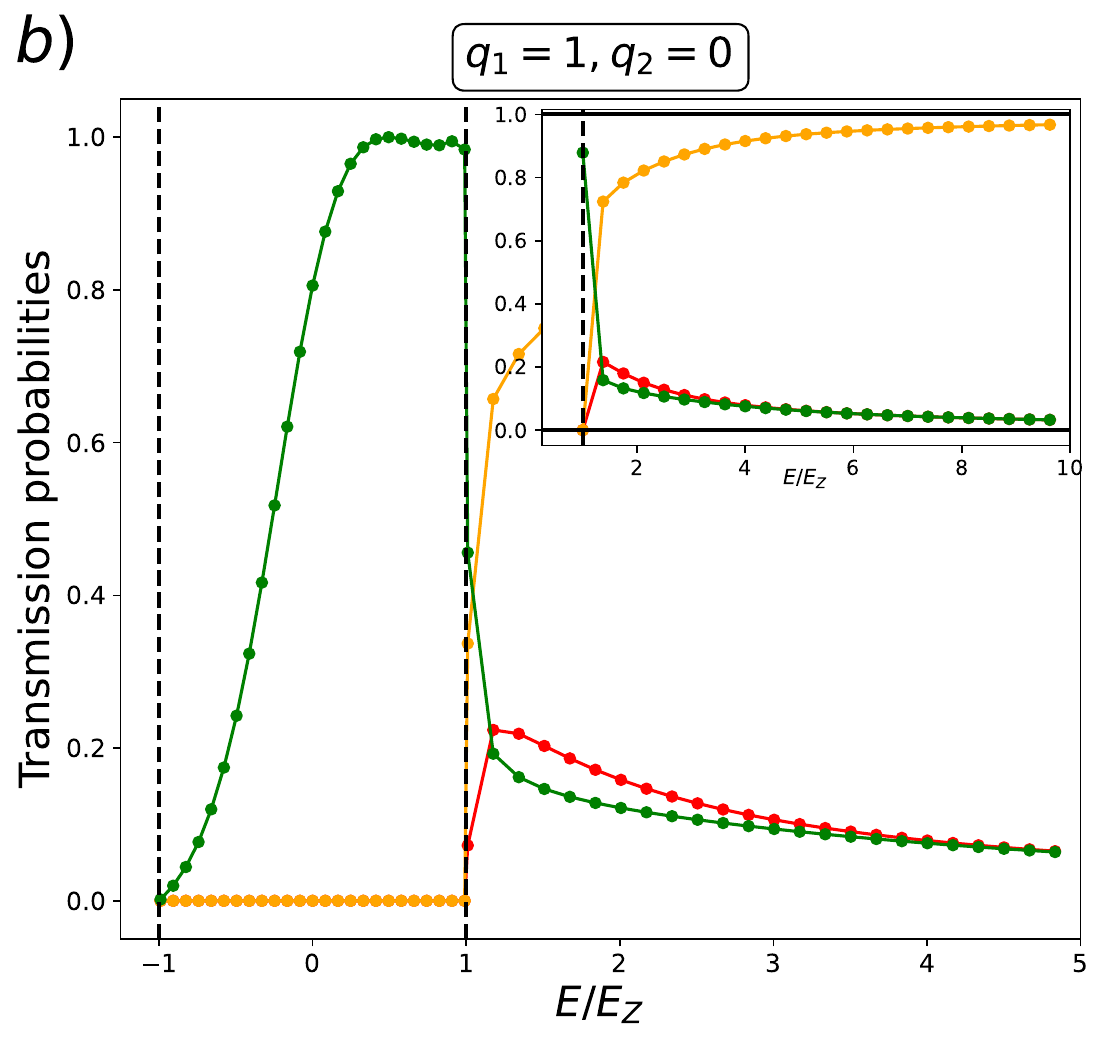}
		\includegraphics[width=0.3\textwidth]{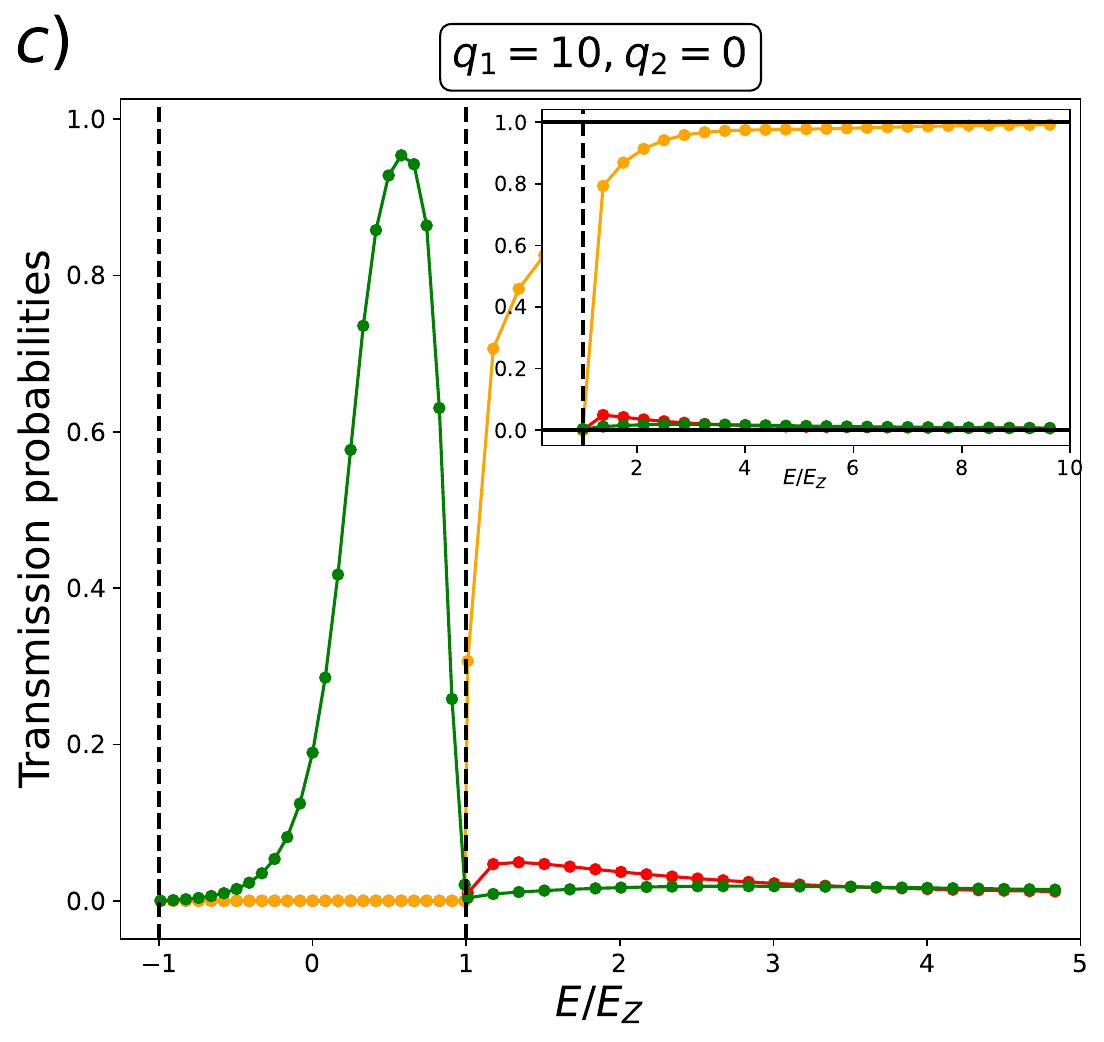}
		\includegraphics[width=0.3\textwidth]{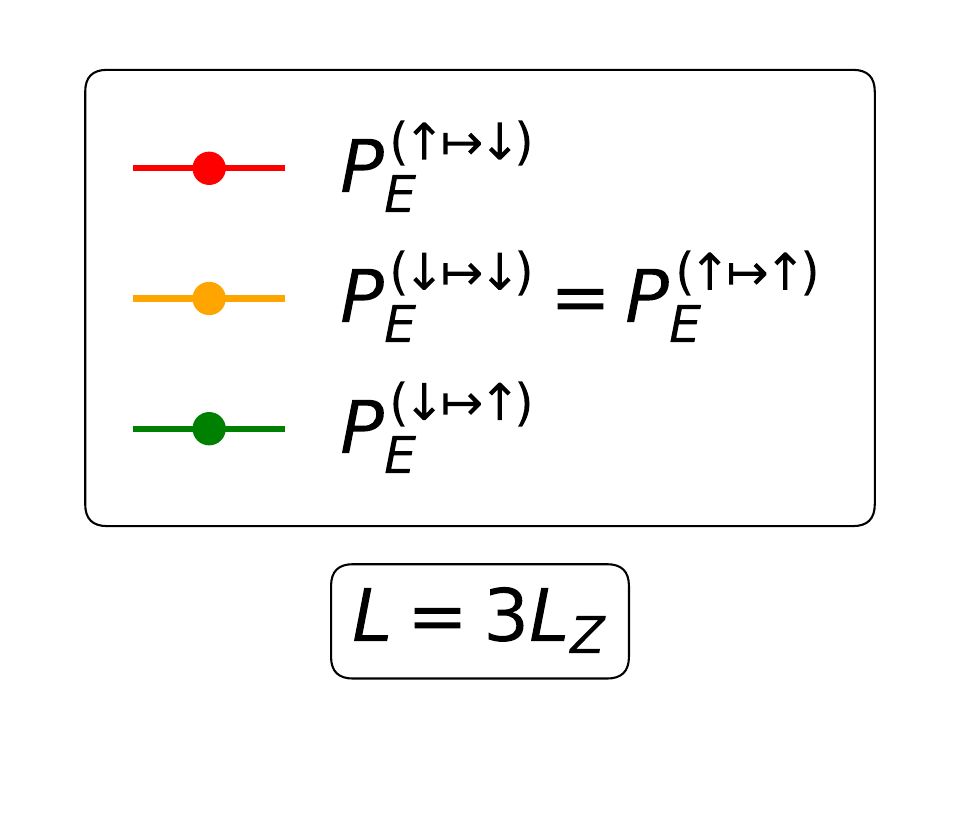}
		\includegraphics[width=0.3\textwidth]{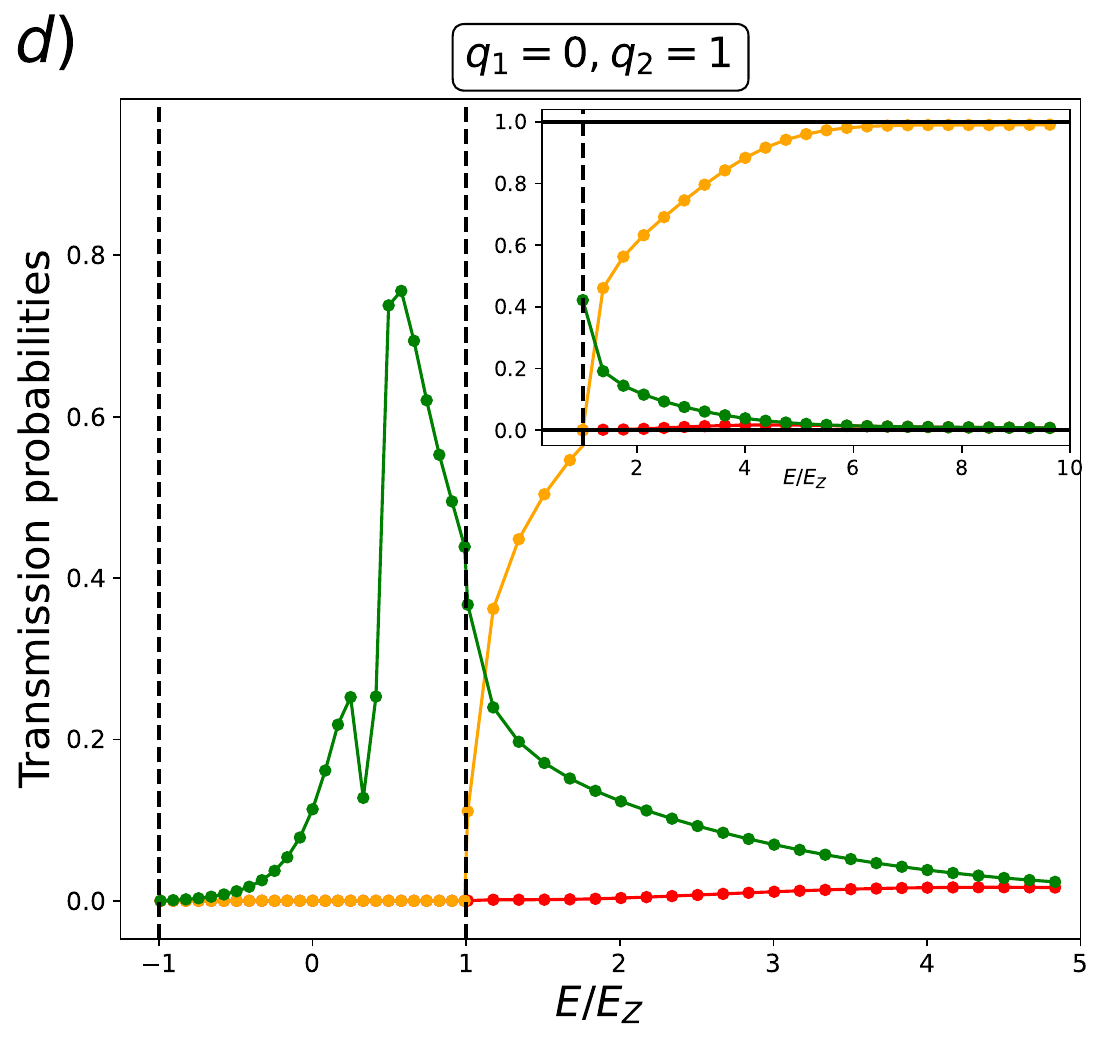}
		\includegraphics[width=0.3\textwidth]{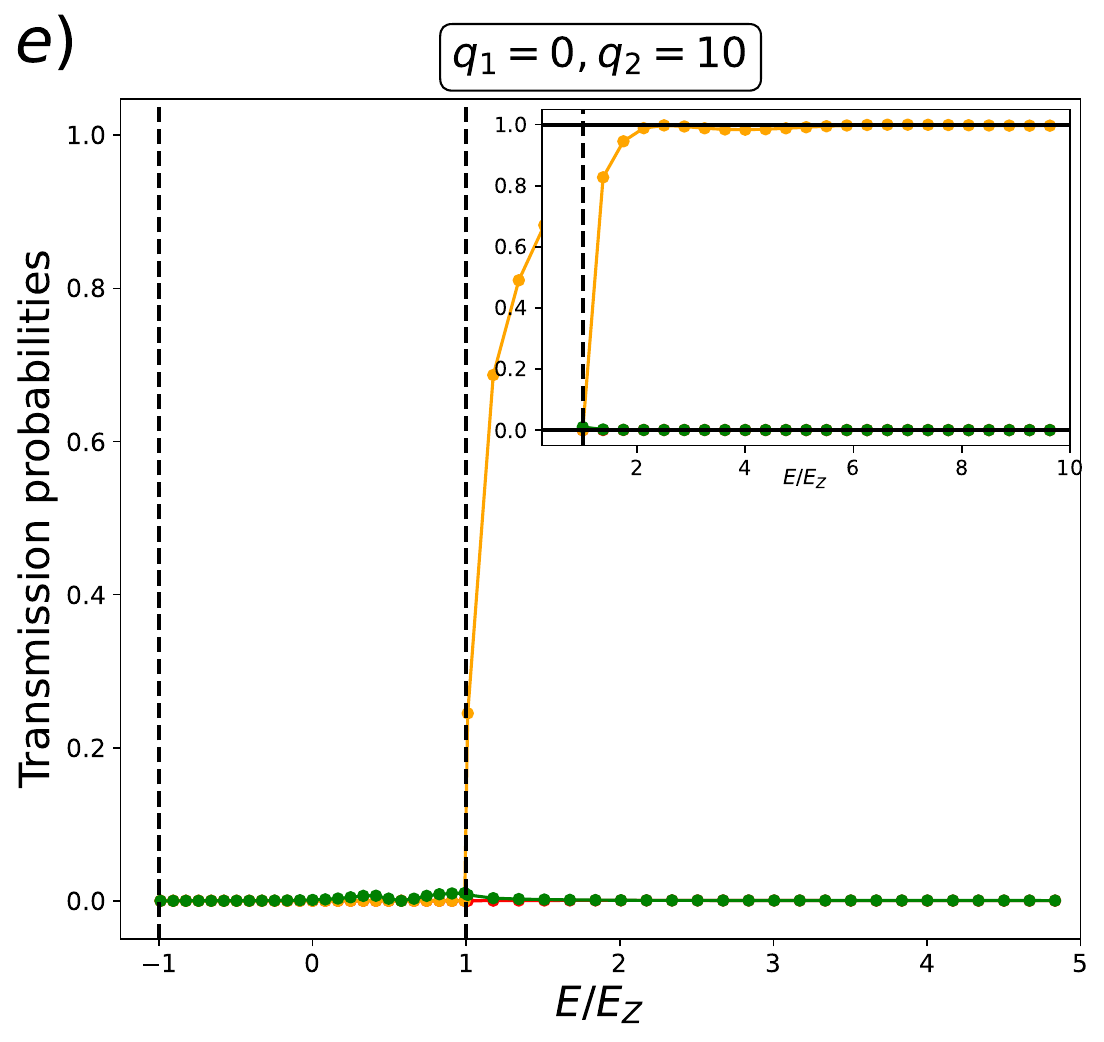}
		\caption{
			Transmission probabilities as a function of injection energy for Scheme I (Sec.~\ref{sec:expI}), with $L=3L_Z$. Each panel corresponds to different values of the parameters $q_1$ and $q_2$, which are specified at the top of each panel.
			The value of $L=3L_Z$ was selected because it enables nearly perfect spin flip ($P_E^{\uparrow \mapsto \downarrow} \approx 1$) over the energy range from 0 to $E_{\rm Z}$ when both parameters are set to zero ($q_1=0$, $q_2=0$), as illustrated in panel \textsl{a)}.
			The insets in all plots display an extended energy range, allowing access to the high-energy limit where $E / E_{\rm Z} \rightarrow \infty$. In this limit, for all cases, the red and green curves ($P_E^{\uparrow \mapsto \downarrow}$ and $P_E^{\downarrow \mapsto \uparrow}$) approach zero, while the orange curve ($P_E^{\uparrow \mapsto \uparrow} = P_E^{\downarrow \mapsto \downarrow}$) approaches 1. This behavior confirms the prediction given by Eq.~\eqref{eq:tsolfin1EINF}, where the transmission matrix at large energies coincides with the Berry operator. Notably, this result is independent of $q_1$ and $q_2$, as the Berry operator depends solely on the magnetic field components in the leads [see Eq.~\eqref{eq:holo}].
		}
		\label{fig:PT_I}
	\end{figure*}
	\begin{figure*}[!htbp]
		\centering
		\includegraphics[width=0.49\textwidth]{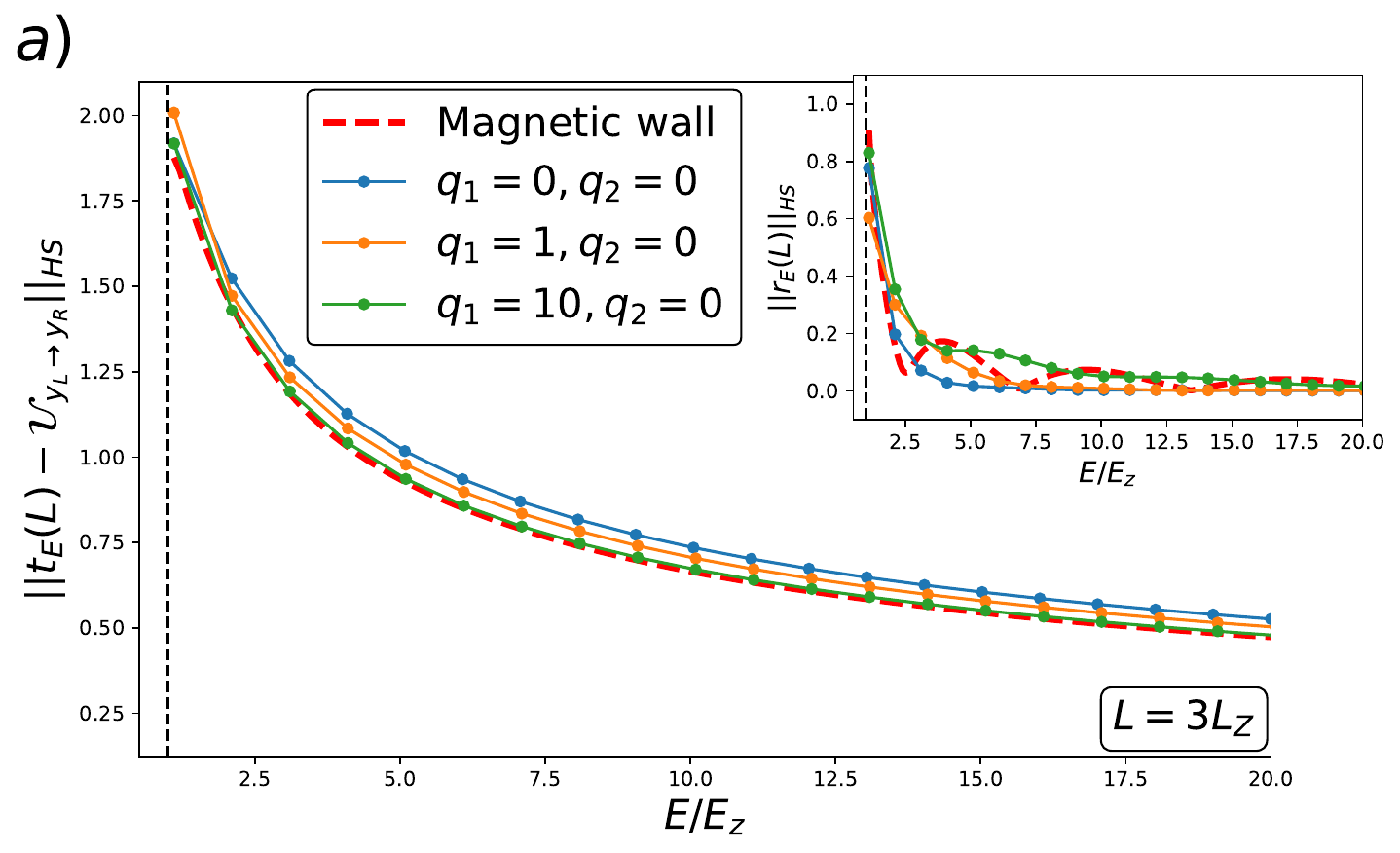}
		\includegraphics[width=0.49\textwidth]{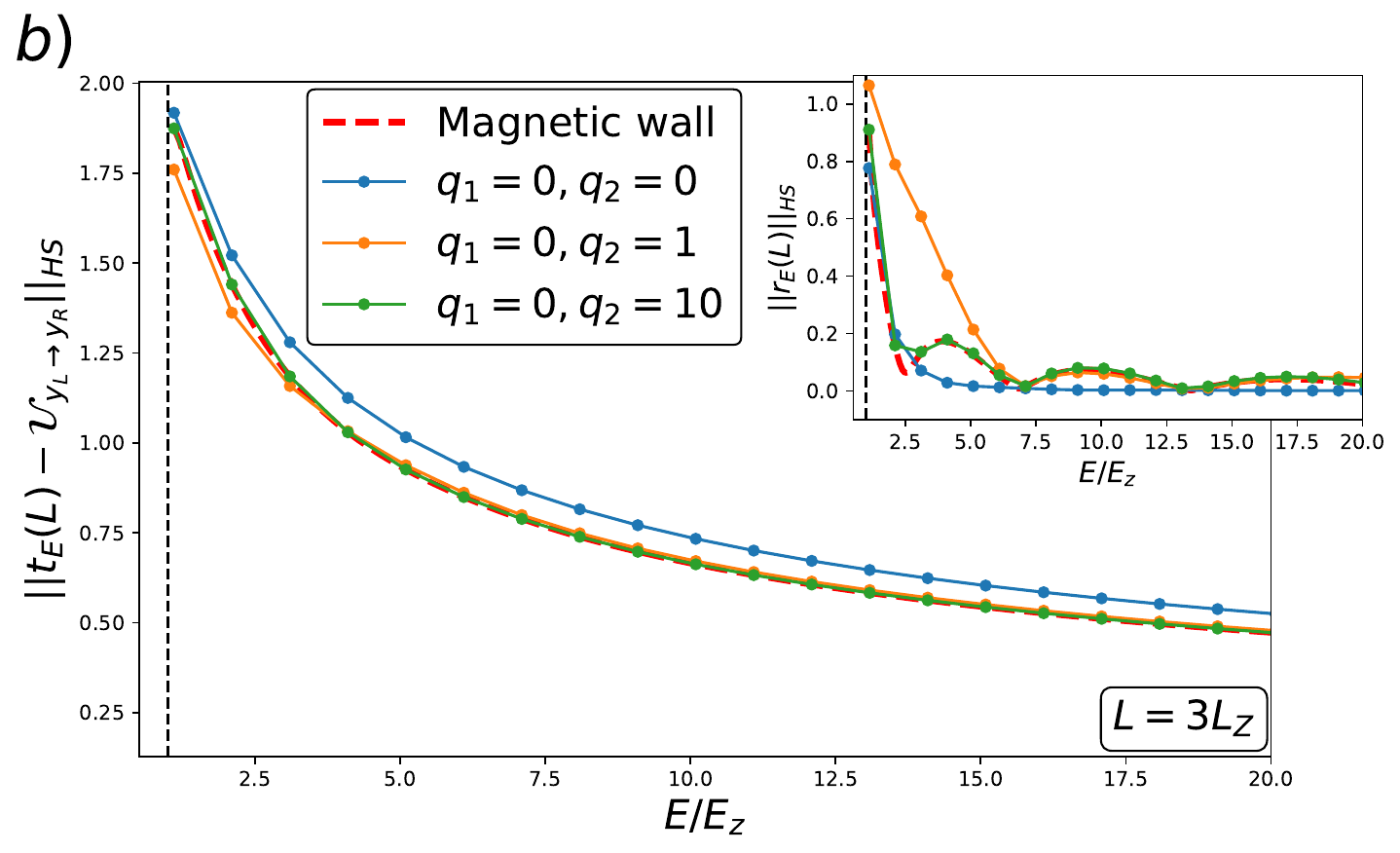}
		\caption{
			Scheme I, numerical results of the Hilbert-Schmidt norms $\|t_E(L)-\mathcal{U}_{y_{\rm L}\to y_{\rm R}}\|_{\rm HS}$ and $\|r_E(L)\|_{\rm HS}$ (blue, orange, and green dotted curves) as a function of injection energy $E$. The dashed red lines represent the exact solution for the magnetic wall configuration in Scheme I (Appx.~\ref{appx:infinitesimalscattering}), showing $\|t^{I}_E(L)-\mathcal{U}_{y_{\rm L} \to y_{\rm R}}\|_{\rm HS}$ and $\|r^{I}_E(L)\|_{\rm HS}$. In panel \textsl{a)} we vary $q_1$ ($q_1 = 0, 1, 10$) with $q_2 = 0$ fixed, while in panel \textsl{b)} we fixes $q_1 = 0$ and varies $q_2$ ($q_2 = 0, 1, 10$). In both cases, the distance between the dashed red curve and the dotted curves decreases as $q_1$ or $q_2$ increases, confirming that Scheme I (Sec.~\ref{sec:expI}) approaches the magnetic wall configuration ($\bm{n}_{\rm L}=\bm{n}_3$, $\bm{n}_{\rm R}=-\bm{n}_3$) in the limit $q_1\to \infty$ (with $q_2=0$) or $q_2 \to \infty$ (with $q_1=0$), as discussed in Appx.~\ref{appx:magneticwall}. For the sake fo completeness, fixing \(q_1=q_2=0\) and \(\beta_0(y)=1\), Equation~\eqref{eq:bzI} gives \(B_1(y) = B_0 \, \sin^2\!\Big(\frac{\pi (y - y_{\rm L})}{y_{\rm R} - y_{\rm L}}\Big)\) and \(B_3(y) = B_0 \, \cos\!\Big(\frac{\pi (y - y_{\rm L})}{y_{\rm R} - y_{\rm L}}\Big)\), which correspond to the magnetic field vector of Fig.~\ref{fig:ex1}\textsl{a)}.}
		\label{fig:distI}    
	\end{figure*}
	
	\begin{figure*}[!htbp]
		\centering
		\includegraphics[width=0.3\textwidth]{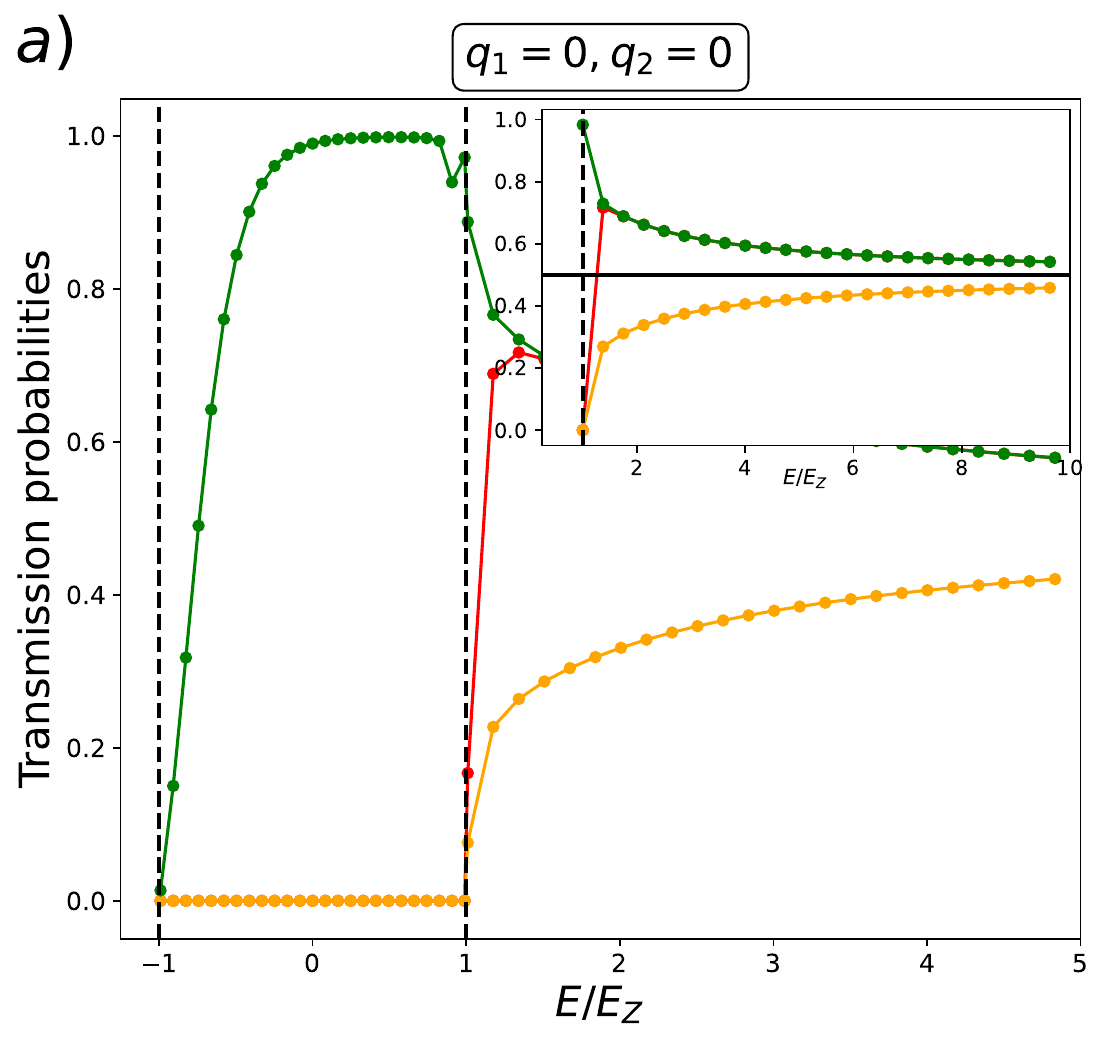}
		\includegraphics[width=0.3\textwidth]{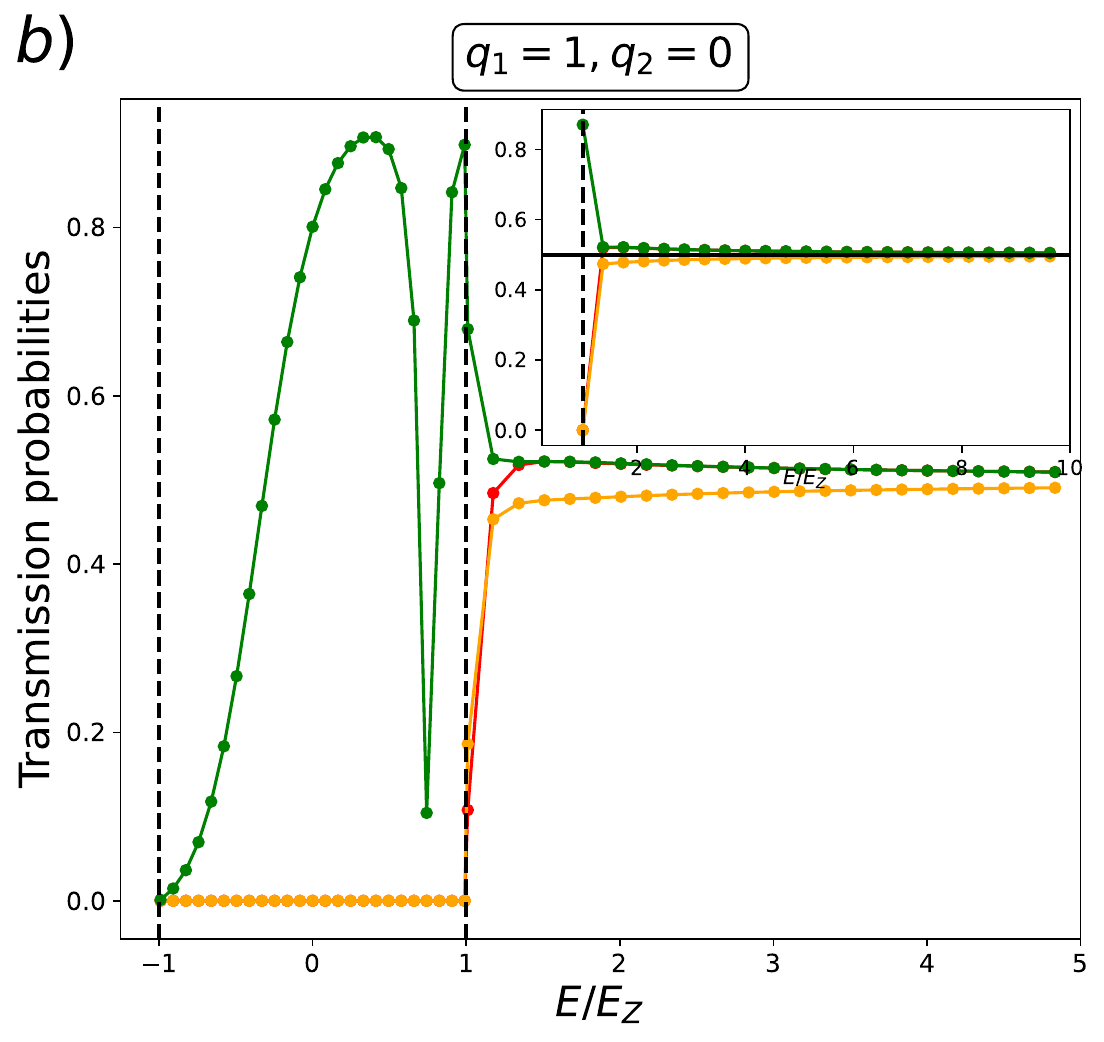}
		\includegraphics[width=0.3\textwidth]{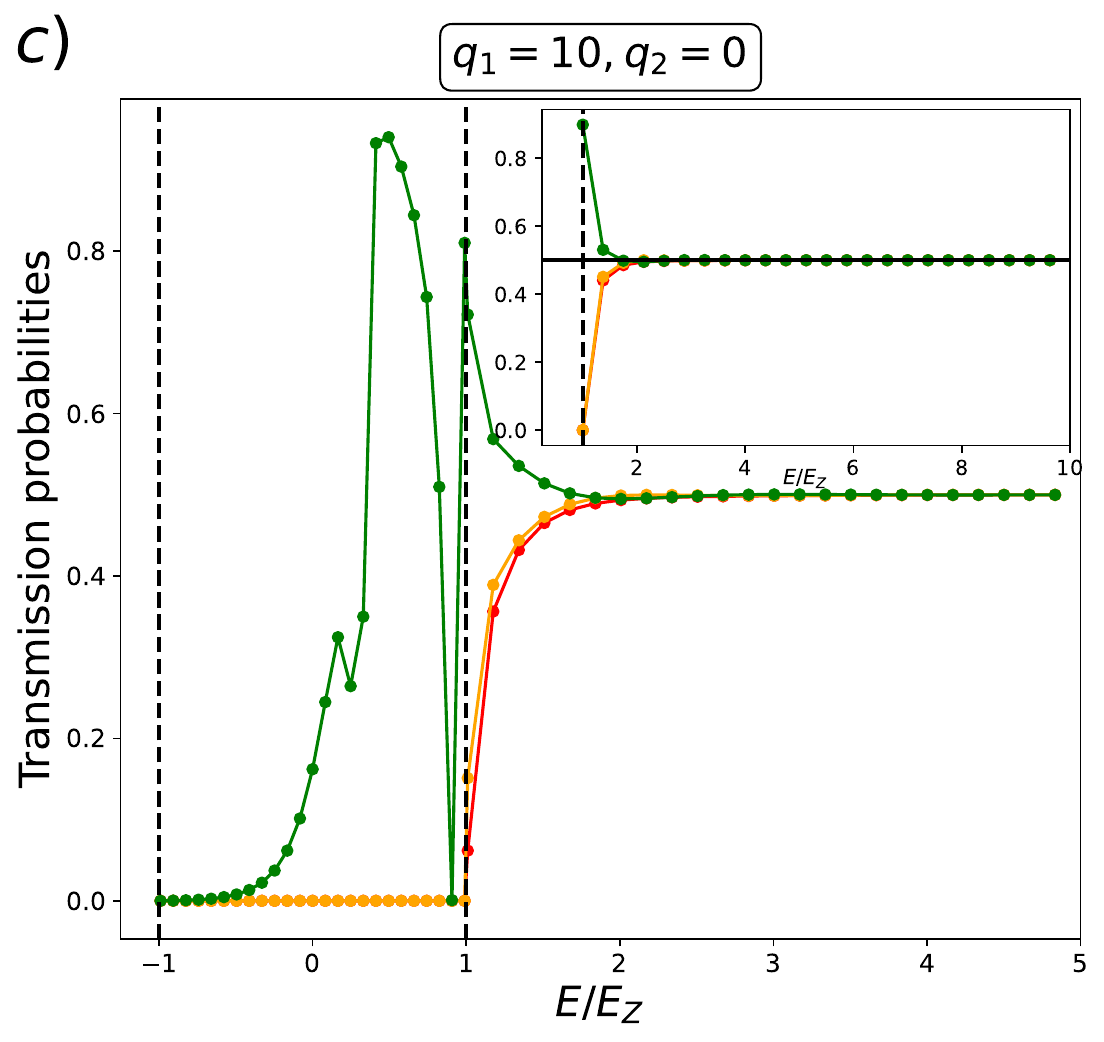}
		\includegraphics[width=0.3\textwidth]{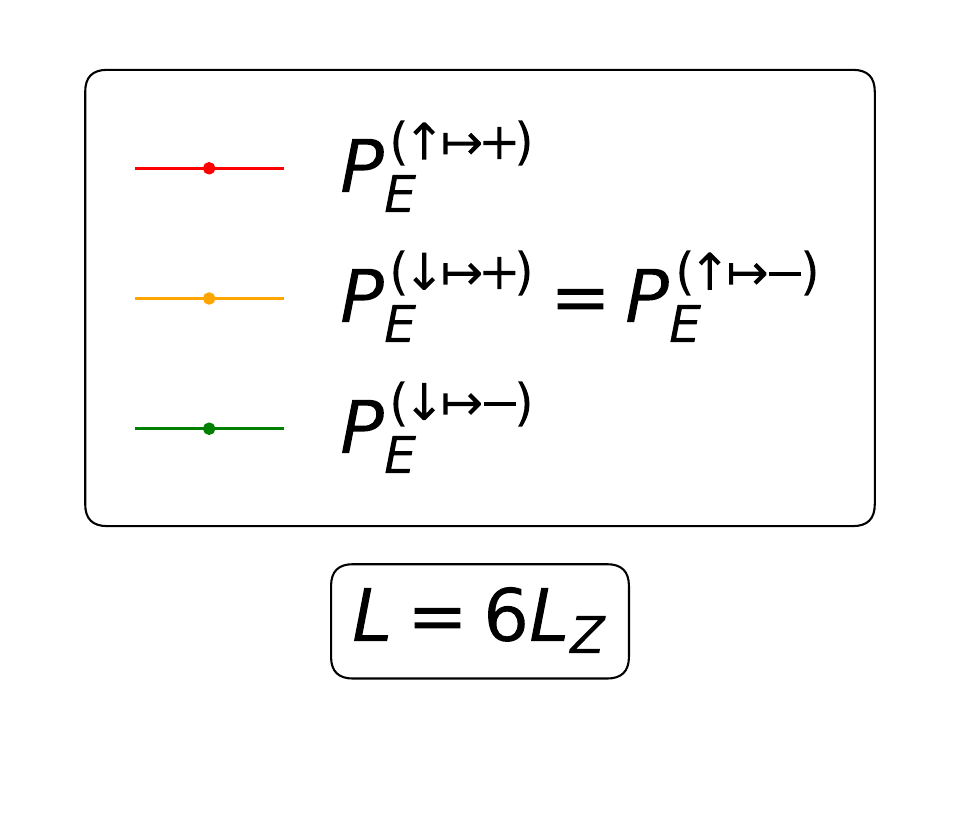}
		\includegraphics[width=0.3\textwidth]{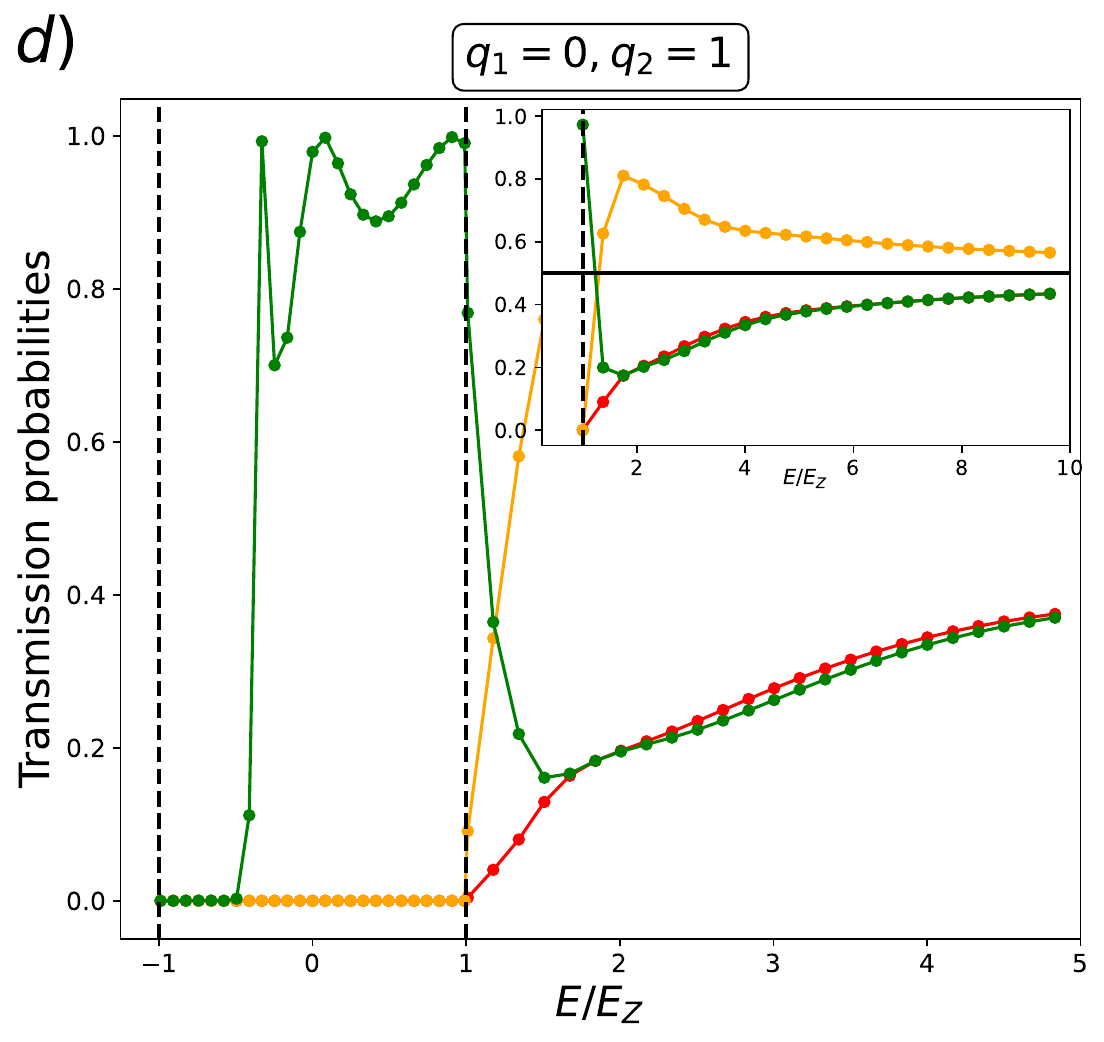}
		\includegraphics[width=0.3\textwidth]{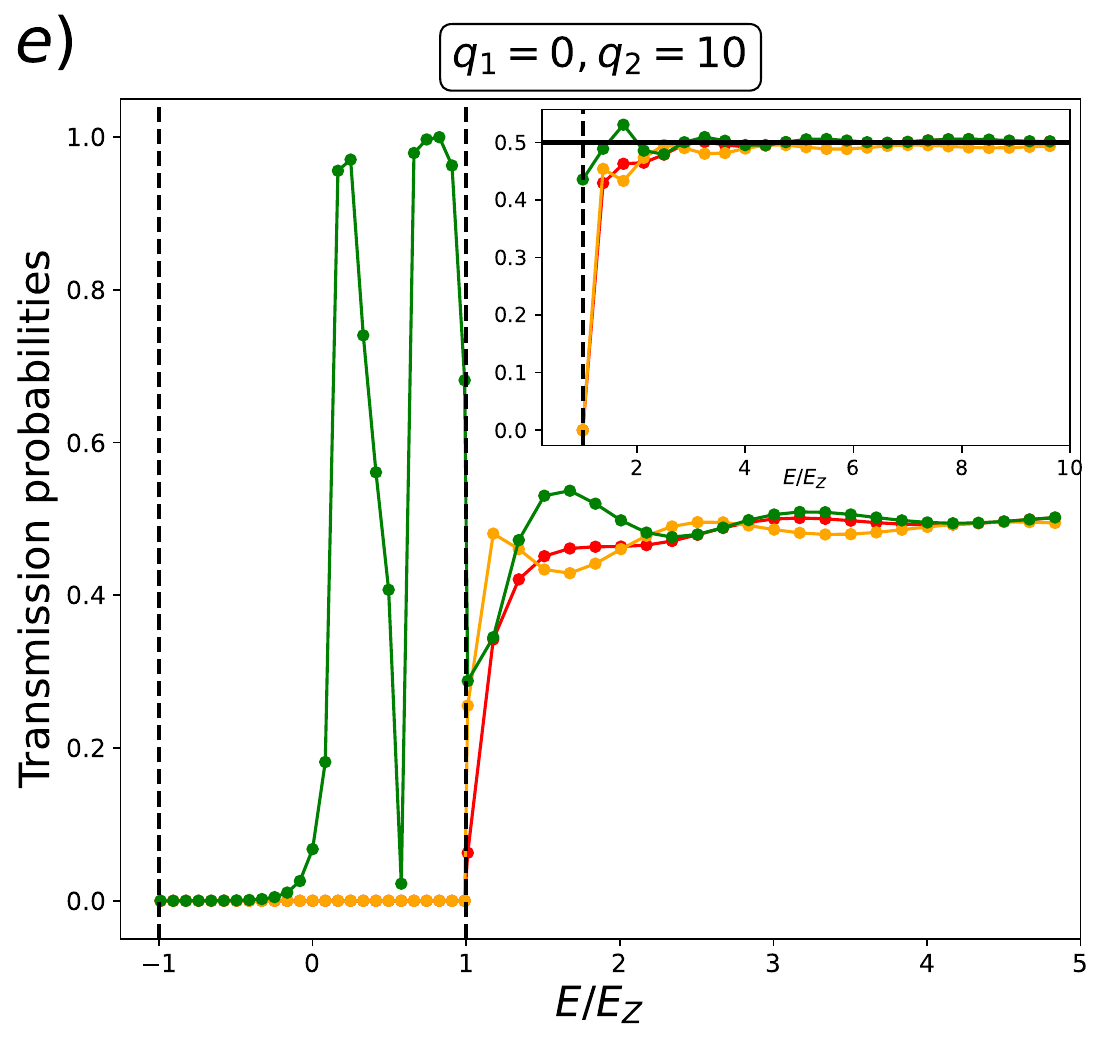}
		\caption{
			Transmission probabilities as a function of injection energy for Scheme II (Sec.~\ref{sec:expII}), with $L=6L_Z$. Each panel corresponds to different values of the parameters $q_1$ and $q_2$, which are specified at the top of each panel.
			The value of $L=6L_Z$ was selected because it enables nearly perfect spin mixing ($P_E^{\downarrow \mapsto -} \approx 1$) over the energy range from 0 to $E_{\rm Z}$ when both parameters are set to zero ($q_1=0$, $q_2=0$), as illustrated in panel \textsl{a)}.
			The insets in all plots display an extended energy range, allowing access to the high-energy limit where $E / E_{\rm Z} \rightarrow \infty$. In this limit, for all cases, the red, green and orange curves ($P_E^{\uparrow \mapsto +}$, $P_E^{\downarrow \mapsto -}$ and $P_E^{\downarrow \mapsto +} = P_E^{\uparrow \mapsto -}$) approach 1/2. This behavior confirms the prediction given by Eq.~\eqref{eq:tsolfin1EINF}, where the transmission matrix at large energies coincides with the Berry operator. Notably, this result is independent of $q_1$ and $q_2$, as the Berry operator depends solely on the magnetic field components in the leads [see Eq.~\eqref{eq:holo}].}
		\label{fig:PT_II}
	\end{figure*}
	\begin{figure*}[!htbp]
		\centering
		\includegraphics[width=0.48\textwidth]{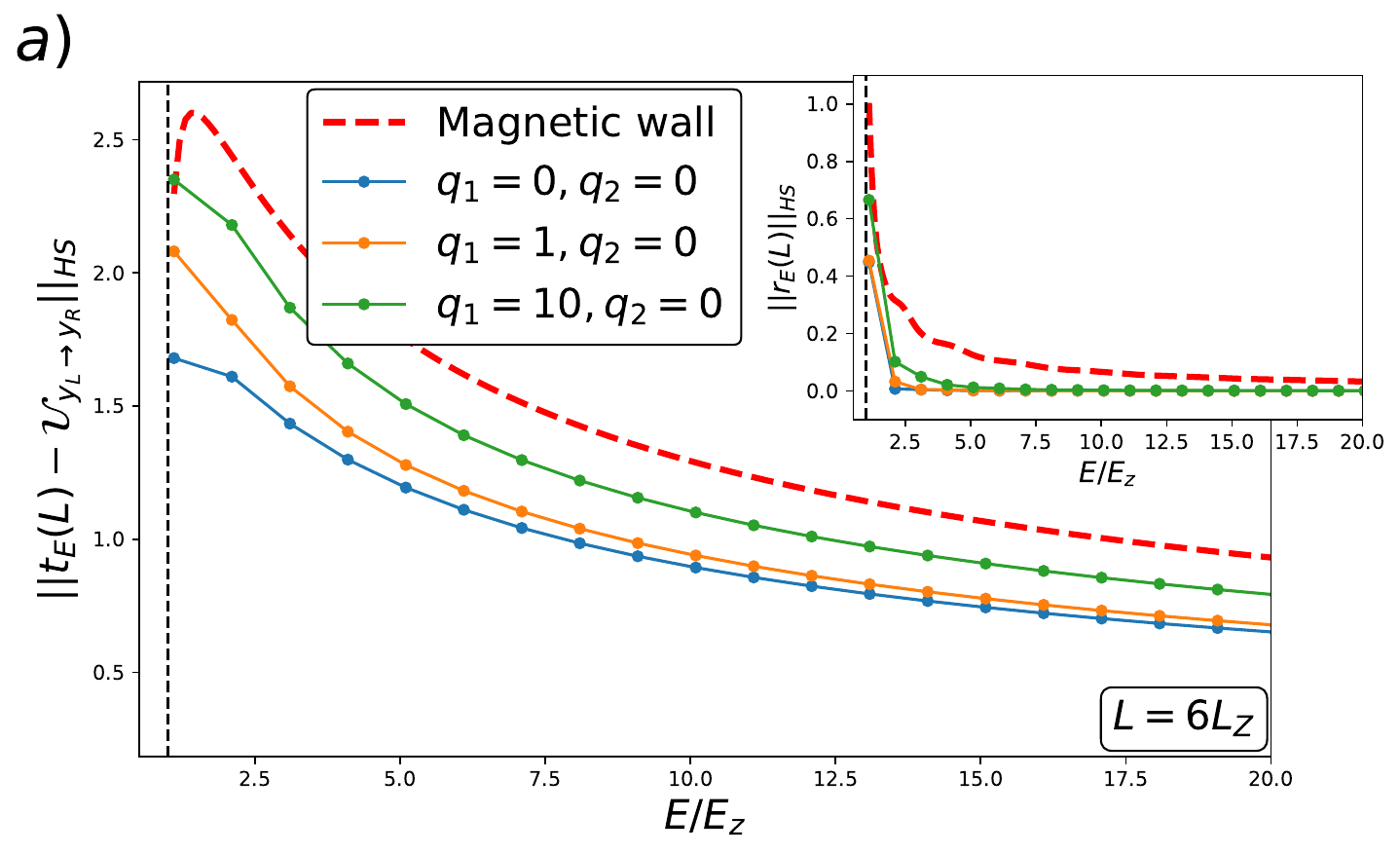}
		\includegraphics[width=0.48\textwidth]{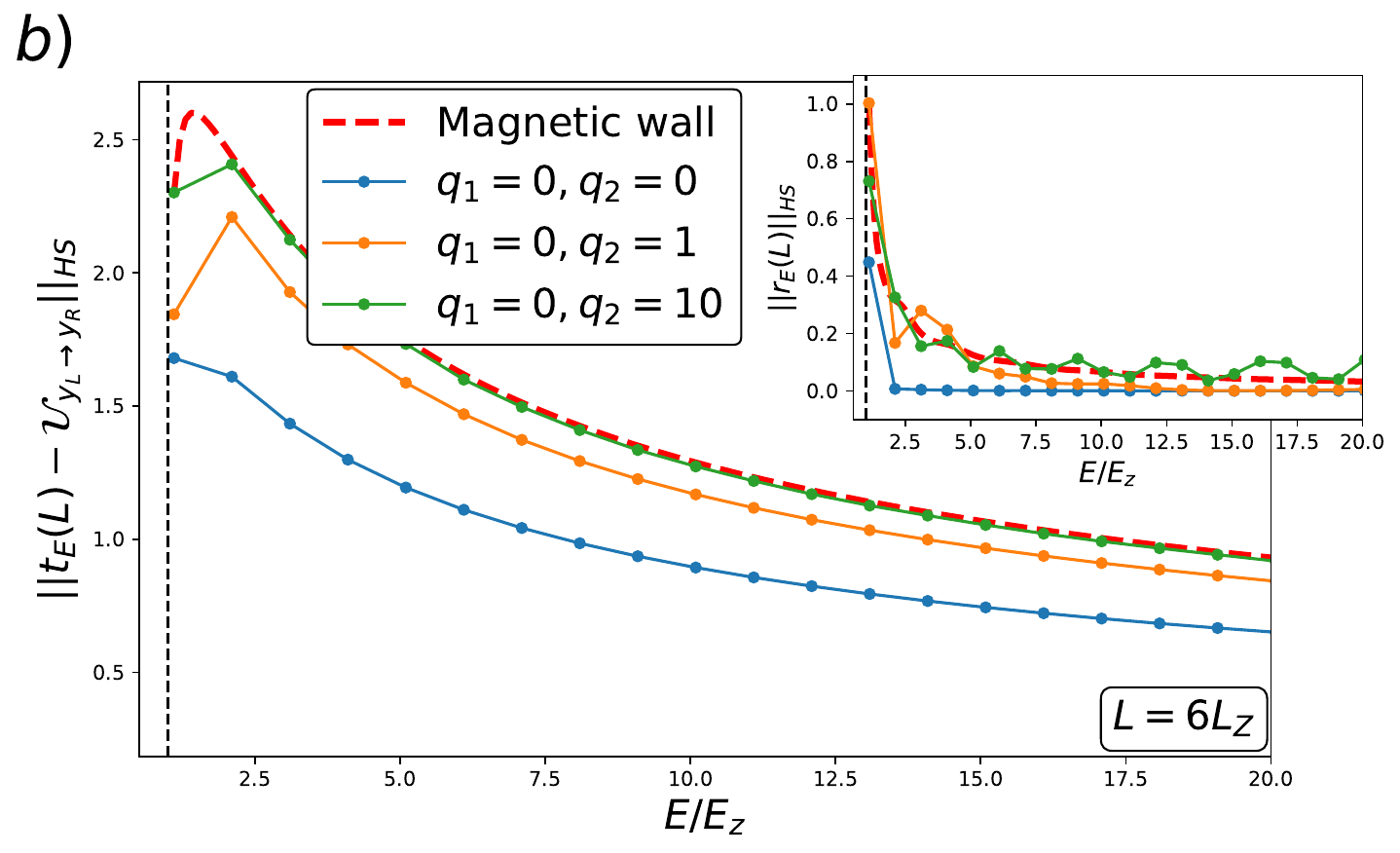}
		\caption{Scheme II, numerical results of the Hilbert-Schmidt norms 
			$\|t_E(L)-\mathcal{U}_{y_{\rm L}\to y_{\rm R}}\|_{\rm HS}$ and 
			$\|r_E(L)\|_{\rm HS}$ (blue, orange, and green dotted curves) as a function of 
			injection energy $E$. The dashed red lines represent the exact solution for the 
			magnetic wall configuration in Scheme II (Appx.~\ref{appx:magneticwall}), 
			showing $\|t^{\text{II}}_E(L)-\mathcal{U}_{y_{\rm L} \to y_{\rm R}}\|_{\rm HS}$ 
			and $\|r^{\text{II}}_E(L)\|_{\rm HS}$. In panel \textsl{a)} we vary the integer 
			$q_1$ ($q_1 = 0, 1, 10$) while keeping $q_2 = 0$ constant. In panel 
			\textsl{b)} we fix $q_1 = 0$ and vary $q_2$ ($q_2 = 0, 1, 10$). In both cases, 
			the distance between the dashed red curve and the dotted curves decreases as 
			either $q_1$ or $q_2$ increases. However, the green curve in the inset of panel 
			\textsl{b)} exhibits noticeable jumps, causing it to diverge from the red dashed 
			curve. Although we did not plot the distance $\|r_E(L)\|_{\rm HS}$ for larger 
			values of $q_2$ with $q_1=0$, we verified that this distance approaches the red 
			dashed curve $\|r^{\text{II}}_E(L)\|_{\rm HS}$ as $q_2 \to \infty$. This behavior 
			supports our earlier discussion in Appx.~\ref{appx:magneticwall}: in the 
			limiting cases where $q_1\to\infty$ (with $q_2=0$) and $q_2 \to \infty$ (with 
			$q_1=0$), Scheme II (Sec.~\ref{sec:expII}) reduces to the magnetic wall 
			configuration with $\bm{n}_{\rm L}=\bm{n}_3$ and $\bm{n}_{\rm R}=\bm{n}_1$. For the sake of completeness, fixing $q_1=q_2=0$ and $\beta_1(y)=\beta_2(y)=0$, 
				Equation~\eqref{eq:bzII} gives $B_1(y) = B_0 \, \sin^2\!\Big(\frac{\pi (y - 
					y_{\rm L})}{2(y_{\rm R} - y_{\rm L})}\Big)$, $B_3(y) = B_0 \, \cos^2\!\Big(
				\frac{\pi (y - y_{\rm L})}{2(y_{\rm R} - y_{\rm L})}\Big)$, which correspond to the magnetic 
				field vector of Fig.~\ref{fig:ex1}\textsl{b)}.}
		\label{fig:distII}
	\end{figure*}
	

\begin{thebibliography}{99}
\bibitem{note-berry-op}
		The Berry operator was firstly introduced in Ref~\cite{adiabatic-nonabelian}. 
\bibitem{windings}{This conclusion remains valid even when the magnetic field undergoes multiple complete rotations (windings) within the scattering region. Conceptually, the scattering region can be divided into segments corresponding to each winding: for each segment, in the high-energy regime, the transmission matrix reduces to the Berry operator of the system, namely the identity matrix. Since the transmission is unitary and no reflection occurs, the total transmission through all segments remains the product of the transmission matrices of all segments, i.e. the identity matrix.} 
\end{thebibliography}
\end{document}